\documentclass[reqno,a4paper,11pt]{article}
\usepackage{amsfonts,amssymb,amsmath,graphicx}
\usepackage{theorem}
\usepackage{url, cite}
\usepackage{enumerate}
\usepackage{vmargin}
\usepackage{todonotes} 
\usepackage[title]{appendix}
\usepackage{cancel}
\usepackage{tikz} 
\usetikzlibrary{arrows}
\usetikzlibrary{positioning,arrows,chains,matrix,scopes,fit,decorations.markings,decorations}
\tikzstyle{block}=[draw opacity=0.7,line width=1.4cm]
\newcommand{\bgreen}{\begin{color}{green}}

\newtheorem{definition}{Definition}[section]

\newtheorem{proposition}[definition]{Proposition}

\newtheorem{theorem}[definition]{Theorem}

\newtheorem{lemma}[definition]{Lemma} \numberwithin{equation}{section}
\newtheorem{rmk}[definition]{Remark}

\newcommand{\beq}{\begin{equation}} \newcommand{\eeq}{\end{equation}}
\newcommand{\bea}{\begin{eqnarray}} \newcommand{\eea}{\end{eqnarray}}
\newcommand{\beano}{\begin{eqnarray*}}
  \newcommand{\eeano}{\end{eqnarray*}}
\newcommand{\bma}{\begin{pmatrix}} \newcommand{\ema}{\end{pmatrix}}

\def\cA{{\cal A}} \def\cB{{\cal B}} \def\cC{{\cal C}}    
    \def\cL{{\cal L}}  
 \def\cP{{\cal P}} \def\cQ{{\cal Q}} \def\cR{{\cal
    R}}

\def\ff{{\mathfrak f}} \def\fg{{\mathfrak g}} \def\fh{{\mathfrak h}}
  \def\fp{{\mathfrak p}}
 \def\fs{{\mathfrak s}} 
\def\fq{{\mathfrak q}}

 \newcommand{\ZZ}{{\mathbb Z}}

\newcommand{\wh}[1]{\widehat{#1}} 
\newcommand{\wt}[1]{\widetilde{#1}}

\newcommand{\prf}{\underline{Proof:}\ } \newcommand{\finprf}{\null
  \hfill {\rule{5pt}{5pt}}} \newcommand{\ie}{{\it i.e.}\ }
\newcommand{\eg}{{\it e.g.}\ }
\newcommand{\cf}{{\it c.f.}\ }

\makeatletter \def\@xfootnote[#1]{%
  \protected@xdef\@thefnmark{#1}%
  \@footnotemark\@footnotetext} \makeatother
  
\newcommand{\be}{\begin{equation}}
\newcommand{\ee}{\end{equation}}
\newcommand{\vf}{\varphi} 

\newcommand{\bz}{\boldsymbol{z}}

\newcommand{\ba}{\boldsymbol{a}}
\newcommand{\bbb}{\boldsymbol{b}}
\newcommand{\bc}{\boldsymbol{c}}
\newcommand{\bdd}{\boldsymbol{d}}

\newcommand{\bff}{\boldsymbol{f}}
\newcommand{\ld}{\lambda}
\newcommand{\bse}{\begin{subequations}}
\newcommand{\bM}{\boldsymbol{M}}
\newcommand{\ese}{\end{subequations}}

\newcommand{\pl}{\partial} 
\newcommand{\dd}{\delta}

\newcommand{\wb}[1]{\overline{#1}}

\newcommand{\bred}{\begin{color}{red}}

\def\hypotilde#1#2{\vrule depth #1 pt width 0pt{\smash{{\mathop{#2}
\limits_{\displaystyle\widetilde{}}}}}}
\def\hypohat#1#2{\vrule depth #1 pt width 0pt{\smash{{\mathop{#2}
\limits_{\displaystyle\widehat{}}}}}}

\def\hypobar#1#2{\vrule depth #1 pt width 0pt{\smash{{\mathop{#2}
\limits_{\displaystyle\bar{}}}}}}

\title{$\mathrm{PGL}(3)$-invariant integrable systems from
factorisation of linear differential and difference operators}
\date{\empty}
\author{Frank Nijhoff
  \,,   ~~~Linyu Peng
  \,, ~~~Cheng Zhang
  \,, ~~~Da-jun Zhang
 }

\begin{document}
\maketitle
\begin{abstract}
In this paper, we present a unified approach to constructing continuous and discrete $\mathrm{PGL}(3)$-invariant integrable systems, formulated in terms of the common dependent variables $z_1,z_2$, from linear spectral problems and their factorisation. Starting from third-order spectral problems, we first provide explicit forms of the differential and difference invariants, generalising the Schwarzian derivative and cross-ratio to the rank-$3$ setting. The factorisation induces dualities among linear spectral problems, underlying the exact discretisation and multi-dimensional consistency of the associated Boussinesq systems. Then, we derive both continuous and discrete $\mathrm{PGL}(3)$-invariant Boussinesq systems, representing natural rank-$3$ generalisations of the Schwarzian KdV and cross-ratio equations. A geometric lifting-decoupling mechanism is developed to explain the reduction of these systems to the $\mathrm{PGL}(2)$-invariant Boussinesq equations. Finally, we derive a  ${\mathrm{PGL}}(3)$-invariant system of {\em generating PDEs} together with
its Lagrangian structure, in which the lattice parameters serve as independent variables, providing the generating PDE system for the Boussinesq hierarchy. 
\medskip

\noindent{\bf Keywords:} projective differential invariants; PGL(3)-invariant systems; multi-dimensional consistency; exact discretisation; Boussinesq system
\end{abstract}

\newcommand{\zc}[1]{$\framebox{\tiny zc}$\ \textbf{\texttt{{\color{blue}\footnotesize#1}}}}

\date{\empty}


\section{Introduction}
Originating in the geometric investigations of Klein, Lie, and Hilbert, the theory of {\em 
differential invariants} sought to characterise geometric objects preserved under group 
actions. In the context of integrable systems, such invariants naturally emerge from linear 
spectral problems associated with Lax pairs, where the underlying symmetry groups act on the 
spaces of solutions. Remarkably, the nonlinear evolution equations that arise as 
``projective formulations'' of the Lax pairs typically inherit this invariance, leading to so-called ``Schwarzian'' or ``projective'' versions 
of well-known integrable models. 

The paradigmatic example is the {\em Schwarzian derivative} arising from the second-order 
Schr\"odinger spectral problem. This fundamental $\mathrm{PGL}(2)$-invariant appears in the 
``projective'' formulation of the celebrated Korteweg--de Vries (KdV) equation, known as the 
{\em Schwarzian KdV} equation \cite{KN1,Weiss}. The {\em composition rule} of the Schwarzian 
derivative offers a projective geometric perspective on the KdV hierarchy, reflecting the 
link between {\em  projective connections} and {\em Virasoro algebra} with the latter coinciding with 
the second Hamiltonian structure of the KdV hierarchy (see, for example, \cite{Dickey,  
Kirillov, segal, OT}). In the discrete realm, the {\em cross-ratio equation} \cite{NC2}\textemdash which reappeared as the Q1$(\delta=0)$ member of the Adler--Bobenko--Suris (ABS) classification \cite{ABS1}\textemdash serves as the discrete Schwarzian KdV equation \cite{N1}. It plays a central role in the theories of discrete surfaces and discrete holomorphic functions \cite{BP, AgBo, BS22}, as well as in reductions to discrete Painlev\'e 
equations \cite{DIGP,NJH}.  

More than a mere geometric reformulation, the Schwarzian KdV equation may be seen as the most fundamental member of a Miura-related chain of KdV equations following Wilson, who also gave the name ``Ur-KdV'' \cite{Wilson1988} (see \cite{N1, DepSchiff} for other ``Ur-type'' equations). Wilson explicitly explained, from a differential Galois theory perspective,  that the KdV potential arises as the $\mathrm{PGL}(2)$-invariant subfield of a differential field generated by the projective variable in terms of the Schwarzian derivative, and that the Miura chain corresponds to a tower of differential field extensions where the projective variable sits at the top. Identifying $\mathrm{PGL}(2)$ as the differential Galois group, a natural Poisson structure on the space of projective curves was constructed using the theory of Poisson--Lie groups \cite{STS}. Remarkably, this framework was extended to the discrete setting, with explicit constructions of difference field extensions associated with discrete spectral problems and Poisson structures on the space of discrete projective curves (polygons) \cite{STS, M1} (see also \cite{BW, IB}).  

Despite these advances in the rank-$2$ KdV setting, a systematic projective formulation for the rank-$3$ Boussinesq (BSQ) equations remains surprisingly underdeveloped.  Within the Gel'fand--Dikii formalism, the BSQ hierarchy is well understood \cite{DS, Dickey, GD}, and even the $\mathrm{PGL}(2)$-invariant forms of BSQ, known as {\em Schwarzian BSQ equations}, have long been known in both continuous and discrete cases \cite{Weiss, N1}. However, a projective formulation of BSQ equations in terms of $\mathrm{PGL}(3)$-invariants, where $\mathrm{PGL}(3)$ arises as the natural group of transformations in the rank-$3$ setting, has not been fully established.  

The relevance of $\mathrm{PGL}(3)$ projective formulation of BSQ is further underscored by recent developments in discrete integrable systems. On the one hand, the construction of {\em pentagram maps} is rooted in the geometry of discrete projective curves in the projective space ${\mathbf P}^2$ 
\cite{Sch1, OST}. Indeed, the BSQ equations and pentagram maps are closely related: continuous limit of the latter recovers the continuous BSQ equation \cite{OST}; both the discrete BSQ equation and the pentagram maps are governed by third-order spectral problems in the Lax formulation.
On the other hand, a $\mathrm{PGL}(3)$-invariant lattice BSQ system was obtained as a reduction of a Q3 analogue of the lattice BSQ equations \cite{SZZN}, though without explicit geometric considerations. Moreover, a remarkable $\mathrm{PGL}(3)$-invariant system of PDEs, known as {\em generating PDEs} for the BSQ hierarchy was introduced in 
\cite{TN1, TN2} (see Section \ref{sec:genPDEs} for details); however, its projective formulation has remained only a conjecture.

The primary objective of the present work is to provide a unified approach to constructing $\mathrm{PGL}(3)$-invariant integrable systems related to the BSQ equations. Our methodology is grounded in linear spectral problems and their factorised forms. The continuous and discrete realms are connected through the notion of {\em dualities} induced by the factorisation. This yields a unified representation of all systems in terms of a common set of dependent variables throughout the paper (denoted as   $z_1, z_2$ that are the inhomogeneous coordinates of the spectral problems), providing a
coherent algebraic and geometric picture across different integrable regimes.
\medskip


Our main results include:
\begin{itemize}
\item Explicit formulae for the {\em generating set} of $\mathrm{PGL}(3)$-differential and difference invariants in the sense of \cite{Olvera,Mansfield2010} (see also Remark \ref{rmk:gi}). They are given in Definitions \ref{def:diffc} and \ref{def:diffi} generalising respectively the Schwarzian derivative and cross-ratio. We also provide {\em composition rules} for the $\mathrm{PGL}(3)$-differential invariants (\cf Theorem \ref{eq:compo}) and continuum limits of $\mathrm{PGL}(3)$-difference invariants (\cf Proposition \ref{eq:conti}). The method used here is essentially classical, 
based on the correspondence between nondegenerate projective curves and scalar 
linear differential or difference equations \cite{Wil,OST}, but adapted to the 
rank-$3$ setting. In particular, the relation between the difference invariants  and the corner invariants of \cite{OST} is made explicit at the end of Section~\ref{sec:gl3ac}.

\item Continuous--discrete duality and self-duality induced by factorisation of linear differential or difference operators (see Section \ref{sec:3}), underlying the exact discretisation and multi-dimensional consistency of the associated BSQ systems; we also provide the associated Darboux--Crum formulae (see Appendix \ref{app:DCF}). This part develops known ideas of factorisation and Darboux-transformation for linear spectral operators, providing explicit derivations adapted to the rank-$3$ case.

\item Derivation of the continuous and discrete $\mathrm{PGL}(3)$-invariant BSQ systems (\cf Sections \ref{sec:41} and \ref{sec:42}). A {\em lifting-decoupling mechanism} is provided to reduce $\mathrm{PGL}(3)$-invariant BSQ systems to (${\mathrm{PGL}}(2)$-invariant) Schwarzian BSQ equations. The lift of the discrete ${\mathrm{PGL}}(3)$-invariant BSQ systems is a three-component quad-system, consistent around a cube, serving as the rank-$3$ analogue of the cross-ratio equation. Although explicit forms of some of these systems were previously known (see Remarks~\ref{rmk:OST} and \ref{rmk:SZN}), our treatment gives an intrinsic derivation from linear spectral problems and their factorised forms, together with a unified projective formulation.

\item Derivation of the {\em ${\mathrm{PGL}}(3)$-invariant generating PDE}---a coupled system of PDEs with the associated lattice parameters serving as the independent variables, containing the entire hierarchy of ${\mathrm{PGL}}(3)$-invariant BSQ equations by systematic expansions---together with its Lagrangian structure (\cf Section \ref{sec:genPDEs}), corresponding to the projective formulation of the BSQ generating systems \cite{TN1, TN2}. This projective formulation of the generating systems and their Lagrangian structure is, to our knowledge, entirely new.
\end{itemize} 

 Our work establishes a unified projective formulation for $\mathrm{PGL}(3)$-invariant integrable systems, positioning the rank-$3$ BSQ setting as the natural generalization of the well-understood rank-$2$ KdV theory. The methods developed here could naturally extend to higher-rank cases offering a  systematic approach to $\mathrm{PGL}(N)$-invariant integrable systems. Note that our constructions are mainly algebraic in nature and apply equally to systems defined over the real field $\mathbb{R}$ or the complex field $\mathbb{C}$. While the complex case is perhaps the most relevant, for the sake of notational simplicity, we shall omit explicit references to the base field and use the general notations such as $\mathrm{PGL}(3)$ and $\mathbf{P}^2$ throughout the paper.

    The paper is organised as follows. Section \ref{sec:2} derives the generating ${\mathrm{PGL}}(3)$-differential 
   and difference invariants (see Remark \ref{rmk:gi} for the definition of generating invariants) from third-order spectral problems and establishes their related properties such as composition rules and continuous limits. Section \ref{sec:3} studies factorisation of the third-order spectral problems, revealing the continuous--discrete duality and self-duality among the associated BSQ equations. Sections \ref{sec:41} and \ref{sec:42} present the continuous, discrete  $\mathrm{PGL}(3)$-invariant BSQ systems together with their geometric interpretation and  decoupling mechanism. Section \ref{sec:genPDEs} presents the $\mathrm{PGL}(3)$-invariant generating PDE and its Lagrangian structure, with some details of the derivation given in Appendix \ref{app:genPDEproofs}. Appendices \ref{app:A} and \ref{app:B} present the general theory of $\mathrm{PGL}(N)$-invariants for arbitrary $N$. Appendix \ref{app:DCF} provides the associated Darboux--Crum formulae.


\section{
Projective invariants from linear spectral problems }  \label{sec:2}
In this section, we provide explicit formulae for $\mathrm{PGL}(3)$-differential and difference invariants derived from third-order linear spectral problems, along with the {\em composition rules} for the $\mathrm{PGL}(3)$-differential invariants. We also present their multivariate extensions, which will be crucial for the development of $\mathrm{PGL}(3)$-invariant BSQ-type integrable systems discussed in subsequent sections. A group-theoretical approach to the invariants and a general framework for $\mathrm{PGL}(N)$-invariants will be presented respectively in Appendices \ref{app:A} and \ref{app:B}. 

It is a classical result in projective differential geometry that a scalar linear differential equation of order $N$ corresponds to a curve in the projective space $\mathbf{P}^{N-1}$  \cite{Wil}. The independent solutions of the linear differential equation serve as homogeneous coordinates of the lifted curve, and the coefficients (or potentials), up to gauge transformations, are the generating differential projective invariants. 
Although this geometric correspondence is well established in the context of classical invariant theory, conformal field theory 
and integrable systems (see, for example, \cite{Wil,  FIZ, Dickey, Dickey2,  OT, Radul}), our focus is to provide explicit derivations of these invariants, and extend this framework to the discrete realm, where generating difference invariants arise from linear difference equations of order $N$ corresponding to discrete projective curves (polygons) in $\mathbf{P}^{N-1}$. Below, we focus on the $N=2$ and $N= 3$ cases, relevant to our focus on BSQ-type equations.

\subsection{Projective differential invariants} 
We first illustrate the well-known method for constructing the $\mathrm{PGL}(2)$-differential invariant, namely the Schwarzian derivative, based on a second-order linear ordinary differential equation. Then, we apply the same technique to a third-order equation, and derive the $\mathrm{PGL}(3)$-differential invariants that are higher-order analogs of the Schwarzian derivative. This procedure can be generalized as a systematic algorithm to obtain $\mathrm{PGL}(N)$-differential invariants based on an $N$th-order linear ordinary differential equation (see Appendix \ref{appgln1}  for further details). 

\subsubsection{$\mathrm{PGL}(2)$-differential invariant and the Schwarzian derivative}
Consider the Schr\"odinger spectral problem
\be\label{eq:2ndord-diff}
(\pl_x^2+u)\vf=\ld\vf\,.
\ee
The potential $u$ is supposed to be a smooth function, and $\lambda$ plays the role of the spectral parameter. This equation is also known as the {\em  Liouville normal form}, as a general second-order linear ordinary differential equation that can always be transformed to \eqref{eq:2ndord-diff} \cite{Wil}. 
From the viewpoint of ordinary differential equations, the spectral parameter $\lambda$ could be absorbed into $u$. However, in the context of integrable systems, the appearance of $\lambda$ is related to the important notion of Lax pair. Therefore, we keep using the spectral problem \eqref{eq:2ndord-diff} in this paper. 

The construction of the invariant is based on the observation that \eqref{eq:2ndord-diff} has two linearly independent solutions, say  $\vf_1,\vf_2$, and that the general solution is a linear combination of those
two solutions: $\vf=c_1\vf_1+c_2\vf_2$. 
Introduce the inhomogeneous coordinate $z$ as the ratio of $\vf_1, \vf_2$, \ie $z:=\vf_1/\vf_2$. Geometrically, $z$ represents an affine coordinate of a class of nondegenerate projectively equivalent curves in $\mathbf{P}^1$ (nondegeneracy means $z_x \neq 0$)\cite{Wil, OT}. For simplicity, we set $\vf_2 = \vf$, and rewrite \eqref{eq:2ndord-diff} for $\vf_1=z\vf$ as
\begin{equation}
z_{xx}\vf+2z_x\vf_x=0\,. 
\end{equation}
This leads to
\be\label{eq:HC}
\pl_x\log \vf=-\frac{z_{xx}}{2z_x}\, ,
\ee
where the left-hand side is a Hopf--Cole transformation for $\vf$. Substituting the above formula into the spectral problem \eqref{eq:2ndord-diff} for $\vf$ to eliminate $\vf$ in favour of $z$, we can express $u$ in terms of $z$ as follows
\be\label{eq:Schde}
2(u-\lambda)= \mathbf{S}[z]\ , \quad  \mathbf{S}[z]:=\frac{z_{xxx}}{z_x}-\frac32\frac{z_{xx}^2}{z_x^2}\ .
\ee
The expression $\mathbf{S}[z]$, (or $\mathbf{S}[z](x)$ if the independent variable needs to be specified), is the {\it Schwarzian derivative}.
 The quantity $u-\lambda$ is invariant under the action of  $\mathrm{GL}(2)$ on the independent solutions $(\vf_1,\vf_2)$, namely 
\be\label{eq:sl2transf} (\vf_1,\vf_2)~\mapsto (\vf_1,\vf_2)\bM \ , \quad \bM=\left(\begin{array}{cc} 
m_{11} & m_{12} \\ m_{21} & m_{22} \end{array}\right) \  ,\quad \det\bM \neq 0\,.    \ee
This induces a  $\mathrm{PGL}(2)$-action on $z$: 
\be\label{eq:sl2transfz} 
z~\mapsto \frac{m_{11}z+m_{21}}{m_{12}z+m_{22}} \ , 
\ee
under which $\mathbf{S}[z]$ is invariant. 

\begin{rmk}
\label{rmk:gi}    For a Lie group acting regularly on a jet space, there exists a finite generating set of differential invariants, such that every differential invariant can be locally written as a function of the generating invariants and their derivatives using invariant differential operators (see, \eg \cite{Olvera,Mansfield2010}). Although the $\mathrm{PGL}$-actions are not necessary regular themselves, they become regular when they are prolonged to acting on jet spaces. In the $\mathrm{PGL}(2)$-action \eqref{eq:sl2transfz} for $(x,z)$, the generating invariant is the Schwarzian derivative $\mathbf{S}[z]$; every differential invariant can be expressed as a function of $\mathbf{S}[z]$ and its derivatives with respect to $x$. An alternative derivation of the Schwarzian derivative $\mathbf{S}[z]$ using infinitesimal generators is provided in Appendix \ref{appa}.
\end{rmk}

 \subsubsection{$\mathrm{PGL}(3)$-differential invariants}\label{sec:pgl3}
Consider the following third-order spectral problem 
 \be\label{eq:3rdord-diff}
(\pl_x^3+u\pl_x+v)\vf=\lambda \vf\ ,
\ee
and let  $\vf_1, \vf_2, \vf_3$ be the set of linearly independent solutions. The natural  $\mathrm{GL}(3)$ symmetry on the solutions
\be\label{eq:sl3transf} (\vf_1,\vf_2,\vf_3)~\mapsto (\vf_1,\vf_2,\vf_3)\bM\  \,, \quad \bM=(m_{ij}) \,,\quad \det \bM \neq 0\,,\ee
induces a  $\mathrm{PGL}(3)$ transformation (a right action) on the 
inhomogeneous coordinates
\begin{equation}
    z_1=\frac{\vf_1}{\vf_3}\ ,\quad  z_2=\frac{\vf_2}{\vf_3}\  , 
\end{equation} 
namely, 
\be\label{eq:projSL3}
(z_1, z_2) \mapsto \left(\frac{\textstyle m_{11}z_1+m_{21}z_2+m_{31}}{\textstyle m_{13}z_1+m_{23}z_2+m_{33}}, \frac{\textstyle m_{12}z_1+m_{22}z_2+m_{32}}{\textstyle m_{13}z_1+m_{23}z_2+m_{33}}\right)\,   .
\ee
Geometrically, $(z_1, z_2)$ serve as affine coordinates of a class of nondegenerate projectively equivalent curves in $\mathbf{P}^2$ (nondegeneracy means $\partial_x ^2z_1 \partial_x z_2 -\partial_x ^2z_2 \partial_x z_1 \neq 0$).

Next, we construct the generating differential invariants in a similar way as the $\mathrm{PGL}(2)$ case. 
Writing the equations for $\vf=\vf_3$ and $\vf_i=z_i\vf$, $i=1,2$, we obtain the following system for both $z_1$ and $z_2$:
\bse\label{eq:BSQuv}\begin{eqnarray}
-u&=&\frac{z_i^{(3)}}{z_i^{(1)}}+3\frac{ z_i^{(2)}}{z_i^{(1)}} \pl_x\log \vf+3\left[ \pl_x^2\log\vf+(\pl_x\log\vf)^2\right]\, , \label{eq:BSQuva}\\
\lambda-v&=&-(\pl_x\log\vf)\left[\frac{z_i^{(3)}}{ z_i^{(1)}}+3\frac{ z_i^{(2)}}{ z_i^{(1)}} \pl_x\log \vf\right]+\pl_x^2\log\vf-2(\pl_x\log \vf)^3\,,\label{eq:BSQuvb}
\end{eqnarray}\ese 
where $
z_i^{(j)}:=\pl_x^jz_i$.
From \eqref{eq:BSQuva} for both $i=1,2$,  we can eliminate $u$ and express $\pl_x\log\vf$ as a ratio of two determinants
\be\label{eq:dlogvf} 
\pl_x\log\vf=-\frac13 \frac{\left|\boldsymbol{z}^{(3)},\boldsymbol{z}^{(1)} \right|}{\left|\boldsymbol{z}^{(2)},\boldsymbol{z}^{(1)}\right|}\,,
\ee
where $\boldsymbol{z}=(z_1,z_2)^\intercal$.
This is then substituted back to \eqref{eq:BSQuv} and we obtain $u$ and $v-\lambda$ in terms of $z_1$ and $z_2$, which turns out to be symmetric with respect to $z_1$ and $ z_2$:
\bse\label{eq:BSQdiffin}
\begin{eqnarray}
u&=& \frac{\left|\boldsymbol{z}^{(4)},\boldsymbol{z}^{(1)} \right|}{\left|\boldsymbol{z}^{(2)},\boldsymbol{z}^{(1)}\right|}+2 \frac{\left|\boldsymbol{z}^{(3)},\boldsymbol{z}^{(2)} \right|}{\left|\boldsymbol{z}^{(2)},\boldsymbol{z}^{(1)}\right|}-\frac43\left( \frac{\left|\boldsymbol{z}^{(3)},\boldsymbol{z}^{(1)} \right|}{\left|\boldsymbol{z}^{(2)},\boldsymbol{z}^{(1)}\right|}\right)^2 \ , \label{eq:BSQdiffina}\\
v-\lambda&=& \frac{1}{3}u_x -\frac{8}{27}\left( \frac{\left|\boldsymbol{z}^{(3)},\boldsymbol{z}^{(1)} \right|}{\left|\boldsymbol{z}^{(2)},\boldsymbol{z}^{(1)}\right|}\right)^3-\frac{1}{3}\frac{\left|\boldsymbol{z}^{(4)},\boldsymbol{z}^{(2)} \right|}{\left|\boldsymbol{z}^{(2)},\boldsymbol{z}^{(1)}\right|}\label{eq:BSQdiffinb}\\
&&+\frac{8}{9} \frac{\left|\boldsymbol{z}^{(3)},\boldsymbol{z}^{(2)} \right|\left|\boldsymbol{z}^{(3)},\boldsymbol{z}^{(1)} \right|}{\left|\boldsymbol{z}^{(2)},\boldsymbol{z}^{(1)}\right|^2}+\frac29 \frac{\left|\boldsymbol{z}^{(4)},\boldsymbol{z}^{(1)} \right|\left|\boldsymbol{z}^{(3)},\boldsymbol{z}^{(1)} \right|}{\left|\boldsymbol{z}^{(2)},\boldsymbol{z}^{(1)}\right|^2} \ .\nonumber
\end{eqnarray}\ese 

\begin{definition}\label{def:diffc}
Let $\left|\boldsymbol{z}^{(2)},\boldsymbol{z}^{(1)}\right|\neq 0$. We define the quantities $
\mathbf{S}_1[z_1,z_2]$ and 
$\mathbf{S}_2[z_1,z_2]$ as 
\bse\label{eq:BSQdiffin1}
\begin{eqnarray}
\mathbf{S}_1[z_1,z_2]&=& \frac{\left|\boldsymbol{z}^{(4)},\boldsymbol{z}^{(1)} \right|}{\left|\boldsymbol{z}^{(2)},\boldsymbol{z}^{(1)}\right|}+2 \frac{\left|\boldsymbol{z}^{(3)},\boldsymbol{z}^{(2)} \right|}{\left|\boldsymbol{z}^{(2)},\boldsymbol{z}^{(1)}\right|}-\frac43\left( \frac{\left|\boldsymbol{z}^{(3)},\boldsymbol{z}^{(1)} \right|}{\left|\boldsymbol{z}^{(2)},\boldsymbol{z}^{(1)}\right|}\right)^2 \ , \label{eq:BSQdiffina1}\\
\mathbf{S}_2[z_1,z_2]&=&\frac{\left|\boldsymbol{z}^{(4)},\boldsymbol{z}^{(2)} \right|}{\left|\boldsymbol{z}^{(2)},\boldsymbol{z}^{(1)}\right|} -\frac23 \frac{\left|\boldsymbol{z}^{(4)},\boldsymbol{z}^{(1)} \right|\left|\boldsymbol{z}^{(3)},\boldsymbol{z}^{(1)} \right|}{\left|\boldsymbol{z}^{(2)},\boldsymbol{z}^{(1)}\right|^2}\label{eq:BSQdiffinb1}\\
&&-\frac{8}{3} \frac{\left|\boldsymbol{z}^{(3)},\boldsymbol{z}^{(2)} \right|\left|\boldsymbol{z}^{(3)},\boldsymbol{z}^{(1)} \right|}{\left|\boldsymbol{z}^{(2)},\boldsymbol{z}^{(1)}\right|^2}+\frac{8}{9}\left( \frac{\left|\boldsymbol{z}^{(3)},\boldsymbol{z}^{(1)} \right|}{\left|\boldsymbol{z}^{(2)},\boldsymbol{z}^{(1)}\right|}\right)^3 \,.  \nonumber
\end{eqnarray}\ese 
\end{definition}
\medskip

Here, $\mathbf{S}_1[z_1,z_2], \mathbf{S}_2[z_1,z_2]$ are related to the potentials $u,v$ as
\begin{equation}
\mathbf{S}_1[z_1,z_2]=u\,,\quad \mathbf{S}_2[z_1,z_2]=-3(v-\lambda)+u_x\,,
\end{equation}
which are clearly $\mathrm{PGL}(3)$-invariant.  They can be used as the set of $\mathrm{PGL}(3)$-\textit{generating differential invariants} of a single independent variable, \cf Remark \ref{rmk:gi}.

\begin{rmk}\label{rem:wei}
   Since all differential invariants constructed in this way are rational functions in the jet space, we can define their weight as the degree of homogeneity in terms of derivatives. Specifically, the weight is the total number of derivatives in the numerator minus the total number of derivatives in the denominator. In this sense, $\mathbf{S}_1$ has a weight of $2$, while $\mathbf{S}_2$ has a weight of $3$; the Schwarzian derivative $\mathbf{S}[z] $ has weight $2$.
\end{rmk}

\subsection{Composition rules for the differential invariants}
The composition rule (also known as connection formula) of Schwarzian derivative is a result of smooth deformation of the second-order linear equation \eqref{eq:2ndord-diff}, and has close connection to important notions of mathematical physics such as Virasoro algebras, KdV hierarchy and conformal field theory \cite{FIZ, Dickey, Dickey2,  OT, Radul}. Here,  we present a derivation from the linear spectal problem to be able to generalise that derivation to the $\mathrm{PGL}(3)$-invariants $\mathbf{S}_1[z_1,z_2], \mathbf{S}_2[z_1,z_2]$, \cf \eqref{eq:BSQdiffin1}. 

Consider the Schwarzian derivative. Let $y=f(x)$ be a smooth function and define $\varphi(x)$, satisfying  \eqref{eq:2ndord-diff}, as a composite  $\varphi(x)=\psi(y)=\psi\circ f(x)$. Substituting it back to the spectral problem yields a linear equation for $\psi(y)$:
\begin{equation}\label{eq:au2}
\partial_y^2\psi+W(y)\partial_y\psi+U(y)\psi=0\ ,
\end{equation}
where
\begin{equation}\label{eq:WU}
W(y)=\frac{f_{xx}}{f_x^2}\ , \quad U(y)=\frac{u-\lambda}{f_x^2}=\frac{\mathbf{S}[z]}{2f_x^2}\ .
\end{equation}
Starting with two independent solutions $\psi_1(y)$ and $\psi_2(y)$, we define their ratio as $Z(y)=\psi_1(y)/\psi_2(y)$; therefore,  the two independent solutions of the original spectral problem are
\begin{equation}
\varphi_1(x)=\psi_1\circ f(x)\ ,\quad \varphi_2(x)=\psi_2\circ f(x)\ ,
\end{equation}
and hence their ratio is
\begin{equation}
z=\frac{\varphi_1}{\varphi_2}=\frac{\psi_1}{\psi_2}\circ f=Z\circ f \ .
\end{equation}
 Substituting $\psi_1=Z\psi$ with  $\psi=\psi_2$ to \eqref{eq:au2} yields
\begin{equation}
\partial_y\log \psi=-\frac{1}{2}\frac{Z_{yy}}{Z_y}-\frac{W}{2}\ .
\end{equation}
This is substituted back to \eqref{eq:au2} and we get the following identity
\begin{equation}\label{eq:comSch0}
 \mathbf{S}[Z]=2U-\partial_yW-\frac{1}{2}W^2\ .
\end{equation}
By using the chain rule
\begin{equation}
\partial_yW=\frac{\partial_xW}{\partial_xy}=\frac{\partial_xW}{f_x}\ ,
\end{equation}
and taking \eqref{eq:WU} into account,  the identity \eqref{eq:comSch0} becomes
\begin{equation}
\mathbf{S}[Z\circ f]=f_x^2 \left(\mathbf{S}[Z]\circ f\right)+\mathbf{S}[f] \ ,
\end{equation}
which is the {\it composition rule of the Schwarzian derivative}. 

Now consider the  $\mathrm{PGL}(3)$-differential invariants $\mathbf{S}_1[z_1,z_2], \mathbf{S}_2[z_1,z_2]$. 
Similarly, let $y=f(x)$ be a smooth function and define $\varphi(x)=\psi(y)=\psi\circ f(x)$. Inserting this to \eqref{eq:3rdord-diff} yields
\begin{equation}\label{eq:au3}
\partial_y^3 \psi+W(y)\partial_y^2\psi+U(y)\partial_y\psi+V(y)\psi =0\ ,
\end{equation}
where
\begin{equation}\label{eq:WUV}
  W(y)=3\frac{f_{xx}}{f_x^2}\,,\quad 
U(y)=\frac{f_{xxx}+uf_x}{f_x^3}
\, ,\quad V(y)=\frac{v-\lambda}{f_x^3} 
\, .
\end{equation}
Let $\psi_1(y)$, $\psi_2(y)$ and $\psi_3(y)$ be three independent solutions of \eqref{eq:au3} and define 
\begin{equation}
Z_1=\frac{\psi_1}{\psi_3} \ ,\quad Z_2=\frac{\psi_2}{\psi_3} \ .
\end{equation}
This gives three independent solutions of the original spectral problem, \ie $\varphi_i=\psi_i\circ f$ ($i=1,2,3$) and hence their ratios are
\begin{equation}
z_1=\frac{\varphi_1}{\varphi_3}=Z_1\circ f \,,\quad  z_2=\frac{\varphi_2}{\varphi_3}=Z_2\circ f\ .
\end{equation}

Let $\psi=\psi_3$. The equations for $\psi_1=Z_1\psi$ and $\psi_2=Z_2\psi$ lead to 
\begin{equation}
\partial_y\log\psi=-\frac{1}{3}\frac{\left|\boldsymbol{Z}^{(3)},\boldsymbol{Z}^{(1)}\right|}{\left|\boldsymbol{Z}^{(2)},\boldsymbol{Z}^{(1)}\right|}-\frac{W}{3}\ .
\end{equation}
Substituting this back to \eqref{eq:au3},  we obtain the following two identities
\bse \label{eq:comPGL30}
\begin{align}
\mathbf{S}_1[Z_1,Z_2]&=U-\partial_yW-\frac{1}{3}W^2\ , \label{eq:comPGL30a}  \\ 
\mathbf{S}_2[Z_1,Z_2]&=-3V+\partial_y\mathbf{S}_1[Z_1,Z_2]+W \mathbf{S}_1[Z_1,Z_2]+\partial^2_yW+W\partial_yW+\frac{1}{9}W^3\ .\label{eq:comPGL30b}
\end{align}
\ese
Taking the functions $W(y)$, $U(y)$ and $V(y)$ into consideration, the identities \eqref{eq:comPGL30} can be rewritten. We summarize the final result as the following theorem.
 
 \begin{theorem}\label{eq:compo}
  The {\it composition rules of the $\mathrm{PGL}(3)$-differential invariants} \eqref{eq:BSQdiffin1}   are 
\begin{subequations}
\label{eq:comPGL3}\begin{align}
\mathbf{S}_1[Z_1\circ f,Z_2\circ f]&= f_x^2\left(\mathbf{S}_1[Z_1,Z_2]\circ f\right) +2\mathbf{S}[f] \ , \\ 
\mathbf{S}_2[Z_1\circ f,Z_2\circ f]&= f_x^3\left(\mathbf{S}_2[Z_1,Z_2]\circ f\right)-f_xf_{xx}\left(\mathbf{S}_1[Z_1,Z_2]\circ f\right) -\partial_x\mathbf{S}[f]\ ,   
\end{align}
\end{subequations}
where  
$\mathbf{S}[f]$ is the Schwarzian derivative.
 \end{theorem}

\begin{rmk}
The above formulae encode the transformation properties of 
$\mathrm{PGL}(3)$-differential invariants under reparametrizations $x\mapsto y =f(x)$, and were central to  classical $W_n$ ($n=3$ here)  algebra \cite{FIZ, Dickey, Dickey2,  Radul}. Our emphasis is put on  explicit formulae in terms of the inhomogeneous coordinates $z_1,z_2$.
In $W_3$ algebra, $\mathbf{S}_1[z_1,z_2]$ corresponds to a {\em quasi-primary field} of weight $2$ (whose infinitesimal version corresponds to the Virasoro algebra), while an alternative invariant 
\begin{equation} \mathcal{S}_2[z_1,z_2] =-\frac{1}{6}\partial_x\mathbf{S}_1[z_1,z_2]- \frac{1}{3}\mathbf{S}_2[z_1,z_2] \, ,
\end{equation}
resulted from $\mathcal{S}_2[z_1,z_2]=  v-\lambda-\frac{u_x}{2}$,  transforms as a genuine weight $3$ differential \cite{FIZ}:  \begin{equation}
     \mathcal{S}_2[Z_1\circ f, Z_2\circ f] =f_x^3 \left(\mathcal{S}_2[Z_1,Z_2]\circ f\right)\,.
 \end{equation}
Then,  $\mathbf{S}_1[z_1,z_2]$ and $ \mathcal{S}_2[z_1, z_2]$ could form another set of generating $\mathrm{PGL}(3)$-differential invariants. 
Geometrically, $\mathbf{S}_1[z_1,z_2]$ coincides with the {\em projective curvature} of a nondegenerate curve, while  $ \mathcal{S}_2[z_1, z_2]$ governs the so-called {\em projective arc-length element} \cite{OT}. 
\end{rmk}

\subsection{Difference projective invariants}
\label{sec:23}
The technique of constructing differential invariants can be well adapted to the discrete realm. Here, we consider functions depending on a discrete variable, say $n\in\ZZ$. We employ $T$ as the forward shift operator, and use the  $~\bar{}~$ notations to denote shifts. For a function $\varphi:=\vf(n)$, one has the forward shifts as
\begin{equation}
  T\,\vf = \overline{\vf}=\vf(n+1)\,,\quad  T^2\,\vf = \overline{\overline{\vf}} =\vf(n+2)\, ,\quad \dots\,,  
\end{equation}
and the backward shifts as
\begin{equation}
T^{-1} \vf = \underline{ \vf }=\vf(n-1)\,, \quad \dots\, 
\end{equation}
\subsubsection{The $\mathrm{PGL}(2)$-difference invariant: cross-ratio}
Consider a second-order linear
difference equation
\be\label{eq:2ndDE}
(T^2+\mathfrak{h} T +\alpha )\vf=\lambda \vf\ . 
\ee 
Here $\vf$ and $\mathfrak{h}$ are functions of  $n\in \ZZ$, $\alpha$ is the lattice parameter and $\lambda$ is the spectral parameter. Let $z$ be the ratio of two linearly independent solutions $\vf_1, \vf_2$ of \eqref{eq:2ndDE}: $z=\vf_1/\vf_2$.
The action of  $\mathrm{GL}(2)$ given by \eqref{eq:sl2transf} on the solution space of the linear problem \eqref{eq:2ndDE} induces fractional linear transformations to $z$ as shown in \eqref{eq:sl2transfz}.  Let $\varphi_2 =\varphi$, and the linear ordinary difference equation 
\eqref{eq:2ndDE} gives rise to 
the system of equations:
\bse\label{eq:2ndsys}\begin{eqnarray} 
\wb{\wb{\vf}}+\mathfrak{h}\wb{\vf} &=&(\ld-\alpha)\vf\  , \\ 
 \wb{\wb{z}}\,\wb{\wb{\vf}}+\mathfrak{h}\wb{z}\, \wb{\vf} &=& (\ld-\alpha)z\vf\  , 
\end{eqnarray}\ese 
which we  can write as:
\begin{equation} 
\left(\begin{array}{cc} 1 & 1 \\ \wb{\wb{z}} & \wb{z}\end{array}\right) 
\left(\begin{array}{c} \wb{\wb{\vf}}/\vf \vspace{0.1cm}\\ \mathfrak{h}\wb{\vf}/\vf\end{array}\right) 
=(\ld-\alpha) \left(\begin{array}{c} 1 \\ z \end{array}\right) \,. 
\end{equation}
Solving this system yields 
\begin{equation} \frac{\wb{\wb{\vf}}}{\vf}=(\ld-\alpha)  \frac{
\left|\begin{array}{cc} 1 & 1 \\ z & \wb{z}\end{array}\right| } 
{\left|\begin{array}{cc} 1 & 1 \\ \wb{\wb{z}} & \wb{z}\end{array}\right|}\ , \quad  
\mathfrak{h}\frac{\wb{\vf}}{\vf}=(\ld-\alpha) \frac{ 
\left|\begin{array}{cc} 1 & 1 \\ \wb{\wb{z}} & z \end{array}\right| } 
{\left|\begin{array}{cc} 1 & 1 \\ \wb{\wb{z}} & \wb{z}\end{array}\right|}\   \,,  \end{equation} 
from which by eliminating $\vf$ (by shifting the second forward, multiplying 
it by its original form, and equating the result with the first expression), 
we obtain the equality:
\be\label{eq:2ndinv}
-\frac{\mathfrak{h}{\wb{\mathfrak{h}}  }}{\ld-\alpha}=\frac{\left(\wb{\wb{\wb{z}}}-\wb{z}\right)\left(\wb{\wb{z}}-z\right)}
{\left(\wb{\wb{\wb{z}}}-\wb{\wb{z}}\right)\left(\wb{z}-z\right)}\,.
\ee 
The right-hand side, as a {\it cross-ratio} of four points $\wb{\wb{\wb{z}}}$, $\wb{\wb{z}}$, $\wb{z}$, $z$,  is manifestly $\mathrm{PGL}(2)$-invariant. Moreover, it is a \textit{generating $\mathrm{PGL}(2)$-difference invariant} such that every difference invariant can be expressed as a function of it and its shifts (see Appendix \ref{appb} for a proof, and Remark \ref{rmk:gi} for the differential counterpart).


\begin{rmk}
In the continuous case, the set of generating differential invariant is essentially unique (up to functional dependence). For instance, for the $\mathrm{PGL}(2)$-action \eqref{eq:sl2transfz}, the only functionally independent third-order differential invariant is the Schwarzian derivative \eqref{eq:Schde}.

In contrast, in the discrete setting the set of generating difference invariant is not unique, since it depends on the choice of stencil. For example, the generating difference invariant \eqref{eq:2ndinv} is defined in terms of $\wb{\wb{\wb{z}}}$, $\wb{\wb{z}}$, $\wb{z}$, $z$, but an equally valid invariant can be defined on shifted stencils such as $\wb{\wb{z}}$, $\wb{z}$, $z$, ${\hypobar 0 z}$, or other four-point configurations. These invariants are related by lattice shifts and therefore generate the same algebra of difference invariants. 

\end{rmk}

  
\begin{rmk}
  Geometrically, \eqref{eq:2ndDE} defines a class of projectively equivalent nondegenerate discrete projective curves (or polygons) in ${\mathbf P}^{1}$. The nondegeneracy means that any $2$ consecutive vertices span the full projective space ${\mathbf P}^{1}$, \ie $\wb{z}-{ z}\neq 0$. This notion can be generalized to the correspondence between an $N$th-order linear difference equation and discrete projective curves in  ${\mathbf P}^{N-1}$ \cite{OST}. 
\end{rmk}

One can take straightforwardly continuum limits by expanding the shifted objects through Taylor expansions to obtain the corresponding  differential invariants. For instance, taking 
\begin{equation}
     z=z(x)=z(x_0+n\dd)\,,\quad \wb{z}=z(x+ \dd)=z+\dd z_x+\frac{1}{2}\dd^2 z_{xx}+\cdots\, ,
\end{equation}
Expanding \eqref{eq:2ndinv} in powers of $\dd$ yields
\begin{equation}
\frac{\left(\wb{\wb{\wb{z}}}-\wb{z}\right)\left(\wb{\wb{z}}-z\right)}
{\left(\wb{\wb{\wb{z}}}-\wb{\wb{z}}\right)\left(\wb{z}-z\right)}
=4\left( 1-\frac{1}{2}\dd^2 {\mathbf S}(z)
+ \cdots \right)\  ,  \end{equation} 
thus recovering the Schwarzian derivative.

\subsubsection{The $\mathrm{PGL}(3)$-difference invariants} \label{sec:gl3ac}
We now consider a third-order linear difference equation 
\be\label{eq:3dDE}
\Lambda\vf=\lambda\vf \ ,\quad \Lambda=T^3+\mathfrak{h}T^2+\mathfrak{g}T +\alpha \,.
\ee 
We have the same $\mathrm{GL}(3)$-action  as in the continuous case given by \eqref{eq:sl3transf} on the space of solutions. Introducing again $z_1=\vf_1/\vf_3$ and $z_2=\vf_2/\vf_3$ and letting $\vf=\vf_3$, we get the following set of equations:
\bse\label{eq:3dsys}
\begin{eqnarray}
 \wb{\wb{\wb{\vf}}}+\mathfrak{h}\wb{\wb{\vf}}+\mathfrak{g}\wb{\vf}&=&(\ld-\alpha)\vf\  , \\ 
 \wb{\wb{\wb{z}}}_i\wb{\wb{\wb{\vf}}}+\mathfrak{h}\wb{\wb{z}}_i\wb{\wb{\vf}}+
\mathfrak{g}\wb{z}_i\wb{\vf}&=&(\ld-\alpha)z_i\vf\ ,\quad (i=1,2)\  , 
\end{eqnarray}
\ese 
which we  can write as:
\begin{equation} \left(\begin{array}{ccc} 1 & 1 & 1\\ 
\wb{\wb{\wb{z}}}_1 & \wb{\wb{z}}_1 & \wb{z}_1 \\ 
\wb{\wb{\wb{z}}}_2 & \wb{\wb{z}}_2 & \wb{z}_2 
\end{array}\right) 
\left(\begin{array}{c} \wb{\wb{\wb{\vf}}}/\vf \vspace{0.1cm} \\ \mathfrak{h}\wb{\wb{\vf}}/\vf \vspace{0.1cm} \\ 
\mathfrak{g}\wb{\vf}/\vf\end{array}\right) 
=(\ld-\alpha) \left(\begin{array}{c} 1 \\ z_1 \\ z_2\end{array}\right) \  . \end{equation}
Solving the system using Cramer's rule we obtain the following expressions: 
\be\label{eq:3dphis}
\frac{\wb{\wb{\wb{\vf}}}}{\vf}= (\ld-\alpha)\frac{\left|\wb{\wb{\bz}}, \wb{\bz} , \bz \right|} 
{\left| \wb{\wb{\wb{\bz}}},\wb{\wb{\bz}},\wb{\bz}\right|} \ ,\quad 
\frac{\wb{\wb{\vf}}}{\vf}= -\frac{\ld-\alpha}{\mathfrak{h}}
\frac{\left|\wb{\wb{\wb{\bz}}}, \wb{\bz} , \bz \right|} 
{\left| \wb{\wb{\wb{\bz}}},\wb{\wb{\bz}},\wb{\bz}\right|} \ ,\quad 
\frac{\wb{\vf}}{\vf}= \frac{\ld-\alpha}{\mathfrak{g}}
\frac{\left|\wb{\wb{\wb{\bz}}}, \wb{\wb{\bz}} , \bz \right|} 
{\left| \wb{\wb{\wb{\bz}}},\wb{\wb{\bz}},\wb{\bz}\right|}\   ,  
\ee 
where $\bz$ denotes a $3$-component vector $\bz=(1,z_1,z_2)^\intercal$\footnote{We retain the notation 
 $\bz$, although its dimension differs from that in Section \ref{sec:pgl3} where $\bz$ denoted a $2$-component vector $(z_1,z_2)^\intercal$. This distinction is apparent from the determinantal structures involved ($2\times 2$ versus $3\times 3$ determinants).}, and the numerators and denominators 
in the above expressions being $3\times3$ determinants consisting of these vectors with the associated 
shifts applied to them. 

Eliminating $\vf$, we obtain two independent invariants: 
\be\label{eq:3dinv}
-\frac{\wb{\mathfrak{g}} {\mathfrak{g}}}{(\ld-\alpha) { \mathfrak{h}}}=\frac{ \left| \wb{\wb{\wb{\wb{\bz}}}}, \wb{\wb{\wb{\bz}}},\wb{\bz} \right|\,
\left| \wb{\wb{\wb{\bz}}}, \wb{\wb{\bz}}, {\bz} \right|}
{ \left| \wb{\wb{\wb{\wb{\bz}}}}, \wb{\wb{\wb{\bz}}},\wb{\wb{\bz}} \right|\,
\left| \wb{\wb{\wb{\bz}}}, \wb{\bz}, {\bz} \right|} \ ,\quad 
-\frac{\wb{\mathfrak{h}}
 {\mathfrak{g}}}{\ld-\alpha}=\frac{ \left| \wb{\wb{\wb{\wb{\bz}}}}, \wb{\wb{\bz}},\wb{\bz} \right|\,
\left| \wb{\wb{\wb{\bz}}}, \wb{\wb{\bz}}, { \bz} \right|}
{ \left| \wb{\wb{\wb{\wb{\bz}}}}, \wb{\wb{\wb{\bz}}},\wb{\wb{\bz}} \right|\,
\left| \wb{\wb{\bz}}, \wb{\bz}, {\bz} \right|}\ .
\ee 
These can be considered as the generating difference invariants corresponding to the $\mathrm{PGL}(3)$-action in the discrete case; see Appendix \ref{appb} for more details. 
\begin{definition}\label{def:diffi}
Let $\bz, \wb{\bz}, \wb{\wb{\bz}}, \wb{\wb{\wb{\bz}}},  \wb{\wb{\wb{\wb{\bz}}}}$ be five consecutive points in general position, meaning no three of the five points are collinear. Define the quantities $
\mathbf{I}_1[z_1,z_2]$ and 
$\mathbf{I}_2[z_1,z_2]$ as 
\begin{equation}
 \label{eq:disBSQdiffin}
  \mathbf{I}_1[z_1,z_2] =\frac{ \left| \wb{\wb{\wb{\wb{\bz}}}}, \wb{\wb{\wb{\bz}}},\wb{\bz} \right|\,
\left| \wb{\wb{\wb{\bz}}}, \wb{\wb{\bz}},  \bz \right|}
{ \left| \wb{\wb{\wb{\wb{\bz}}}}, \wb{\wb{\wb{\bz}}},\wb{\wb{\bz}} \right|\,
\left| \wb{\wb{\wb{\bz}}}, \wb{\bz},  \bz \right|}\,,\quad
\mathbf{I}_2[z_1,z_2] =\frac{ \left| \wb{\wb{\wb{\wb{\bz}}}}, \wb{\wb{\bz}},\wb{\bz} \right|\,
\left| \wb{\wb{\wb{\bz}}}, \wb{\wb{\bz}},  \bz \right|}
{ \left| \wb{\wb{\wb{\wb{\bz}}}}, \wb{\wb{\wb{\bz}}},\wb{\wb{\bz}} \right|\,
\left| \wb{\wb{\bz}}, \wb{\bz}, {\bz} \right|} \, .
\end{equation} 
\end{definition}

The $ \mathbf{I}_1[z_1,z_2],   \mathbf{I}_2[z_1,z_2]$  can be chosen as the  \textit{generating  $\mathrm{PGL}(3)$-difference invariants}. 
It should be noted that the generating invariants may have various equivalent expressions due to the Pl\"ucker relation for 3D vectors as follows 
\be\label{eq:Pluck2}
\left| \ba,\bbb,\bdd\right|\,\left|\bc,\bdd,\bff \right| =
\left| \ba,\bc,\bdd\right|\,\left|\bbb,\bdd,\bff \right|-\left| \bbb,\bc,\bdd\right|\,\left|\ba,\bdd,\bff \right|\ .
\ee

\begin{rmk}In contrast to the continuous case where the potentials in the linear differential operators are rational expressions of the inhomogeneous coordinates (see respectively \eqref{eq:Schde} and \eqref{eq:BSQdiffin} for the $\mathrm{PGL}(2)$ and $\mathrm{PGL}(3)$ cases), one has quadratic combinations of the discrete potentials as rational expressions of the inhomogeneous coordinates, \cf \eqref{eq:2ndinv} and \eqref{eq:3dinv}. From a differential/difference Galois theory perspective, the difference projective invariants are indeed quadratic extension of the difference potentials (see, \eg \cite{STS}).
\end{rmk}

Analogous to the cross-ratio's continuum limit, the generating $\mathrm{PGL}(3)$-difference invariants \eqref{eq:disBSQdiffin} admit the following continuum limits.

\begin{proposition}\label{eq:conti}
Let $ {\mathbf I}_1[z_1,z_2]$,  ${\mathbf  I}_2[z_1,z_2]$ be the  $\mathrm{PGL}(3)$-difference invariants given by \eqref{eq:disBSQdiffin}, and let $z_1$, $z_2$ be smooth functions of $x=x_0+n \delta$, where $n$ is the discrete variable and $\delta$ is an infinitesimal quantity, \ie
\begin{equation}
  z_1: = z_1(x)=z_1(x_0+ n \delta)\,,\quad   z_2: = z_2(x)=z_2(x_0+ n \delta)\,.
\end{equation}
Then, by performing Taylor expansion in $\delta$, one gets
\bse\begin{align}
  {\mathbf   I}_1[z_1,z_2] &= 3 -{\mathbf   S}_1[z_1,z_2] \delta^2-\left(2 \partial_x{\mathbf   S}_1[z_1,z_2]+\frac{1}{2} {\mathbf   S}_2[z_1,z_2] \right)\delta^3+O(\delta^4)\,, \\
    {\mathbf   I}_2 [z_1,z_2]&=9- 6{\mathbf   S}_1[z_1,z_2] \delta^2-12 \partial_x {\mathbf   S}_1[z_1,z_2]\delta^3+ O(\delta^4)\,. 
\end{align}
\ese

\end{proposition}
\prf This relies on series expansion of  ${\mathbf I}_1[z_1,z_2],   {\mathbf  I}_2[z_1,z_2] $ in $\delta$.  By identifying the highest-order derivatives in the coefficients of the expansions, one can identify those coefficients with the differential invariants ${\mathbf   S}_1[z_1,z_2]$  and ${\mathbf   S}_2[z_1,z_2]$.
\finprf

\begin{rmk}
It is also interesting to note that these $\mathrm{PGL}(3)$-invariants, as a generalisation of the cross-ratio, have been developed independently in the community of image recognition and computer vision; see, \eg  \cite{NS1991, IWeiss1993}.
\end{rmk}

The third-order linear difference equation \eqref{eq:3dDE} was used in the Lax formulation of pentagram maps \cite{OST, FSol, Izo}, where the coefficients  $\fh$ and $\fg$ were employed as {\em dual coordinates} to represent the maps. 
Another choice of $\mathrm{PGL}(3)$-invariant coordinates, called \emph{corner invariants} (or
\emph{corner cross-ratios}), was also introduced in \cite{OST}. 
The following is an explicit description of the corner invariants in terms of $I_1, I_2$\footnote{We thank the anonymous referee for pointing out this connection to us.}. 

Let us recall the definition of corner invariants following \cite{OST}. For convenience, set
\begin{equation}
\bz_\ell=(1,z_1(n+\ell),z_2(n+\ell))^\intercal\,,\quad [ijk]=\det(\bz_i,\bz_j,\bz_k)\,.     
\end{equation}
Then, the invariants \eqref{eq:BSQdiffin1} can be written as \begin{equation}\label{eq:I12}
I_1[z_1,z_2]=\frac{[431][320]}{[432][310]}\,,\quad
I_2[z_1,z_2]=\frac{[421][320]}{[432][210]}\,. 
\end{equation}
Consider five consecutive points $\bz_0,\bz_1,\dots, \bz_4$ in generic position in $\mathbb P^2$ so that no three of them are collinear.  The \emph{left corner cross-ratio} denoted by $\bf x$ (respectively, the right corner cross-ratio denoted by $\bf y$) is defined as the cross-ratio of four points on the line $(\bz_0\bz_1)$ (respectively, on $(\bz_3\bz_4)$) as (see Figure \ref{fig:placeholder}):  
\begin{subequations}
    \begin{align}
{\bf x}=&[\bz_0,\bz_1,(\bz_0\bz_1)\cap(\bz_2\bz_3),(\bz_0\bz_1)\cap(\bz_3\bz_4)]\,,\\
{\bf y}=&[(\bz_0\bz_1)\cap(\bz_3\bz_4),(\bz_1\bz_2)\cap(\bz_3\bz_4),\bz_3,\bz_4]\,.
\end{align}
Here, the cross-ratio is fixed as  \begin{equation}\label{eq:cr}
[t_1,t_2,t_3,t_4]=\frac{(t_1-t_2)(t_3-t_4)}{(t_1-t_3)(t_2-t_4)}\,, 
\end{equation} 
with $t_j$, $j =1,2,3,4$, being the affine coordinates of the  four collinear points. 
 \end{subequations}
  A direct computation gives
\begin{equation}
{\bf x}=\frac{[432][310]}{[431][320]}\,,\quad
{\bf y}=\frac{[431][210]}{[421][310]}\,.    
\end{equation}
 Comparing this with \eqref{eq:I12}, one obtains
\begin{equation}
{\bf x}=1/I_1\,,\quad
{\bf y}=I_1/I_2\,.    
\end{equation}

\begin{figure}
    \centering
\begin{tikzpicture}[scale=0.9, font=\small]

  \tikzset{
    vtx/.style={circle, fill=black, inner sep=1.4pt},
    int/.style={circle, fill=black, inner sep=1.4pt},
    aux/.style={dashed}
  }

  \coordinate (vm2) at (0,0);     
  \coordinate (vm1) at (3,0);     
  \coordinate (v0)  at (4,1);     
  \coordinate (vp1) at (2,3);     
  \coordinate (vp2) at (0,4);     
  \coordinate (vm3) at (-1,.6);     
  \coordinate (v3) at (-.6,3.4);     
 \draw[thick] (vm2) -- (vm3)  ;  
  \draw[thick] (vp2) -- (v3);
  %
  \coordinate (P) at (5,0);
  \coordinate (Q) at (8,0);
  \coordinate (R) at (4.667,1.667);


  \draw[aux] (vm2) -- (Q);
  \draw[aux] (P) -- (vp1);
  \draw[aux] (Q) -- (vp2);

  \draw[aux] (vm1) -- (R);

  \draw[thick] (vm2) -- (vm1) -- (v0) -- (vp1) -- (vp2);

  \node[vtx] at (vm2) {};
  \node[vtx] at (vm1) {};
  \node[vtx] at (v0)  {};
  \node[vtx] at (vp1) {};
  \node[vtx] at (vp2) {};

  \node[int] at (P) {};
  \node[int] at (Q) {};
  \node[int] at (R) {};

  \node[below left]  at (vm2) {$\bz_{0}$};
  \node[below]       at (vm1) {$\bz_{1}$};
  \node[right]       at (v0)  {$\bz_2$};
  \node[above]        at (vp1) {$\bz_{3}$};
  \node[above]       at (vp2) {$\bz_{4
  }$};

  \node[below]       at (P) {$(\bz_0\bz_1)\cap(\bz_2\bz_3)$};
  \node[below right]       at (Q) {$(\bz_0\bz_1)\cap(\bz_3\bz_4)$};
  \node[above right] at (R) {$(\bz_1\bz_2)\cap(\bz_3\bz_4)$};


\end{tikzpicture}
    \caption{Five consecutive points $\bz_0,\bz_1,\dots,\bz_4$ of a discrete projective curve in
$\mathbb P^2$, together with the three intersection points
$(\bz_0\bz_1)\cap(\bz_2\bz_3)$, $(\bz_0\bz_1)\cap(\bz_3\bz_4)$ and
$(\bz_1\bz_2)\cap(\bz_3\bz_4)$.}
    \label{fig:placeholder}
\end{figure}

\subsection{Partial differential \& difference invariants} 

The invariant derived in the previous sections for a single independent variable can be naturally extended to a two-variable setting---that is, from curves to surfaces. Such extensions are essential for constructing projective-invariant integrable  systems (see Section \ref{sec:4}).  Here, we state the results for $\mathrm{PGL}(2)$- and $\mathrm{PGL}(3)$-invariants with two independent variables, but omit the details, as they can be derived through systematic methods. These include the use of infinitesimal generators (Appendix \ref{app:A}; see, also, \cite{Hydon2000,Olverb}), symbolic computations (\eg \cite{HK2007a,HK2007b, Kogan2023}), or moving frames (\eg \cite{FO1999,Mansfield2010,MRHP2019,WH2024,MMW2013}).

\paragraph{Partial differential invariants.} In addition to the $x$, let us introduce an extra independent variable $t$. 

\begin{itemize}
\item $\mathrm{PGL}(2)$-action on $(x,t,z)$. A new generating invariant is 
\begin{equation}
\frac{z_t}{z_x}\ .
\end{equation}
 The invariant $z_t/z_x$ and the Schwarzian derivative ${\mathbf S}[z]$ given by \eqref{eq:Schde} form the set of generating invariants for the $\mathrm{PGL}(2)$-action with two independent variables and one dependent variable. They generate all other $\mathrm{PGL}(2)$-differential invariants as functions of them as well as their derivatives with respect to $x$ and $t$. 
\item $\mathrm{PGL}(3)$-action on $(x,t,z_1,z_2)$. The new generating invariants can be chosen 
\begin{equation}\label{eq:PGL3tinv}
{\cal R}_1 = \frac{\left|\boldsymbol{z}_t,\boldsymbol{z}^{(1)}\right|}{\left|\boldsymbol{z}^{(2)},\boldsymbol{z}^{(1)}\right|}\ , \quad {\cal R}_2=\frac{\left|\boldsymbol{z}_t,\boldsymbol{z}^{(2)}\right|}{\left|\boldsymbol{z}_t,\boldsymbol{z}^{(1)}\right|}-\frac23\frac{\left|\boldsymbol{z}^{(3)},\boldsymbol{z}^{(1)}\right|}{\left|\boldsymbol{z}^{(2)},\boldsymbol{z}^{(1)}\right|} \ ,
\end{equation}
where $\bz=(z_1,z_2)^\intercal$, and $\bz^{(j)}$ denotes $j$-th derivatives with respect to $x$. Similarly, together with the invariants ${\mathbf S}_1[z_1,z_2], {\mathbf S}_2[z_1,z_2]$ defined in Definition \ref{def:diffc}, they generate all $\mathrm{PGL}(3)$-differential invariants with two independent variables and two dependent variables. 

\end{itemize}



\paragraph{Partial difference invariants.}
The derivation of difference invariants for multiple independent variables is more straightforward than in the differential case. It simply involves replacing certain variables with their shifted counterparts in the new direction, a fact that follows directly from the infinitesimal invariance condition (see Appendix \ref{appb}).  This is generalisable to $\mathrm{PGL}(N)$ ($N>3$) case:  new generating difference invariants when a second discrete variable $m$ is introduced can be obtained by replacing all $N$-th forward shifts $\boldsymbol{z}^{(N)}$ by $\widetilde{z}$ in \eqref{eq:PGLNdifferenceinv}. 

Let us introduce a new discrete variable, say $m$, associated to a $~\widetilde{}~$ shift operator.

\begin{itemize}
    \item $\mathrm{PGL}(2)$-action on $(n,m,z)$. A new generating invariant can be chosen as, for instance, one of the following invariants:
    \begin{equation}\label{eq:fundinvsl2tx}
\frac{(\widetilde{z}-{\hypobar 0 z})(\wb{z}-z)}{(\widetilde{z}-z)(\wb{z}-{\hypobar 0 z})}\ ,\quad \frac{(\widetilde{z}-z)(\wb{\wb{z}}-\wb{z})}{(\widetilde{z}-\wb{z})(\wb{\wb{z}}-z)}\ ,\quad  \frac{(\widetilde{z}-z)(\wb{z}-{\hypobar 0 z})}
{(\widetilde{z}-\wb{z})(z-{\hypobar 0 z})}\ ,   \quad  \frac{(\widetilde{z}-\wb{z})(\wb{\wb{z}}- z)}
{(\widetilde{z}-\wb{\wb{z}})(\wb{z}- z)} \ .
\end{equation}
These invariants are connected through the determinantal relation for $2$-component vectors 
\be\label{eq:Pluck0}
\left| \ba,\bbb\right|\,\left|\bc,\bdd\right| =
\left| \ba,\bc\right|\,\left|\bbb,\bdd \right|-\left|\ba,\bdd \right|\, \left| \bbb,\bc\right|\ .
\ee
If we consider the first-order Taylor expansions for the shifts
\begin{equation}
\wb{z}=z(x+\delta,t)=z+\delta z_x+\cdots\ ,\quad \widetilde{z}=z(x,t+\varepsilon)=z+\varepsilon z_t+\cdots \  ,
\end{equation}
the first invariant in \eqref{eq:fundinvsl2tx}, for instance, becomes
\begin{equation}
\frac{(\widetilde{z}-{\hypobar 0 z})(\wb{z}-z)}{(\widetilde{z}-z)(\wb{z}-{\hypobar 0 z})}=\frac{1}{2}\left(1+\frac{\delta}{\varepsilon}\frac{z_x}{z_t}+\cdots\right)\ ,
\end{equation}
giving the differential invariant $z_t/z_x$.
\item $\mathrm{PGL}(3)$-action on $(n,m,z_1,z_2)$.  We can immediately obtain the new generating difference invariants, for instance, by replacing all the $\wb{\wb{\wb{\wb{\bz}}}}$ with $\widetilde{\bz}$ in \eqref{eq:3dinv}:
\be\label{eq:fundinvsl3tx}
\frac{ \left| \widetilde{\bz}, \wb{\wb{\wb{\bz}}},\wb{\bz} \right|\,
\left| \wb{\wb{\wb{\bz}}}, \wb{\wb{\bz}},  \bz \right|}
{ \left| \widetilde{\bz}, \wb{\wb{\wb{\bz}}},\wb{\wb{\bz}} \right|\,
\left| \wb{\wb{\wb{\bz}}}, \wb{\bz},  \bz \right|}\,,\quad
\frac{ \left| \widetilde{\bz}, \wb{\wb{\bz}},\wb{\bz} \right|\,
\left| \wb{\wb{\wb{\bz}}}, \wb{\wb{\bz}},  \bz \right|}
{ \left| \widetilde{\bz}, \wb{\wb{\wb{\bz}}},\wb{\wb{\bz}} \right|\,
\left| \wb{\wb{\bz}}, \wb{\bz}, {\bz} \right|} \ , 
 \ee 
 where $\bz=(1,z_1,z_2)^\intercal$.
\end{itemize}

Similarly to the differential case, these new difference invariants, together with those obtained in Section \ref{sec:23}, constitute the corresponding generating set of difference invariants, from which all difference invariants can be expressed as functions of these invariants and their shifts. For instance, the cross-ratio \eqref{eq:2ndinv} and any one of \eqref{eq:fundinvsl2tx} form the generating invariants of the $\mathrm{PGL}(2)$-action with two discrete independent variables and one dependent variable, while the invariants \eqref{eq:disBSQdiffin} and \eqref{eq:fundinvsl3tx} form the generating invariants of the $\mathrm{PGL}(3)$-action with two discrete independent variables and two dependent variables.


\paragraph{ Differential-difference invariants.}  Now, let $t$ be the continuous independent variable and $n$ be the discrete one associated to the shift  $~\bar{}~$. 
\begin{itemize}
    \item $\mathrm{PGL}(2)$-action on $(t,n,z)$. The new generating differential-difference invariant can be chosen as
\begin{equation}\label{eq:ddin2}
\frac{(\wb{z} - {\hypobar 0 z})}{(z-{\hypobar 0 z})(\wb{z}-z)}z_t\ .
\end{equation}
    \item $\mathrm{PGL}(3)$-action on $(t,n,z_1,z_2)$.   
    The new generating invariants can be chosen as
\begin{equation}\label{eq:ddinv}
\frac{\left|\bz_t,\bz,\wb{\bz}\right|\;\left|\wb{\bz},\wb{\wb{\bz}},\wb{\wb{\wb{\bz}}}\right|}
             {\left|\bz,\wb{\bz},\wb{\wb{\bz}}\right|\;\left|\bz,\wb{\bz},\wb{\wb{\wb{\bz}}}\right|}\,,\quad 
             \frac{ \left| {\bz}_t, {\bz},\wb{\bz} \right|\,
\left| {\bz}, \wb{\wb{\bz}},  \wb{\wb{\wb{\bz}}} \right|}
{ \left| {\bz}_t, {\bz},\wb{\wb{\bz}} \right|\,
\left| {\bz}, \wb{\bz},  \wb{\wb{\wb{\bz}}}\right|}
\ .
\end{equation} 
    
\end{itemize}

Again, these new differential-difference invariants, together with the difference invariants obtained in Section \ref{sec:23}, can be chosen as generating invariants: the cross-ratio \eqref{eq:2ndinv} and the differential-difference invariant \eqref{eq:ddin2} form a generating set for the $\mathrm{PGL}(3)$-action with two independent variables, one continuous and one discrete, and one dependent variable, while the difference invariants   
\eqref{eq:disBSQdiffin} and  the differential-difference invariants \eqref{eq:ddinv} form a generating set for the $\mathrm{PGL}(3)$ action with two independent variables, one continuous and one discrete, and two dependent variables; see \cite{WH2024} for more examples. It should also be noted that the differential-difference invariant theory is not merely a simple combination of differential and difference theories (see, \eg \cite{P2017,PH2022}); rather, it requires additional conditions that reflect the prolongation structure, and a comprehensive general theory has yet to be developed.

Analogous to the continuum limits established in Proposition \eqref{eq:conti}, the partial difference and semi-discrete invariants introduced above admit limits that recover the corresponding multi-variable differential invariants. A detailed analysis is beyond the scope of the present work.


\section{Factorisation, duality, and exact discretisation of third-order spectral problems}\label{sec:3}
In this section, we explore the interplay between continuous and discrete third-order linear spectral problems through factorisation of operators and Darboux transformations. Starting from the continuous spectral problem \eqref{eq:3rdord-diff}, the associated Darboux transformation induces a third-order discrete spectral problem \eqref{eq:3dDE} and gives rise to a {\em continuous--discrete duality}, where the discrete problem constitutes an {\em exact discretisation} of the continuous one in the sense of Shabat \cite{Shabat1, Shabat2}.
One repeats this procedure in the discrete setting by applying a discrete Darboux transformation to the discrete spectral problem. This induces another covariant discrete operator in a second discrete direction and leads to a {\em self-duality} for the discrete operators.
Finally, we present the underlying nonlinear integrable BSQ equations expressed in terms of the $\mathrm{PGL}(3)$ invariants introduced in Section \ref{sec:2}. The new ingredient is that we also consider deformation of the discrete operator \eqref{eq:3dDE}  with respect to the lattice parameter yielding a non-autonomous semi-discrete BSQ system encoding the complete hierarchy of semi-discrete BSQ equations  \cite{TN1, TN2}.  
Note that this section is focusing on connections between continuous and discrete spectral problems through factorisation of operators. Extensions to  $\mathrm{PGL}(3)$-invariant BSQ-type systems will be developed subsequently in Section \ref{sec:4}.

\paragraph{Notation convention:} In Section \ref{sec:2}, (partial) derivatives were denoted using subscripts, \eg $z_x, z_t$. In this Section and Sections \ref{sec:41}-\ref{sec:42}, when considering continuous BSQ-type PDEs, one uses primes for derivatives with respect to the space variable and dots for derivatives with respect to the time variable. The subscripts are reserved to present derivatives with respect to lattice parameters, \cf Section \ref{sec:genPDEs}.

\subsection{From continuous to discrete: exact discretisation via Darboux transformation}
 The factorisation of linear differential operators was initially studied in the context of Hamiltonian structures and Miura transformations for integrable PDEs \cite{A1, Fordy, Fordy2}. This framework allows for constructions of Darboux transformations, and enables the passage from the continuous spectral problem \eqref{eq:3rdord-diff} to its discrete counterpart \eqref{eq:3dDE} through the exact discretisation procedure \cite{Shabat1, Shabat2}.  

Assume  $L$ admits a factorised form
\begin{equation}
  \label{eq:3dsp}
  L:=  \partial_x^3 + u \partial_x+v =  (\partial_x-r_3)(\partial_x-r_2)(\partial_x-r_1) + \alpha\,,
\end{equation}
where $\alpha$ is a Darboux--B\"acklund parameter. 
Expanding this yields the constraints
\bse\label{eq:cmt01}
\begin{align}
    0&=r_1+r_2+r_3 \,,     \\
  u&=  r_1 r_2 +  r_1 r_3 + r_2 r_3-r'_{2}  - 2 r'_{1}\,, \\
  v &= r_1'r_{2}+ r_1 r'_{2} +r_1'r_3-r_1 r_2 r_3    - r''_{1} + \alpha \,. 
\end{align}
\ese
Eliminating $r_3 = -r_1-r_2$ gives a  Miura-type transformation:\bse
\label{eq:cmt1}
\begin{align}
  u&=  -r_1^2  - r_2^2 - r_1 r_2- 2 r'_{1}-r'_{2} \,,\\
  v&=  r_1 r_2(r_1+r_2) -  r_1 (r'_{1}- r'_{2}) - r''_{1} + \alpha\,. 
\end{align}
\ese
Let  $\vf_1$ be a fixed solution of $ L \varphi = \lambda\varphi$ at $\lambda = \alpha$, and define $r_1=(\log \vf_1)'$. Substituting this into the spectral problem at $\lambda=\alpha$ yields a second-order Riccati-type equation\begin{equation}
  \label{eq:addUV}
r_1''+    3 r_1 r_1' +ur_1 +r_1^3  + v= \alpha\,. 
\end{equation}

A one-step Darboux transformation  is obtained by cyclically permuting the factors in \eqref{eq:3dsp}, leading to the  map in the following lemma.

\begin{lemma}[One-step Darboux transformation] Let $\vf_1$ be a fixed solution of  $ L \varphi = \lambda\varphi$ at $\lambda = \alpha$, and let $r_1=(\log \vf_1)'$. Then the transformation  $\{\varphi,u,v\}\mapsto \{\overline{\varphi},\overline{u},\overline{v}\}$:
  \begin{equation}\label{eq:cDT2}
 \overline{\varphi} = \varphi'-r_1\vf\,,\quad
    \overline{u} = u+3r_1'\,,\quad
     \overline{v} = v+u' + 3(r_1 r_1'+r_1'') \,,
  \end{equation}
maps  $L \varphi = \lambda\varphi$ to $\overline{L}\overline{\varphi}    = \lambda \overline{\vf} $ where $\overline{ L}  = \partial_x^3 + \overline{u} \partial_x + \overline{v}$. 
\end{lemma}

This follows from reordering the factors as $
  \overline{ L}=  (\partial_x-r_1) (\partial_x-r_3)(\partial_x-r_2) + \alpha$. The action of iterated Darboux transformations leads to compact forms for multi-soliton-type solutions, known as Darboux--Crum formulae. They will be presented in Appendix \ref{app:DCF}. Note that Darboux transformations for third-order spectral problems were considered previously in \cite{AN1, LU}, but from a different perspective than the one adopted here.

\begin{rmk}Alternatively, one has another one-step Darboux transformation by letting
  \begin{equation}
  \wh{ L}= (\partial_x-r_2) (\partial_x-r_1)(\partial_x-r_3)+ \alpha\,, \quad \wh{\varphi} = (\partial_x-r_2)(\partial_x-r_1)\varphi\,,
\end{equation}
which amounts to
\begin{equation}
\wh{u} =u-3r_3'\,, \quad \wh{v} =v-u'-3r_3r_3'\,, 
\end{equation}
with $\wh{ L}=  \partial^3_x +\wh{u}\partial_x +\wh{v}$.
 One can express $r_3=-r_1-r_2$ with
\begin{equation}
  \label{eq:g1g2} 
r_1=(\log \vf_{1})'\,, \quad r_2 =(\log \overline{\vf}_{2})' \,,\quad \overline{\vf}_{2} = (\partial_x-r_1 )\vf_2\,.
\end{equation}
Here $\vf_1$, $\vf_2$ are two linearly independent solutions of $L \vf = \lambda \vf$  at $\lambda = \alpha$. \end{rmk}

Now interpret the Darboux map as a shift operator $T$ acting on the eigenfunction:
\begin{equation}\label{eq:dbcovariance}
  T\vf: = \overline{\vf} = \partial_x \vf - r \vf\,,
\end{equation}
where we suppress the subscript on $r$ and allow it to depend on the discrete variable $n$, \ie $r=r(n,x)$. Iterating this map yields
\begin{equation}\label{eq:DTsss}
\overline{  \varphi} = \varphi'-r\, \varphi\,, \quad   \overline{\overline{\varphi}} = \overline{\varphi}'-\overline{r}\, \overline{\varphi}\,,\quad   \overline{\overline{\overline{\varphi}}} = \overline{\overline{\varphi}}'-\overline{\overline{r}} \,\overline{\overline{\varphi}}\,. 
\end{equation}
Computing $\overline{\overline{\varphi}}'$ using the first two equations and $\overline{\varphi}'$ using the first equation, one gets
\begin{equation}\label{eq:matrix3x3}
  \bma \overline{\overline{\overline{\varphi}}} \\ \overline{\overline{\varphi}} \\ \overline{\varphi} \\  \varphi \ema  =
  \bma 1 & -r- \overline{r} -\overline{\overline{r}} & \ast & \star \\
  0 & 1 & -r- \overline{r} &  \overline{r} \,r-r'\\
  0 & 0 & 1 & -r \\
      0 & 0 & 0 & 1
  \ema  \bma \varphi''' \\\varphi'' \\ \varphi' \\ \varphi\ema\,,
\end{equation}
where
\begin{equation}
 \ast=\overline{\overline{r}}\, (\overline{r}+r)+\overline{r}\, r - \overline{r}' - 2  r' \,,\quad  \star= -\overline{\overline{r}}\,\overline{r}\, r + \overline{\overline{r}}r' + \overline{r}'r + \overline{r} r'  - r''\,.
\end{equation}
Expressing $\vf, \vf', \vf'', \vf''' $ in terms of $\vf$, $\overline{\vf}$, $\overline{\overline{\vf}}$, $\overline{\overline{\overline{\vf}}}$, and using $L\vf=\lambda\vf$,  we obtain a third-order linear difference equation:
\begin{equation}\label{eq:discrete3T}
  \Lambda\,\vf = \overline{\overline{\overline{\vf}}}+\mathfrak{h} \overline{\overline{\vf}}+\mathfrak{g} \overline{\vf}+\alpha\vf=\lambda\vf\,,
\end{equation}
where $\Lambda = T^3 + \mathfrak{h} \, T^2 +\mathfrak{g} \,T+ \alpha$ with  
\begin{equation}
  \mathfrak{h}= \overline{\overline{r}}+\overline{r}+r\,,\quad \mathfrak{g} =\overline{r}^2+r^2+ \overline{r} r +  \overline{r}'  + 2 r'   +  u\,.
\end{equation}
Thus, the Darboux iteration induces the discrete spectral problem  \eqref{eq:3dDE}. Inversely, the linear problem   \eqref{eq:discrete3T} allows us to obtain the continuous spectral problem  $L\vf=\lambda\vf$ through the linear system  \eqref{eq:matrix3x3}. In this sense, the continuous operator $L$ and the discrete operator $\Lambda$ are dual: the former governs the $x$-evolution, and the latter the $n$-evolution; they are connected by  \eqref{eq:dbcovariance}, and both share the same solution space spanned by $\vf_1,\vf_2,\vf_3$.

\subsection{Discrete self-duality and multi-dimensional consistency}\label{sec:3.5}

Having established exact discretisation in one lattice direction, we now apply the same factorisation technique to the discrete operator itself, which will generate a second lattice direction. Assume the discrete operator $\Lambda$ \eqref{eq:discrete3T} admits a factorised form
\be
\label{eq:factordop2}
\Lambda:=T^3 + \mathfrak{h}  \, T^2 +\mathfrak{g}  \,T+ \alpha = (T-f_3)(T-f_2)(T-f_1)+\beta\,, 
\ee where $\beta$ is another Darboux--B\"acklund parameter, $\beta\neq \alpha$. Expanding this yields
\bse\label{eq:facint11}
\begin{align}\label{eq:facint1a}
\mathfrak{h}&=-f_3-\overline{f}_2-\overline{\overline{f}}_1 = -\ff - \overline{\overline{f}}_1\, ,\\
\mathfrak{g}&= f_2f_3+\overline{f}_1f_3+\overline{f}_1\overline{f}_2 =  f_2 f_3+\overline{f}_1\ff  \, ,\label{eq:facint1b}\\
\alpha&=\beta{} -f_1f_2f_3\ \label{eq:facint1c}\, , 
\end{align}
\ese
where $\ff:= f_3+\overline{f}_2$. Eliminating $f_2f_3 = (\beta-\alpha)/f_1$ gives a Miura-type transformation
\begin{equation}
  \mathfrak{h} = -\overline{\overline{f}}_1 -\ff, \quad \mathfrak{g} = (\beta-\alpha)/f_1+  \overline{f}_1 \ff\,. 
\end{equation}

Let $\vf_1$ be a fixed solution of $\Lambda \vf = \lambda\vf$ at $\lambda = \beta$, and define  $f_1 = \overline{\vf}_1/\vf_1$. Substituting it into the spectral problem  at $\lambda=\beta$  yields a second-order discrete Riccati-type equation
\begin{equation}
  \label{eq:fds1}
\overline{\overline{f}}_1\overline{f}_1f_1 + \mathfrak{h}\overline{f}_1f_1+ \mathfrak{g}f_1+\alpha-\beta =0\,.
\end{equation}
A one-step discrete Darboux transformation is obtained by cyclically permuting the factors in \eqref{eq:factordop2}, leading to the following theorem.

\begin{theorem}[One-step Darboux transformation]
  \label{th:DDT}
Let $\vf_1$ be a fixed solution of $\Lambda \vf  = \lambda \vf$ at $\lambda =\beta$, and let $f_1 =  \overline{\vf}_1/\vf_1$. Then the transformation $\{\varphi,\fh,\fg\}\mapsto \{\widetilde{\varphi},\widetilde{\fh},\widetilde{\fg}\}$:
  \begin{equation}\label{eq:dspDT1}
 \widetilde{\varphi} = (T-f_1)\varphi\,,~~ 
    \widetilde{\fh} = \overline{\fh}+\overline{\overline{\overline{f}}}_1-f_1\,,~~
     \widetilde{\fg} = \overline{\fg}+\overline{\overline{f}}_1\,\overline{\fh}-f_1 \fh+\overline{\overline{f}}_1(\overline{\overline{\overline{f}}}_1-f_1)\,, 
  \end{equation}
maps $\Lambda\vf = \lambda \vf$ to $\wt{\Lambda }\wt{\vf} = \lambda \wt{\vf}$ where $\wt{\Lambda} = T^3 +\wt{\fh}T^2 +\wt{\fg}T + \alpha$. 
\end{theorem}

The above theorem follows by setting $\wt{\Lambda} = (T-f_1)(T-f_3)(T-f_2)+\beta$. The action of
iterated Darboux transformations leads to compact forms 
known as Darboux--Crum formulae. They will be presented in Appendix \ref{app:DCF}.

\begin{rmk}Alternatively, one has another one-step Darboux transformation by letting
\begin{equation}
    \wh{\Lambda} = (T-f_2)(T-f_1)(T-f_3)+\beta\,,\quad \wh{\vf} =  (T-f_2)(T-f_1)\vf\,, 
\end{equation}
which amounts to
\begin{equation}
    \wh{\fh} = -f_2-\overline{f}_1-\overline{\overline{f}}_3\,,\quad \wh{\fg}  =f_1 f_2+ f_2\overline{f}_3 +\overline{f}_1 \overline{f}_3\,,\quad \alpha=\beta -f_1f_2f_3\,,
\end{equation}
with $\wh{\Lambda}  = T^3 +\wh{\fh}T^2 +\wh{\fg}T+\alpha$. Comparing to \eqref{eq:facint11}, one has
\begin{equation}
    \wh{\overline{\fh}} - \fh  =f_3 -\overline{\overline{f}}_3\,,\quad  \wh{\overline{\fg}} - \fg = f_3  \underline{\fh} -\overline{\overline{f}}_3\fh -f_3\overline{\overline{f}}_3+f_3 \underline{f}_3\,.
\end{equation}
\end{rmk}

Interpret the Darboux map as a shift operator $S$ acting on the eigenfunction:
\begin{equation}
  \label{eq:SS}
  S \vf: = \wt{\vf} =(T-f)\vf \,,
\end{equation}
where we suppress the subscript on $f$ and allow it to depend on the discrete variable $m$, \ie $f = f(n, m)$. Iterating this map yields
\begin{equation}\label{eq:aa}
  \wt{\vf} = \overline{\vf} -f\vf\,,\quad   \wt{\wt{\vf}} = \wt{\overline{\vf}} -\wt{f}\wt{\vf}\,,\quad   \wt{\wt{\wt{\vf}}} = \wt{\wt{\overline{\vf}}} -\wt{\wt{f}}\wt{\wt{\vf}}\,. 
\end{equation}
As in  \eqref{eq:matrix3x3}, one obtains a third-order linear difference equation in the $S$ shift: \begin{equation}
  \label{eq:dspd3}
(S^3+ \fp S^2 +\fq S +\beta)\vf  =\widetilde{\widetilde{\widetilde{ \varphi}}} + \fp\widetilde{\widetilde{\varphi}}+\fq  \widetilde{\varphi}+ {\beta} \varphi =\lambda \varphi\,,
\end{equation}
where
\begin{equation}
  \label{eq:6cuv}
\fp= \overline{\overline{f}} + \widetilde{\overline{f}} +\widetilde{\widetilde{f}} +\fh\,,\quad \fq = (\overline{\overline{f}}+\fh) (\overline{f} + \widetilde{f})+ \widetilde{\overline{f}}\widetilde{f}   + \fg\,,\quad  {\beta} =  (\overline{\overline{f}}+\fh)\overline{f}f +  f \fg +a\,.
\end{equation}

Now compare the two spectral problems \eqref{eq:factordop2} and \eqref{eq:dspd3}  connected by \eqref{eq:SS}. Introduce a potential variable $w$ such that
\begin{equation}\label{eq:fh}
    \fh = \overline{\overline{\overline{w}}} - w\,,\quad f = \wt{w} -\overline{w}\,,
\end{equation}
which is a  result of $    \widetilde{\fh} = \overline{\fh}+\overline{\overline{\overline{f}}}-f$  in the Darboux transformation \eqref{eq:dspDT1}. In terms of $w$,  $\fg$ becomes
\begin{equation}\label{eq:fg}
\fg =  \frac{\beta-\alpha}{f} - (\overline{\overline{f}}+\fh)\overline{f} = (\wt{\overline{w}} - \overline{\overline{w}})(w -  \wt{\overline{\overline{w}}} )+\frac{\beta-\alpha}{\wt{w} - \overline{w}}\,.   
\end{equation}
Remarkably,  in terms of $w$, the coefficients  $\fp$ and $\fq$  in the induced spectral problem \eqref{eq:dspd3} take exactly the same forms as $\fh$ and $\fg$  respectively  by interchanging $\{~\wt{~}~, \alpha \}  $ with $\{~\bar{~}~, \beta\}$. This reveals a self-duality between  \eqref{eq:factordop2} and \eqref{eq:dspd3} through \eqref{eq:SS}.  The self-duality is encoded in the symmetry:
\begin{equation}
\{T,\alpha\}   \leftrightarrow  \{ S, \beta \}\,,
\end{equation}
which exchanges the roles of the two lattice directions and their associated parameters.

The self-duality extends naturally to any number of lattice directions: given a third discrete direction associated with a shift operator, say, $ 
R$ associated with the  parameter $\gamma$,  the induced spectral problem retains the same structure as  \eqref{eq:factordop2} with coefficients taking the same forms as $\fh$ and $\fg$  through $\{T,\alpha\}   \leftrightarrow  \{ R, \gamma \}$.
\newcommand{\Sg}{\Sigma}

\subsection{BSQ equations and deformation with respect to lattice parameter}\label{sec:3.4}
Integrable BSQ-type equations arise as compatibility conditions of linear spectral problems. The time deformations could involve infinitely many continuous ``time variables'' corresponding to hierarchy of BSQ equations (see, for instance, \cite{Dickey}). The shift deformations are induced by Darboux transformations, and, in particular, the discrete Darboux transformations will give rise to discrete BSQ equations consistent on a multi-dimensional lattice. We also  consider deformation of the discrete spectral problem with respect to the lattice parameter. This yields a non-autonomous semi-discrete BSQ system which encodes the complete hierarchy of semi-discrete BSQ equations  \cite{TN1, TN2}

\paragraph{Continuous BSQ equation:} Consider the Lax pair (see, for instance, \cite{Dickey})
\begin{equation}
  \label{eq:lapBsq}
   \varphi''' + u \varphi'+v\varphi = \lambda\varphi\,, \quad  \dot{\varphi}  =\varphi'' + \frac{2}{3}u \varphi\,,
 \end{equation}
 where ~$\dot{~}$~ denotes derivative with respect to a time variable. Compatibility yields
\begin{equation}\label{eq:compcbsqe1}
\dot{v}-v'' +\frac{2}{3}u u'+\frac{2}{3}u'''=0\,, \quad  \dot{u}+u''-2v' = 0\, .
\end{equation}
Eliminating $v$ and their derivatives, one gets the standard BSQ equation
\begin{equation}
  \label{eq:cbsq}
  3\ddot{u}+4  (uu'+  u''')'=0\,.
\end{equation}
\paragraph{Semi-discrete BSQ equation:} Compatibility of the following system
\begin{equation}\label{eq:semidiscreteBSQ}
   \varphi''' + u \varphi'+v\varphi = \lambda\varphi\,, \quad  \overline{\varphi}  =\varphi' -r \vf\,,
   \end{equation}
with ~$\bar{~}$~ shift induced by the Darboux transformation \eqref{eq:cDT2} yields  
\begin{equation}
  \label{eq:calL3}
\overline{u} = u +3 r'\,,\quad \overline{v} = v+u'+3rr'+3r''\,,\quad   v+ur+r''+r^3+3rr'+\alpha=0\,. 
\end{equation}
By introducing a potential variable $w$ such that
\begin{equation}
  \label{eq:3ww}
  \overline{w} - w = r \,,\quad 3w' =u \,, 
\end{equation}
the above system can be reduced to a  semi-discrete equation \begin{equation}
  \label{eq:semidbsq1}
(w+\overline{w}+\overline{\overline{w}})'' = (\overline{w}-w)^3 + (\overline{w}-\overline{\overline{w}})^3 +3(\overline{w}-\overline{\overline{w}})\overline{\overline{w}}' +3(\overline{w}-w)w'=0\,,
\end{equation}
known as a semi-discrete potential BSQ equation \cite{Shabat2, W}. Note that the above $w$ is the same $w$ as in \eqref{eq:fh}. 
\paragraph{Discrete BSQ equation:} 
Taking the compatibility of 
  \begin{equation}\label{eq:laxlpBSQ}    \overline{\overline{\overline{\vf}}}+\fh\overline{\overline{\vf}}+\fg\overline{\vf}+\alpha\vf=\lambda\vf\,,\quad  \widetilde{\varphi} = \overline{\varphi} -f \vf
  \end{equation}
yields
  \begin{equation}
    \label{eq:diffbsqsys1}
\widetilde{\fh} =\overline{\fh} + \overline{\overline{\overline{f}}}-f\,,\quad \widetilde{\fg}=\overline{\fg}+\overline{\overline{f}}\, \overline{\fh}-f \fh+\overline{\overline{f}}(\overline{\overline{\overline{f}}}+f)\,, \quad \fg =\frac{\beta-\alpha}{f}-(\overline{\overline{f}}+\fh)\overline{f}\,. 
\end{equation}
In terms of the potential variable $w$, the above system leads to the lattice potential BSQ equation defined on a nine-point stencil \cite{GD}
  \begin{equation}\label{eq:latpotbsq}
\frac{\alpha-\beta}{\overline{\overline{w}} - \widetilde{\overline{w}}} -\frac{\alpha-{\beta}}{ \widetilde{\overline{w}}-\widetilde{\widetilde{w}} }= (\widetilde{\overline{\overline{w}}}-\widetilde{\widetilde{\overline{w}}})\widetilde{\widetilde{\overline{\overline{w}}}}-(\widetilde{\overline{\overline{w}}}-w)\overline{w}+(\widetilde{\widetilde{\overline{w}}}-w)\widetilde{w}\,. 
\end{equation}
The multi-dimensional consistency of  \eqref{eq:latpotbsq} could be understood as a result of the self-dual formulation of discrete spectral problems. 
Note that, recently,  the multi-dimensional consistency property of the nine-point lattice potential BSQ equation \eqref{eq:latpotbsq} was explicitly verified on a $3\times 3\times 3$-vertex cubic lattice comprising $27$ lattice vertices and $8$ elementary cubes \cite{SZZN}. 
\paragraph{Non-autonomous semi-discrete BSQ system:} Consider the compatibility of 
\begin{equation}\label{eq:genvf}
  \wb{\wb{\wb{\vf}}}+\fh \wb{\wb{\vf}}+\fg \wb{\vf} + s \vf = 0\,, \quad      \vf_s  =-w_s{\underline{ \vf}}\, , 
\end{equation}
where ${\underline{\cdot}}$ denotes the backward shift and  ${\cdot}_s$ denotes partial derivative with respect to $s$. Comparing to \eqref{eq:discrete3T}, the parameter $s$ is set as $ s = \alpha -\lambda$ with the spectral parameter  $\lambda$  being fixed, so that \begin{equation}
  \partial_s = {\partial_\alpha}\,.\end{equation}
Therefore, $\vf_s$ means deformation of $\vf$ with respect to the lattice parameter $\alpha$. The compatibility yields the following set of semi-discrete BSQ equations
\begin{equation}\label{eq:semisyst}
  \fg_s = \fh \wb{\wb{w}}_s - \underline{\fh}w_s\,,\quad w_s =\frac{n}{\underline{\fg}} \,. 
\end{equation}
Using \eqref{eq:fh} as in the self-dual formulation of the spectral problem  \eqref{eq:discrete3T} and eliminating $\fg$ in the above system yields a non-autonomous semi-discrete equation in $w$ only:
\begin{equation}\label{eq:genratingw}
  (n+1)\frac{\wb{w}_{ss}}{\wb{w}^2_s}= \left(\overline{\overline{\overline{w}}} - w\right) \wb{\wb{w}}_s - \left(\overline{\overline{w}} - \underline{w}\right)w_s\,, 
\end{equation}
which was known as  the {\em semi-discrete generating equation} for the (potential) BSQ hierarchy in the sense that it encodes the complete hierarchy of semi-discrete BSQ equations (including \eqref{eq:semidbsq1} and higher-order equations) \cite{TN1, TN2}. The pair \eqref{eq:genvf} will play a crucial role in later derivation of the  $\mathrm{PGL}(3)$-invariant generating systems (see Section \ref{sec:genPDEs}). 

\begin{rmk}
Soliton-type solutions for the above BSQ equations can be expressed in closed form via the Darboux--Crum formulae that are given in Appendix  \ref{app:DCF}. 
 
\end{rmk}

\section{$\mathrm{PGL}(3)$-invariant BSQ systems and generating systems}\label{sec:4}

This section first presents the continuous and discrete $\mathrm{PGL}(3)$-invariant BSQ systems (see \eqref{eq:z1z2PDE} and \eqref{eq:pgl3invsys} respectively) in terms of the common dependent variables  $z_1, z_2$ that are inhomogeneous coordinates of the Lax pairs of BSQ equations. 
These systems are natural rank-$3$ analogues of continuous and discrete Schwarzian KdV equations. A remarkable feature is that the $\mathrm{PGL}(2)$-invariant (Schwarzian) BSQ equations\textemdash both continuous and discrete\textemdash can be recovered from their $\mathrm{PGL}(3)$ counterparts via a {\em lifting-decoupling mechanism}.
By considering deformations with respect to the lattice parameters, we derive two additional  
$\mathrm{PGL}(3)$-invariant systems in $z_1, z_2$: a non-autonomous semi-discrete system \eqref{eq:gensemi} and a coupled system of fourth-order PDEs with the lattice parameters serving as independent variables \eqref{eq:genPDEs}. We refer to them as $\mathrm{PGL}(3)$-invariant generating systems as they encode the complete hierarchies of semi-discrete and continuous BSQ equations, respectively. 

\subsection{Continuous $\mathrm{PGL}(3)$-invariant BSQ system}
\label{sec:41}
We begin by deriving the continuous  $\mathrm{PGL}(3)$-invariant BSQ system from the Lax pair \eqref{eq:lapBsq}.  Let $\varphi_1, \varphi_2, \varphi$ be a basis of solutions to the spectral problem. In terms of the inhomogeneous coordinates
\begin{equation}\label{eq:inhomog}
z_1 = \varphi_1/\varphi, \quad z_2 =\varphi_2/\varphi  \,,
\end{equation}
  the compatibility of Lax pair yields a coupled system:
\be\label{eq:z1z2PDE} 
\frac{\dot{z}_1-z_1''}{z_1'}=\frac{\dot{z}_2-z_2''}{z_2'}=-\frac{2}{3}\frac{z_1'z_2'''-z_2'z_1'''}{z_1'z_2''-z_2'z_1''}\,. 
\ee
This system possesses the remarkable property of being invariant under the $\mathrm{PGL}(3)$ action \eqref{eq:projSL3}. This can be verified by comparing with the generating invariants $\cR_1, \cR_2$ in \eqref{eq:PGL3tinv}: the system \eqref{eq:z1z2PDE} is equivalent to the conditions
${\cal R}_1 = 1$ and $ {\cal R}_2 = 0$,  which are manifestly $\mathrm{PGL}(3)$-invariant. 
Therefore, we refer to   \eqref{eq:z1z2PDE} as the {\em $\mathrm{PGL}(3)$-invariant BSQ system}.

The derivation of \eqref{eq:z1z2PDE} is straightforward. Insert $\vf_1 =z_1\vf$, $\vf_2 =z_2 \vf$ into \eqref{eq:lapBsq}. It follows from  the third-order linear spectral problem (the first equation in \eqref{eq:lapBsq}) that \eqref{eq:dlogvf} holds. The second equation in  \eqref{eq:lapBsq} (the time evolution part of the Lax pair)  yields
   \begin{equation}\label{eq:z1z2ctins}
  \dot{z}_i=z_i''+2(\vf'/\vf)z_i\, , \quad i=1,2\, .   
  \end{equation}
  Combining  \eqref{eq:dlogvf} and \eqref{eq:z1z2ctins} leads to \eqref{eq:z1z2PDE}.

  There is a well-known one-to-one correspondence between the third-order linear differential operator  and classes of projectively equivalent nondegenerate curves in  $ {\mathbf P}^2 $ \cite{OT}. Viewing $z_1 ,z_2$ as an affine chart of a nondegenerate projective curve in  $ {\mathbf P}^2 $  (nondegeneracy means $z_1'z_2''-z_2'z_1'' \neq 0$), the lift to $(\varphi_1, \varphi_2, \varphi)$ corresponds precisely to the Lax pair of the BSQ equation \eqref{eq:lapBsq}. 
Therefore,  system \eqref{eq:z1z2PDE} governs a class of projectively equivalent nondegenerate curves in $ {\mathbf P}^2 $ evolving along the BSQ flow. 
It represents a natural rank-$3$ analogue of the Schwarzian KdV equation.

\begin{rmk}\label{rmk:OST}
The system \eqref{eq:z1z2PDE} is the affine curve flow given in
Remark~$6.6$ of \cite{OST}, which was obtained as the continuum limit of the pentagram map: 
  \begin{equation}
 \bma \dot{z_1} \\\dot{z_2}\ema =  \bma z_1'' \\z_2''\ema-\frac{2}{3}\frac{z_1'z_2'''-z_2'z_1'''}{z_1'z_2''-z_2'z_1''}  \bma z_1' \\z_2'\ema     \,. 
   \end{equation}
Equivalently, lifting $z_1, z_2$ to $\Gamma = ( z_1,\, z_2, \,1)^\intercal$,   one obtains the evolution equation   \begin{equation}
    \dot{\Gamma} = \Gamma'' - \frac{2}{3}\frac{ \det(\Gamma', \Gamma''', \Gamma)}{\det(\Gamma', \Gamma'', \Gamma)}\,\Gamma' \,.
\end{equation}
\end{rmk}

Another remarkable property of the $\mathrm{PGL}(3)$-invariant BSQ system \eqref{eq:z1z2PDE} is that each component independently satisfies a $\mathrm{PGL}(2)$-invariant equation, which is known as the {\em Schwarzian BSQ equation} \cite{Weiss}.

\begin{proposition}
Let $z_1, z_2$ be smooth solutions of the $\mathrm{PGL}(3)$-invariant BSQ system \eqref{eq:z1z2PDE}. Then, each component independently satisfies the $\mathrm{PGL}(2)$-invariant equation 
\begin{equation}\label{eq:SBSQ1}
3\partial_t\left(\frac{\dot{z}}{z'}\right)+\partial_x\left({\mathbf S}[z]+\frac{3}{2}\left(\frac{\dot{z}}{z'}\right)^2\right)=0\, , 
\end{equation} 
known as the Schwarzian BSQ equation \cite{Weiss},  where $ {\mathbf S}[z]$ denotes the Schwarzian derivative.
\end{proposition}

\prf
The proof relies on lifting the system \eqref{eq:z1z2PDE} to a three-component system via an auxiliary variable
\begin{equation}\label{eq:xi}
  \xi = \frac{z'_2}{z'_1}\,.  
\end{equation}
One can show that the resulting three-component system involving ($z_1,z_2,\xi$)  decouples into two two-component systems involving  $(z_1,\xi)$ and $(z_2, \xi)$, respectively. Each of these  can be  further reduced to a single-variable equation. A natural geometric interpretation of this mechanism is  presented in Remark \ref{rem:ld} below. 

Precisely, inserting \eqref{eq:xi} into \eqref{eq:z1z2PDE} yields a three-component system in $(z_1,z_2,\xi)$:
\begin{equation}
\label{eq:xi1}  \frac{z_2'}{z_1'}   = \frac{\dot{z}_2-z_2''}{\dot{z}_1-z_1''}  =\xi\,,
\quad   \frac{\dot{z}_1}{z_1'}=-\frac{2}{3}\frac{\xi''}{\xi'}-\frac{1}{3}\frac{z_1''}{z_1'}\,. 
\end{equation}
Differentiating the first equation in \eqref{eq:xi1} with respect to $x$ allows us to eliminate $z_2$:
\begin{equation}\label{eq:xi3}
  \dot{\xi} = \xi''+\left(\frac{\dot{z}_1+z_1''}{z_1'}\right)\xi' \,,
\end{equation}
which together with the second equation in \eqref{eq:xi1} forms a system in $(z_1,\xi)$. One can express this system as
\begin{equation} \label{eq:zxi}
   \frac{ \dot{z}_1}{z_1'}  = - \frac{\dot{\xi} +\xi''}{2 \xi'}\,, \quad       \frac{z''_1}{z_1'}  = \frac{3\dot{\xi}-\xi''}{2 \xi'}\,. 
   \end{equation}
Eliminating  $\xi$ or $z_1$ yields a single-variable equation in $z_1$ or $\xi$. Each of the equations is precisely the Schwarzian BSQ equation \eqref{eq:SBSQ1}. Similar procedures can be applied to  $z_2$. 
 \finprf
\medskip
\begin{rmk}[Geometric interpretation.]\label{rem:ld}
  The {\em lifting and decoupling} mechanism admits the following geometric picture. Consider the lift of $z_1, z_2$ to a curve $\Gamma = (z_1, z_2,1)$, then the lifted dual curve $\Xi$ is given by
  \begin{equation}
    \Xi =  \Gamma \times \Gamma' =\left( -z_2',   z_1' ,  z_1 z_2'-z_2 z_1'\right) \,, 
  \end{equation}
which,  by definition, is incident to $\Gamma$. The auxiliary variable $\xi$ can be identified with one of the affine coordinates of the dual projective curve, and the triple $(z_1,z_2;\xi)$  parametrises an {\em incident manifold} with a vanishing form $dz_2 - \xi dz_1 = 0 $. This form is preserved by the BSQ flow, \ie \eqref{eq:xi1}, and allows elimination of all derivatives of $z_2$ in favour of $z_1$ and $\xi$.

\end{rmk}

  Now let us see the natural group of transformations on the two-component system \eqref{eq:zxi}.
 Let $\xi_1, \xi_2$ be an affine chart of the dual curve $\Xi$  with
 \begin{equation}
\label{eq:xixi}\xi_1 =\xi \,,\quad \xi_2 = \frac{z_2' z_1-z_1' z_2}{z_1'} = z_1\xi_1 -z_2\,. 
 \end{equation}
 A $\mathrm{PGL}(3)$ action \eqref{eq:projSL3}  induces a $\mathrm{PGL}(3)$ transformation on $\xi_1, \xi_2$.   Consider a  standard embedding of $\mathrm{PGL}(2)$  into $\mathrm{PGL}(3)$ acting on $z_1, z_2$ as
 \begin{equation}\label{eq:pgl2emb}
(z_1, z_2)  \mapsto    \left(\frac{\textstyle m_{11}z_1+m_{31}}{\textstyle m_{13}z_1+m_{33}}\,, \frac{\textstyle z_2}{\textstyle m_{13}z_1+m_{33}} \right)\,, 
 \end{equation}
 with $\Delta = m_{11}m_{33} - m_{13}m_{31} \neq 0$. This yields
 \begin{equation}\label{eq:pgl2emb2}
   (\xi_1, \xi_2) \mapsto \left(  m_{33}\xi_1 +m_{13}\xi_2\,, m_{31}\xi_1 +m_{11}\xi_2 \right)\,.  
 \end{equation}
One can check that \eqref{eq:zxi} is invariant under the above  $\mathrm{PGL}(2)$ action by taking account of \eqref{eq:xixi}. This implies that the single-variable equation in $z_1$, \ie the Schwarzian BSQ equation \eqref{eq:SBSQ1},  is $\mathrm{PGL}(2)$-invariant.

\subsection{Discrete  $\mathrm{PGL}(3)$-invariant BSQ system} 
\label{sec:42}We now turn to the  exact discretisation of \eqref{eq:z1z2PDE} of the continuous system \eqref{eq:z1z2PDE}.  Recall the Lax pair of the discrete BSQ equation \eqref{eq:laxlpBSQ}. 
Taking  the inhomogeneous coordinates $z_1, z_2$  from the basis solutions $\varphi_1, \varphi_2, \varphi$ as in \eqref{eq:inhomog}, the compatibility condition yields a coupled system for 
$z_1, z_2$ of the form
\begin{equation}\label{eq:pgl3invsys}
  \frac{\wb{z}_1-z_1}{\wt{z}_1-\wb{z}_1}=  \frac{\wb{z}_2-z_2}{\wt{z}_2-\wb{z}_2}=\frac{\beta-\lambda}{\alpha-\beta}\frac{\left|\wt{\wb{\wb{\bz}}}, \wt{\wb{\bz}} , \bz \right|} 
  {\left| \wt{\wb{\wb{\bz}}},\wt{\wb{\bz}},\wt{\bz}\right|}\,, \quad  \bz  = (1,z_1,z_2)^\intercal \, ,
\end{equation}
defined on a five-point stencil in a square lattice. 
Note that the  equality of the first two terms in  \eqref{eq:pgl3invsys}  can  be equivalently written as
\begin{equation}\label{eqq:1}
  \left| \bz,\wb{\bz},\wt{\bz}\right| =0\,, 
\end{equation}
 defined on a three-point stencil. This equation yields the equality
\begin{equation}
  \frac{\wb{z}_1-z_1}{\wt{z}_1-\wb{z}_1}=    \frac{\wb{z}_2-z_2}{\wt{z}_2-\wb{z}_2} = \frac{\left| \wb{\wb{\bz}}, \bz , \wb{\bz}\right |}{\left|\wb{\wb{\bz}}, \wt{\bz} , \wb{\bz} \right |}\,, \end{equation}
which allows us to express the second equality of \eqref{eq:pgl3invsys} as 
\begin{equation}\label{eqq:2}
  \frac{\alpha-\beta}{\beta-\lambda} = \frac{\left|\wt{\wb{\wb{\bz}}}, \wt{\wb{\bz}} , \bz \right|} 
  {\left| \wt{\wb{\wb{\bz}}},\wt{\wb{\bz}},\wt{\bz}\right|}\frac{\left|\wb{\wb{\bz}}, \wt{\bz} , \wb{\bz} \right |}{\left| \wb{\wb{\bz}}, \bz , \wb{\bz}\right |}
=\frac{\left| \wt{\wb{\wb{\bz}}},\wt{\wb{\bz}},{\bz} \right|  \left|  {\wt{\wb{\bz}}},{\wt{\bz}},\wb{\wb{\bz}} \right| }{\left| \wt{\wb{\wb{\bz}}},{\wt{\wb{\bz}}},{\wt{\bz}} \right|   \left| {\wt{\wb{\bz}}},{{\bz}},\wb{\wb{\bz}} \right| } \cdot  \frac{\left| \wt{\wb{\bz}},{{\bz}},\wb{\wb{\bz}} \right|  \left|  {\wb{\wb{\bz}}},{\wt{\bz}},\wb{\bz} \right| }{\left| \wt{\wb{\bz}},\wt{\bz},\wb{\wb{\bz}} \right|   \left| {\wb{\wb{\bz}}},{{\bz}},\wb{\bz} \right| }
  \, . 
\end{equation}
 Comparing with the generating invariants \eqref{eq:fundinvsl3tx}, \eqref{eqq:2} is manifestly $\mathrm{PGL}(3)$-invariant. Therefore, we refer to \eqref{eq:pgl3invsys} as the {\em discrete $\mathrm{PGL}(3)$-invariant BSQ system}. It constitutes an exact discretisation of the continuous system \eqref{eq:z1z2PDE} in the sense of Shabat \cite{Shabat1, Shabat2}.

The system \eqref{eq:pgl3invsys} can be derived as follows: substituting $\vf_1 = z_1\vf, \vf_2 = z_2\vf$ into the discrete deformation part, \ie the second equation in \eqref{eq:laxlpBSQ}, yields
\begin{equation}
      \frac{\wt{\vf}}{\vf}=\frac{\wb{\vf}}{\vf}-f\   , \quad \wt{z}_i\frac{\wt{\vf}}{\vf}=\wb{z}_i\frac{\wb{\vf}}{\vf}-z_if\,   , \quad i=1,2\,.
\end{equation}
Arranging the above equations leads to
the following expressions for $f$
\begin{equation}\label{eq:fff111}
  f = \frac{\wt{z}_1-\wb{z}_1}{\wt{z}_1-z_1}\,\frac{\wb{\vf}}{\vf} = \frac{\wt{z}_2-\wb{z}_2}{\wt{z}_2-z_2}\,\frac{\wb{\vf}}{\vf}= \frac{\wt{z}_1-\wb{z}_1}{\wb{z}_1-z_1}\,\frac{\wt{\vf}}{\vf}=\frac{\wt{z}_2-\wb{z}_2}{\wb{z}_2-z_2}\,\frac{\wt{\vf}}{\vf}\, . 
\end{equation}
Also insert $\vf_1 = z_1\vf, \vf_2 = z_2\vf$ into the first equation in \eqref{eq:laxlpBSQ} which is the linear difference equation studied in Section \ref{sec:3.5}. It follows from the factorization  \eqref{eq:factordop2} that \bse
\begin{align}
(\lambda- \beta)\vf &= \wt{\wb{\wb{\vf}}} -(f_3+\wb{f}_2)\wt{\wb{\vf}}+f_2f_3\wt{\vf}\  , \\ 
(\lambda- \beta)z_i\vf &= \wt{\wb{\wb{z}}}_i\,\wt{\wb{\wb{\vf}}} -(f_3+\wb{f}_2)\wt{\wb{z}}_i\,\wt{\wb{\vf}}+f_2f_3\wt{z}_i\wt{\vf}\, ,  \quad i  =1,2\,, 
\end{align} 
\ese
with  $f_2f_3=(\beta-\alpha)/f$. This allows us to obtain
\begin{equation}\label{eq:lb111}
  \frac{1}{f}\frac{\wt{\vf}}{\vf }= \frac{\beta-\lambda}{\alpha-\beta}\frac{\left|\wt{\wb{\wb{\bz}}}, \wt{\wb{\bz}} , \bz \right|} 
{\left| \wt{\wb{\wb{\bz}}},\wt{\wb{\bz}},\wt{\bz}\right|}\, .   
\end{equation}
Combining \eqref{eq:fff111} and \eqref{eq:lb111} leads to \eqref{eq:pgl3invsys}. 

Beyond its  $\mathrm{PGL}(3)$-invariance, the system \eqref{eq:pgl3invsys} possesses two remarkable features: first, it can be ``lifted'' to a three-component quad-system that is multi-dimensional consistent; second, each of its components satisfies independently the discrete $\mathrm{PGL}(2)$-invariant BSQ equation known as the Schwarzian BSQ equation \cite{Weiss}.

The lift of  \eqref{eq:pgl3invsys}  can be done by introducing an auxiliary field $y$
  \begin{subequations} \label{eq:eq:3com3} \begin{equation}\label{eq:side11}
    y = \frac{\widetilde{z}_2-z_2}{\widetilde{z}_1-z_1} = \frac{\overline{z}_2-z_2}{\overline{z}_1-z_1}\,.     \end{equation}
  This allows us to get a quad-equation involving $z_1,y$ in the form
  \begin{equation}\label{eq:3com3}
         \frac{ \widetilde{\overline{y}} -\overline{y}}{\widetilde{\overline{y}} -\widetilde{y}} =\frac{(\alpha-\lambda)}{(\beta-\lambda)}\frac{(\overline{z}_1-z_1)}{(\widetilde{z}_1-z_1)}\frac{(\widetilde{\overline{z}}_1-\widetilde{z}_1)}{(\widetilde{\overline{z}}_1-\overline{z}_1)}\,.
  \end{equation}
  \end{subequations}
We regard  \eqref{eq:eq:3com3} as a three-component system. Alternative forms of \eqref{eq:side11} 
 \begin{equation}\label{eqq:side1}
      (\widetilde{z}_1-z_1)  y = \widetilde{z}_2-z_2\,,\quad (\overline{z}_1-z_1)y =\overline{z}_2-z_2
   \,, \end{equation}
    can be seen as ``side'' equations defined on edges of a quadrilateral. Taking respectively  ~$\bar{~}$~ and ~$\wt{~}$~ shifts of the first and second equations in \eqref{eqq:side1} yields expressions of $\wt{\wb{z}}_1, \wt{\wb{z}}_2$:
        \begin{equation}\label{eqq:quad3}
  \wt{\wb{z}}_1         =\frac{ \wb{y}\, \wb{z}_1  - \wt{y} \,\wt{z}_1 + \wt{z}_2- \wb{z}_2}{
 \wb{y} - \wt{y}}\,,\quad  \wt{\wb{z}}_2  = \frac{ \wb{y}\wt{y}(\wb{z}_1-\wt{z}_1) +\wb{y} \,\wt{z}_2 - \wt{y} \,\wb{z}_2    }{\wb{y} - \wt{y}}\,.  
        \end{equation}
Together with \eqref{eq:3com3}, one could express $\wt{\wb{y}}$ in terms of $z_1,z_2,y$ and their first-order shifts. This allows us to set up an initial-value problem on an elementary quadrilateral, and also on an elementary cube in a multi-dimensional lattice. 
       \begin{theorem}[Multi-dimensional consistency.]
The three-component quad-system  \eqref{eq:eq:3com3}  is consistent around an elementary cube in a multi-dimensional lattice. 
\end{theorem}

Reversely, by inserting the expressions of $\wt{\wb{y}}$ obtained respectively by shifts of \eqref{eqq:side1} into \eqref{eq:3com3}, one could recover \eqref{eq:pgl3invsys}, and its natural companion  $\mathrm{PGL}(3)$-invariant system \begin{equation}\label{eq:pgl3invsys2}
  \frac{\wt{z}_1-z_1}{\wb{z}_1-\wt{z}_1}=  \frac{\wt{z}_2-z_2}{\wb{z}_2-\wt{z}_2}=\frac{\alpha-\lambda}{\beta-\alpha}\frac{\left|\wt{\wt{\wb{\bz}}}, \wt{\wb{\bz}} , \bz \right|} 
{\left| \wt{\wt{\wb{\bz}}},\wt{\wb{\bz}},\wb{\bz}\right|}\,, 
\end{equation}
which can also be derived from \eqref{eq:pgl3invsys}  by interchanging  $(~\widetilde{~}~,\alpha)$ and  $(~\bar{~}~,\beta)$. Therefore, the three-component quad-system  \eqref{eq:eq:3com3} ``covers'' both   \eqref{eq:pgl3invsys} and  \eqref{eq:pgl3invsys2}. 

Similar to the continuous case, each
component of    \eqref{eq:pgl3invsys}  independently satisfies a nine-point $\mathrm{PGL}(2)$-invariant equation known as the lattice Schwarzian BSQ equation \cite{N1}.

\begin{proposition} Each component $z_1, z_2$ of the discrete $\mathrm{PGL}(3)$-invariant BSQ system \eqref{eq:pgl3invsys}  independently satisfies the discrete $\mathrm{PGL}(2)$-invariant equation 
  \begin{equation} \label{eq:lattSBSQ} 
\frac{(\alpha-\lambda) (\wt{\wt{\wb{z}}} - \wt{\wt{z}})(\wt{\wb{z}}-\wt{z})-(\beta-\lambda) (\wt{\wt{\wb{z}}} - \wt{\wb{z}})(\wt{\wt{z}}-\wt{z})}
{ (\beta-\lambda) (\wt{\wb{\wb{z}}} - \wb{\wb{z}})(\wt{\wb{z}}-\wb{z})-(\alpha-\lambda) (\wt{\wb{\wb{z}}} - \wt{\wb{z}})(\wb{\wb{z}}-\wb{z})} 
= \frac{(\wt{\wt{\wb{\wb{z}}}} -\wt{\wt{\wb{z}}})(\wb{\wt{z}}-\wt{\wt{z}})(\wt{z}-z)}{(\wt{\wt{\wb{\wb{z}}}}-\wt{\wb{\wb{z}}})(\wb{\wt{z}}-\wb{\wb{z}})(\wb{z}-z)}\, ,
\end{equation}
 known as the lattice Schwarzian BSQ equation \cite{N1}. 
\end{proposition}
\prf Using \eqref{eqq:side1} and the first equation in  \eqref{eqq:quad3}, one could eliminate $z_2$ and get a two-component system in $(z_1,y)$:
\begin{equation}\label{eq:ssys2com}
  \wt{\wb{z}}_1 = \frac{(\wb{y} -y)\wb{z}_1-(\wt{y} -y)\wt{z}_1 }{\wb{y} - \wt{y}}\,,\quad          \frac{ \widetilde{\overline{y}} -\overline{y}}{\widetilde{\overline{y}} -\widetilde{y}} =\frac{(\alpha-\lambda)}{(\beta-\lambda)}\frac{(\overline{z}_1-z_1)}{(\widetilde{z}_1-z_1)}\frac{(\widetilde{\overline{z}}_1-\widetilde{z}_1)}{(\widetilde{\overline{z}}_1-\overline{z}_1)}\,.
\end{equation}
For simplicity, let $\cA = \wb{y} -y, \cB =\wt{y} -y$, then $\wt{\cA}-\wb{\cB} =\cA-\cB $. Also let $\cP = \wb{z}_1-z_1 , \cQ = \wt{z}_1-z_1$ and  $\cC= \frac{\cP \wt{\cP}}{\cQ\wb{\cQ}} $, then \eqref{eq:ssys2com} can be written as
\begin{equation}\label{eq:inter1}
  \cA \,\wb{\cQ} =  \cB \, \wt{\cP} \,,\quad \frac{\wb{\cB}}{\wt{\cA}} = \frac{\alpha-\lambda}{\beta-\lambda}\cC\,, 
\end{equation}
and one can also get
\begin{equation}
  \cA = \frac{(\wb{\cB} - \wt{\cA}) \wt{\cP}}{\wb{\cQ} - \wt{\cP}}, \quad \cB =\frac{(\wb{\cB} - \wt{\cA}) \wb{\cQ}}{\wb{\cQ} - \wt{\cP}}\,,\quad \frac{\cA}{\cB} = \frac{\wt{\cP}}{\wb{\cQ}}\,.
\end{equation}
Taking backward shifts of \eqref{eq:inter1}:
\begin{equation}
  {\hypotilde 0 {\wb{\cB}}} = \frac{\alpha-\lambda}{\beta-\lambda} \cA\, {\hypotilde 0 {\cC}}\,,\quad {\underline{\wt{\cA}}} = \frac{\beta-\lambda}{\alpha-\lambda} \frac{\cB}{  {\underline{\cC}}}\,,\quad \frac{\alpha-\lambda}{\beta-\lambda}{\hypotilde 0 {\underline{\cC}}} = \frac{ {\hypotilde 0 \cB}}{{\underline{\cA}}} =  \frac{({{\hypotilde 0 {\wb{\cB}}}-\cA})  }{(\cB -  {\underline{\wt{\cA}}})}\frac{{\hypotilde 0 {\wb{\cQ}}} (\cQ  -  {\underline{\wt{\cP}}})}{        {\underline{\wt{\cP}}}({\hypotilde 0 {\wb{\cQ}}}      - \cP)} \,, 
\end{equation}
yields
\begin{equation}
  \frac{{{\hypotilde 0 {\wb{\cB}}}-\cA}  }{\cB -  {\underline{\wt{\cA}}}}  =\frac{\cA (\alpha-\lambda) \underline{\cC} ((\alpha-\lambda)  {\hypotilde 0 {\cC}} -(\beta-\lambda) )}{\cB (\beta-\lambda) ((\alpha-\lambda) \underline{\cC} -(\beta-\lambda))}\,. 
\end{equation}
Using the above relations leads to an equation involving $\cP,\cQ,\cC$  only which is precisely a backward-shifted copy of \eqref{eq:lattSBSQ}. Similarly, one could eliminate $z_1$ which leads to the same nine-point single-variable equation in $y$. From \eqref{eq:eq:3com3}, one can also obtain a two-component system in $z_2,y$ which can be decoupled to a single-variable equation in $z_2$. 
\finprf

\begin{rmk}\label{rmk:SZN}
  To the best of our knowledge, the three-component quad-equation \eqref{eq:eq:3com3}, as well as its ``projections''  \eqref{eq:pgl3invsys} and \eqref{eq:pgl3invsys2}, were first identified in \cite{SZZN} as reductions of a Q3 analogue of the lattice BSQ equation.  It appeared as a natural rank‑$3$ extension of the cross‑ratio equation (also known as the discrete Schwarzian KdV equation, or Q1$(\delta=0)$ in the Adler--Bobenko--Suris classification \cite{ABS1}), and represented a genuine extension of discrete integrable system of BSQ type generalizing previous known examples  \cite{GD, H, HZ, TN1, TN2, ZZN}. While the system \eqref{eq:eq:3com3} was first derived in \cite{SZZN} through reductions, our current approach demonstrates it emerges intrinsically from the factorised spectral problem \eqref{eq:laxlpBSQ} naturally associated with the $\mathrm{PGL}(3)$ symmetry. \end{rmk}

The discrete $\mathrm{PGL}(3)$-invariant system \eqref{eq:pgl3invsys} admits a similar geometric interpretation analogous to its continuous counterpart \eqref{eq:z1z2PDE}. The inhomogeneous coordinates $z_1,z_2$ which arise from the third-order spectral problem, characterize  a class of projectively equivalent nondegenerate {\em discrete projective curves} (or polygons) in ${\mathbf P}^2$ \cite{OST}. The lift to $(\vf_1, \vf_2, \vf) $  evolves precisely according to the discrete Lax pair \eqref{eq:laxlpBSQ}. Due to the covariance of the discrete spectral problem (as discussed in  Section \ref{sec:3.5}), the companion system \eqref{eq:pgl3invsys2} admits the same interpretation simply by interchanging the lattice directions.

A {\em dual discrete curve} $\Xi$ to $\Gamma = (z_1, z_2, 1)$ can be obtained through the following incident relation
\begin{equation}\label{eq:inci}
\Xi = \Gamma \times \overline{\Gamma}  = (z_2-\wb{z}_2 , \wb{z}_1 - z_1, \wb{z}_2 z_1 - \wb{z}_1 z_2)\,,
\end{equation}
which makes the auxiliary variable $y$  given by \eqref{eq:eq:3com3} as one of the inhomogeneous coordinates of $\Xi$.  This implies that  $(z_1, z_2;y)$  are living in a ``discrete incident space'' on ${\mathbf P}^2$, and the incident relation \eqref{eq:inci} is preserved by the discrete dynamics \eqref{eq:eq:3com3}. Let
\begin{equation}
y_1 = y, \quad y_2 = \frac{\wb{z}_1 z_2 -\wb{z}_2 z_1}{\wb{z}_1 - z_1} = \wb{z}_2 - \wb{z}_1 y_1
\end{equation}
be the inhomogeneous coordinates of $\Xi$. One can show that under the $\mathrm{PGL}(2)$ action  \eqref{eq:pgl2emb}, $(y_1, y_2)$ transform exactly as \eqref{eq:pgl2emb2}, and  the two-component system \eqref{eq:ssys2com} is invariant. Therefore, \eqref{eq:ssys2com} possesses a natural $\mathrm{PGL}(2)$ symmetry implying the   $\mathrm{PGL}(2)$ symmetry of the lattice Schwarzian BSQ equation \eqref{eq:lattSBSQ}.

\begin{rmk}
Using the Lax pairs of semi-discrete BSQ equations, one could derive semi-discrete $\mathrm{PGL}(3)$-invariant BSQ systems in an analogous manner. For instance,  \eqref{eq:semidiscreteBSQ} leads to 
\begin{equation}
       \frac{z_1'z_2'''-z_2'z_1'''}{z_1'z_2''-z_2'z_1''}-\frac{\overline{z}_1'\overline{z}_2'''-\overline{z}_2'\overline{z}_1'''}{\overline{z}_1'\overline{z}_2''-\overline{z}_2'\overline{z}_1''}        = 3\left(\log \zeta\right)'\,, \quad \zeta= \frac{z'_1}{\overline{z}_1-z_1}=\frac{z'_2}{\overline{z}_2-z_2}\,. 
\end{equation}
 \end{rmk}

\subsection{$\mathrm{PGL}(3)$-invariant generating systems}\label{sec:genPDEs}
The notion of a \textit{generating PDE} for the KdV hierarchy was first introduced in \cite{NHJ}. It appeared as a covariant (with respect to the independent variables) scalar integrable $\mathrm{PGL}(2)$-invariant PDE in which the lattice parameters of the corresponding lattice KdV equation serve as independent variables, while the discrete lattice variables play the role of parameters. Through a systematic expansion, this generating PDE yields the complete hierarchy of KdV equations. A {\em Schwarzian KdV (or $\mathrm{PGL}(2)$-KdV) generating PDE}, serving as a ``projective'' formulation in terms of the inhomogeneous variable was also developed in \cite{NHJ}, encoding the complete hierarchy of Schwarzian KdV equations. 
In \cite{TN1, TN2}, an analogous system of generating PDEs was presented for the BSQ hierarchy together with its Lagrangian structure. 
This coupled system is $\mathrm{PGL}(3)$-invariant and is associated with the complete hierarchy of BSQ equations. 

The generating PDEs possess significant physical relevance: the generating PDE for the KdV hierarchy is a generalisation of the Ernst equation of General Relativity describing gravitational waves, 
while the one for the BSQ hierarchy coincides with 
the Ernst equations arising in the Einstein--Maxwell--Weyl theory, which describes gravitational waves in the presence of neutrino fields subject to electromagnetic 
interaction.  This remarkable connection between water wave theory (KdV, BSQ) and gravitational wave theory  remains to be fully understood. 

In this subsection, we derive the projective formulation (in terms of  $z_1, z_2$) of the BSQ generating PDEs. 
The derivation relies on deformations of the discrete spectral problems with respect to the lattice parameters, \cf \eqref{eq:genvf}.
We begin by deriving a non-autonomous  semi-discrete system \eqref{eq:gensemi} in Section \ref{sec:4.31}, referred to as the {\em semi-discrete $\mathrm{PGL}(3)$-invariant generating  system}, playing a role analogous to the semi-discrete BSQ generating  equation \eqref{eq:genratingw}. 
Then in Section \ref{sec:deriv}, we present the $\mathrm{PGL}(3)$-invariant generating PDE, 
first as a coupled system of PDEs for determinants, and subsequently in vectorial 
form with its Lagrangian structure. The generating PDE corresponds to 
the entire BSQ hierarchy. 


While the explicit forms of the $\mathrm{PGL}(3)$-invariant generating systems are established
here, throughout analysis of their further properties is beyond the scope of the present
work and will be reserved for future investigations. These includes the hierarchy-generating mechanism; possible geometric interpretations; symmetry analysis and reductions to higher-rank Painlev\'e-type equations; explicit connections to physical models such as the Einstein--Maxwell--Weyl theory.

\subsubsection{A semi-discrete $\mathrm{PGL}(3)$-invariant generating system}\label{sec:4.31}
Let $\varphi_1, \vf_2, \vf$ be a basis of solutions to  \eqref{eq:genvf}. Inserting $z_i\vf =\vf_i $ ($i=1,2$) into the deformation part $\vf_s= -w_s\underline{ \vf}$ yields
\begin{equation}\label{eq:433} \bz_s\vf+\bz\vf_s=-w_s\underline{ \bz}\,\underline{ \vf} = \underline{ \bz} \vf_s~~   \Rightarrow ~~ 
  \partial_s \log \vf = \frac{\partial_s z_1}{\underline{z}_1-z_1} = \frac{\partial_sz_2}{\underline{z}_2-z_2}\,,\end{equation}
where ${\underline{\cdot}}$ denotes the backward shift in the lattice variable $n$, and   $\bz$ denotes the three-component vector $\bz =(1, z_1, z_2)^\intercal$. 
Taking a backward shift of the spectral problem, one obtains
\begin{equation}
 \wb{\wb{\bz}}\,\wb{\wb{\vf}}+\underline{ \fh}\,\wb{\bz}\,\wb{\vf}+ \underline{ \fg}\,\bz\,\vf+ s \underline{ \bz}\,\underline{ \vf}=0\,,
\end{equation}
which implies
\begin{equation}
  \underline{ \fg}\left| \wb{\wb{\bz}},\wb{\bz}, \bz \right|\vf+ s \left| \wb{\wb{\bz}},\wb{\bz},\underline{\bz}\right|\underline{ \vf}=0\,.
\end{equation}
Using again $\vf_s= -w_s\underline{ \vf}$ and taking $ w_s =\frac{n}{\underline{\fg}}$  in \eqref{eq:semisyst}, one obtains
\begin{equation}
  \label{eq:fds}
 \partial_s\log \vf = \frac{n}{s}\frac{\left| \wb{\wb{\bz}},\wb{\bz},{\bz}\right|}{\left| \wb{\wb{\bz}},\wb{\bz},\underline{ \bz} \right|}\,.
\end{equation}
Combining \eqref{eq:433} and \eqref{eq:fds} leads to 
\begin{equation}\label{eq:gensemi} 
   \frac{\partial_s z_1}{\underline{z}_1-z_1} = \frac{\partial_sz_2}{\underline{z}_2-z_2} =  \frac{n}{s}\frac{\left| \wb{\wb{\bz}},\wb{\bz},{\bz}\right|}{\left| \wb{\wb{\bz}},\wb{\bz},\underline{ \bz} \right|}\,,
\end{equation}
which is a $\mathrm{PGL}(3)$-invariant non-autonomous differential-difference system in $z_1, z_2$ with $s,n$ being the independent variables.
Similarly, one can obtain a covariant formulation of \eqref{eq:genvf} using the dual spectral problem  \eqref{eq:dspd3}:
\begin{equation}\label{eq:genvf2}
    \wt{\wt{\wt{\vf}}}+\fp \wt{\wt{\vf}}+\fq \wt{\vf} + t \vf = 0\,, \quad      \vf_t  =-w_t{\underline{ \vf}}\, , 
  \end{equation}
with $t = \beta - \lambda$. This leads to a dual system  
\begin{equation}\label{eq:gensemi2} 
\frac{\partial_t z_1}{{\hypotilde 0 z}_1- z_1 }= \frac{\partial_tz_2}{{\hypotilde 0 z}_2-z_2}= \frac{m}{t}\,
\frac{\big|\,\wt{\wt{\bz}},\wt{\bz},\bz\,\big|}{\big|\,\wt{\wt{\bz}},\wt{\bz},{\hypotilde 0 \bz}\,\big|}\,,   
\end{equation} 
with  independent variables $t,m$. 

\begin{rmk}
In addition to the non-autonomous differential-difference systems \eqref{eq:gensemi} and 
\eqref{eq:gensemi2}, we have also the compatible autonomous differential-difference systems 
\begin{subequations}
\label{eq:semiSL3} 
\begin{align}
\label{eq:semiSL3a} 
& \frac{\pl_\tau{z}_1}{z_1-\underline{z}_1}= \frac{\pl_\tau{z}_2}{z_2-\underline{z}_2}= \frac{\alpha'}{\alpha-\lambda}\,
\frac{\big|\,\wb{\wb{\bz}},\wb{\bz},\bz\,\big|}{\big|\,\wb{\wb{\bz}},\wb{\bz},\underline{\bz}\,\big|}\, ,  \\ 
\label{eq:semiSL3b} 
&\frac{\pl_\sigma{z}_1}{z_1-{\hypotilde 0 z}_1}= \frac{\pl_\sigma{z}_2}{z_2-{\hypotilde 0 z}_2}= \frac{\beta'}{\beta-\lambda}\,
\frac{\big|\,\wt{\wt{\bz}},\wt{\bz},\bz\,\big|}{\big|\,\wt{\wt{\bz}},\wt{\bz},{\hypotilde 0 \bz}\,\big|}\,,  
\end{align} 
\end{subequations}
in which $\alpha'=3\alpha^{2/3}$, $\beta'=3\beta^{2/3}$, and where $\tau$ and $\sigma$ are 
{\em Miwa variables} whose vector fields contain the differential vector fields of the 
entire hierarchy of flows, \ie 
\begin{equation}\label{eq:ab}
    \pl_\tau=\sum_{j=0}^\infty \alpha^{-(j+1)/3}\pl_{t_j}\ , \quad 
\pl_\sigma=\sum_{j=0}^\infty \beta^{-(j+1)/3}\pl_{t_j}\ ,  
\end{equation}  where the $t_j$ are the higher time-variables of the BSQ hierarchy. 
The equations \eqref{eq:semiSL3} are derived in a similar way as the non-autonomous system, 
namely by equipping the third-order difference spectral problems \eqref{eq:genvf} and 
\eqref{eq:genvf2} with the time-deformation equations 
\begin{equation}
    \pl_\tau\vf=-w_\tau\underline{\vf}\quad {\rm and}\quad \pl_\sigma\vf=-w_\sigma{\hypotilde 0 \vf}\ . 
\end{equation} 
The difference with the $s$ and $t$-derivatives is that there is no explicit $\tau$ or 
$\sigma$-dependence in the spectral problems, unlike \eqref{eq:genvf} and 
\eqref{eq:genvf2} which depend explicitly on $s$ or $t$. 

Alternatively, Eqs. \eqref{eq:semiSL3} can be derived by applying a ``skew continuum limit" (\cf
\cite{HNJ1}) of the difference-difference system \eqref{eq:pgl3invsys}, but we omit the details. 
The vector fields associated with the non-autonomous differential-difference equations \eqref{eq:gensemi} and \eqref{eq:gensemi2} can be regarded as ``master symmetries" of the $\mathrm{PGL}(3)$-invariant differential-difference hierarchy, \cf \cite{Xenitidis}. The hierarchy of continuous higher-order symmetries of the fully discrete system 
\eqref{eq:pgl3invsys} can be generated by expanding \eqref{eq:semiSL3} in powers of $\alpha$ or $\beta$ as in \eqref{eq:ab}.
\end{rmk}

  \subsubsection{A $\mathrm{PGL}(3)$-invariant generating PDE}\label{sec:deriv}

From here on, the 2-component vector $\bz=(z_1,z_2)^\intercal$ will be used rather than the 3-component 
vector $(1,z_1,z_2)^\intercal$ that was used in the previous subsection. 
In terms of their derivatives, and their $2\times 2$ determinants, the $\mathrm{PGL}(3)$-invariant generating PDE 
associated with the previous semi-discrete and fully discrete equations 
can be described as a coupled system involving two auxiliary scalar variables $P$ and $Q$ 
as follows.

\begin{theorem}
The generating PDE, a coupled system of PDEs in terms of the variables $s$ and $t$, 
is obtained from the following coupled system by eliminating the two auxiliary scalar variables $P, Q$:
\bse\label{eq:PQsystem}
\begin{align}
  P_s &= \left(\frac{1-n}{s} -\frac{|\bz_t,\bz_{ss}|}{|\bz_s,\bz_{t}|}\right) P  +\frac{s}{t} \frac{|\bz_s,\bz_{ss}|}{|\bz_s,\bz_{t}|}  Q\,,  \label{eq:PQsystema} \\
  Q_t &=\left(\frac{1-m}{t} +\frac{|\bz_s,\bz_{tt}|}{|\bz_s,\bz_{t}|}\right)  Q -\frac{t}{s} \frac{|\bz_t,\bz_{tt}|}{|\bz_s,\bz_{t}|}  P  \,,  \label{eq:PQsystemb} 
  \end{align}
and
  \begin{equation}
      P_t+Q_s =2\frac{|\bz_s,\bz_{st}|}{|\bz_s,\bz_{t}|} Q -2\frac{|\bz_t,\bz_{st}|}{|\bz_s,\bz_{t}|}   P   \,. \label{eq:PQsystemc} 
  \end{equation}
  \ese
  \end{theorem}


  
By eliminating the auxiliary variables $P, Q$ from the system \eqref{eq:PQsystem}, according 
to the computation outlined in Appendix \ref{app:genPDEproofs}, we can derive the following 
coupled system in terms of the $2\times 2$ determinants of the derivatives of the 
vector $\bz$: 
\bse
\label{eq:fullgenPDEs}
\begin{align}
 & \partial_s\left[2 \left(\frac{|\bz_t,\bz_{st}|}{|\bz_s,\bz_{t}|}\right)^2 + \left(\frac{|\bz_s,\bz_{tt}|}{|\bz_s,\bz_{t}|}-\frac{m-1}{t}\right)^2-\left(\frac{|\bz_t,\bz_{ss}|}{|\bz_s,\bz_{t}|} +2  \frac{|\bz_s,\bz_{st}|}{|\bz_s,\bz_{t}|}+\frac{n-1}{s}\right)  \frac{t}{s} \frac{|\bz_t,\bz_{tt}|}{|\bz_s,\bz_{t}|}\right]  \nonumber  \\
- & 2 \left(
    \frac{|\bz_s,\bz_{tt}|}{|\bz_s,\bz_{t}|}-\frac{m-1}{t}\right)\partial_t\left(
    \frac{|\bz_s,\bz_{st}|}{|\bz_s,\bz_{t}|}\right)    -  \partial_t\left( \frac{|\bz_t,\bz_{ss}|}{|\bz_s,\bz_{t}|}\right)
    \left(2  \frac{|\bz_t,\bz_{st}|}{|\bz_s,\bz_{t}|} -\frac{|\bz_s,\bz_{tt}|}{|\bz_s,\bz_{t}|}+\frac{m-1}{t}\right) \nonumber \\   
    + &2\partial_t\left( \frac{s}{t} \frac{|\bz_s,\bz_{ss}|}{|\bz_s,\bz_{t}|} \right) \frac{t}{s} \frac{|\bz_t,\bz_{tt}|}{|\bz_s,\bz_{t}|}  +
    \frac{s}{t} \frac{|\bz_s,\bz_{ss}|}{|\bz_s,\bz_{t}|} 
    \partial_t\left( \frac{t}{s} \frac{|\bz_t,\bz_{tt}|}{|\bz_s,\bz_{t}|} \right)   \nonumber \\
&= 2 \pl_s\pl_t\left(  \frac{|\bz_t,\bz_{st}|}{|\bz_s,\bz_{t}|}\right) -\partial_{s}^2\left(\frac{t}{s} \frac{|\bz_t,\bz_{tt}|}{|\bz_s,\bz_{t}|} \right)
       -\partial_{t}^2\left(\frac{|\bz_t,\bz_{ss}|}{|\bz_s,\bz_{t}|} \right)  \ , \label{eq:genPDE1}\\
&  \partial_t  \left[2 \left(\frac{|\bz_s,\bz_{st}|}{|\bz_s,\bz_{t}|}\right)^2 + \left(\frac{|\bz_t,\bz_{ss}|}{|\bz_s,\bz_{t}|}+\frac{n-1}{s}\right)^2-\left( \frac{|\bz_s,\bz_{tt}|}{|\bz_s,\bz_{t}|} +2  \frac{|\bz_t,\bz_{st}|}{|\bz_s,\bz_{t}|}-\frac{m-1}{t}\right)  \frac{s}{t} \frac{|\bz_s,\bz_{ss}|}{|\bz_s,\bz_{t}|}\right]   \nonumber \\
 - &2 \left(\frac{|\bz_t,\bz_{ss}|}{|\bz_s,\bz_{t}|}+\frac{n-1}{s}\right)\partial_s\left(
    \frac{|\bz_t,\bz_{st}|}{|\bz_s,\bz_{t}|}\right)    -  \partial_t\left( \frac{|\bz_t,\bz_{ss}|}{|\bz_s,\bz_{t}|}\right)
       \left(2 \frac{|\bz_s,\bz_{st}|}{|\bz_s,\bz_{t}|}
    -\frac{|\bz_t,\bz_{ss}|}{|\bz_s,\bz_{t}|}-\frac{n-1}{s}\right)  \nonumber\\   
    +&  2\partial_s\left(
    \frac{t}{s} \frac{|\bz_t,\bz_{tt}|}{|\bz_s,\bz_{t}|} \right)
    \frac{s}{t} \frac{|\bz_s,\bz_{ss}|}{|\bz_s,\bz_{t}|} +
 \frac{t}{s} \frac{|\bz_t,\bz_{tt}|}{|\bz_s,\bz_{t}|} 
    \partial_s\left(     \frac{s}{t} \frac{|\bz_s,\bz_{ss}|}{|\bz_s,\bz_{t}|}  \right) \nonumber\\
    &= -2 \pl_s\pl_t\left(
    \frac{|\bz_s,\bz_{st}|}{|\bz_s,\bz_{t}|}\right) +\partial_{s}^2\left(
    \frac{|\bz_s,\bz_{tt}|}{|\bz_s,\bz_{t}|}\right)
    +\partial_{t}^2\left(
    \frac{s}{t} \frac{|\bz_s,\bz_{ss}|}{|\bz_s,\bz_{t}|} \right)\ . \label{eq:genPDE2}
     \end{align}
\ese
This is one form of the $\mathrm{PGL}(3)$-invariant generating PDE; its vector form, which admits a Lagrangian structure, will be introduced later.
Before proceeding, a detailed derivation of the system \eqref{eq:PQsystem} is presented first. The derivation 
is quite nontrivial, relying on intricate compatibility conditions of the factorised 
\textit{discrete} 
spectral problems and the interplay with the underlying discrete structure of lattice shifts. 

\paragraph{Derivation of \eqref{eq:PQsystem}:}Collect the following set of linear equations for $\vf$:
\bse\label{eq:1}
\begin{align}
\label{eq:11}  \wb{\wb{\wb{\vf}}}+\fh \wb{\wb{\vf}}+\fg \wb{\vf} + s \vf = 0\,, \quad   \fh = \wb{\wb{\wb{w}}}-w\,,\quad       & \vf_s  =-w_s{\underline{ \vf}}\, ,  \\
\label{eq:12}    \wt{\wt{\wt{\vf}}}+\fp \wt{\wt{\vf}}+\fq \wt{\vf} + t \vf = 0\,, \quad \fp=\wt{\wt{\wt{w}}} - w\,,\quad  &     \vf_t  =-w_t {\hypotilde  0\vf}\, ,\end{align} 
 and
  \begin{equation}
    \label{eq:13} \wt{\vf} +\wt{w}\,{\vf} = \wb{\vf} + \wb{w}\,\vf \,, 
  \end{equation}
  \ese
  where the first two equations are just copies of \eqref{eq:genvf} and \eqref{eq:genvf2} with the deformation variables $s, t$ related to the lattice parameters $\alpha, \beta$ as
  \begin{equation}
  s = \alpha -\lambda\,,\quad t  = \beta -\lambda \quad \Rightarrow \quad   \partial_s = \partial_\alpha\,,\quad   \partial_t = \partial_\beta\,.
\end{equation}
Their compatibility yields (\cf Section \ref{sec:3.4})
\begin{equation}
      \fg_s = \fh \wb{\wb{w}}_s - \underline{\fh} w_s\,,\quad  w_s  \underline{\fg} = n\,,  \quad 
       \fq_t = \fp \wt{\wt{ w}}_t - {\hypotilde 0 \fp} w_s\,,\quad  w_t   {\hypotilde  0\fq} = m\,  .
\end{equation}
Shifts of $\vf$ in different directions are linked via \eqref{eq:13} (\cf Section \ref{sec:3.5}). It follows from \eqref{eq:13} that (by shifting \eqref{eq:13} and using \eqref{eq:13} for lower-order shifts substitutions)
\begin{equation}\label{eq:14}
\wt{\wt{\vf}} + \wt{\wt{w}} \,\wt{\vf}  +  \wt{w}\, \wt{\wb{w}}\,\vf =   \wb{\wb{\vf}} + \wb{\wb{w}} \,\wb{\vf} + \wb{w} \,\wt{\wb{w}}\,\vf \,.
\end{equation}
 The above set of equations  from \eqref{eq:1}  to \eqref{eq:14} is all we need to derive the system \eqref{eq:PQsystem}, using the following steps:
 \begin{enumerate}
 \item Compute $ \vf_{st}$ and $\vf_{ts}$ using the fundamental deformations $\vf_s  =-w_s{\underline{ \vf}} $ and $\vf_t  =-w_t {\hypotilde  0\vf}$. Combine expressions of $\underline{\vf}_s$ and ${\hypohat 0 \vf}_t$ as backward shifts of the fundamental deformations, and use \eqref{eq:13} to get ${\hypotilde 0 {\underline{\vf}}} $. The compatibility yields
\bse\begin{equation}\label{eq:weq}
w_{st}=\frac{w_t{\hypotilde 0 w}_s-w_s \underline{w}_t}{ \underline{w}- {\hypotilde 0 w}}\ , 
\end{equation} 
and 
\begin{equation}\label{eq:vfeq}
\vf_{st}=\frac{w_t {\hypotilde 0 w}_s }{w_s( \underline{w}- {\hypotilde 0 w})}\vf_s  -\frac{w_s\underline{w}_t}{w_t( \underline{w}- {\hypotilde 0 w})} \vf_t\, .   
\end{equation}\ese
\item Let $\vf_1, \vf_2, \vf$ be a basis of solutions for \eqref{eq:1}. Let $\bz=(z_1,z_2)^\intercal$, and insert $z_1  = \vf_1/\vf$, $z_2 = \vf_2/\vf$ into $\vf_s  =-w_s{\underline{ \vf}} $, $\vf_t  =-w_t {\hypotilde  0\vf}$:
\begin{equation}\label{eq:zz1eq}
\bz_s=(\underline{ \bz}-\bz)\,\pl_s\log\vf\ , \quad \bz_t=({\hypotilde 0 \bz}-\bz)\,\pl_t\log\vf\ .
\end{equation}
Insert $z_1  = \vf_1/\vf$, $z_2 = \vf_2/\vf$ into \eqref{eq:13}: 
\begin{equation}\label{eq:zz2eq} 
 (\wb{\bz}-\bz)\frac{\wb{\vf}}{\vf}=(\wt{\bz}-\bz)\frac{\wt{\vf}}{\vf}\  . 
\end{equation} 
Insert $z_1  = \vf_1/\vf$, $z_2 = \vf_2/\vf$ into  \eqref{eq:vfeq}: 
\begin{equation}\label{eq:zeq}
\bz_{st}+(\pl_t\log\vf)\bz_s+(\pl_s\log\vf)\bz_t=\frac{w_t {\hypotilde 0 w}_s }{w_s( \underline{w}- {\hypotilde 0 w})}\bz_s  -\frac{w_s\underline{w}_t}{w_t( \underline{w}- {\hypotilde 0 w})} \bz_t\,.
\end{equation}
Contracting \eqref{eq:zeq} with $\bz_s$ and  $\bz_t$ respectively using $2\times 2$ determinants yields 
\begin{equation} 
\frac{|\bz_s,\bz_{st}|}{|\bz_s,\bz_{t}|}=-\pl_s\log\vf - \frac{w_s\underline{w}_t}{w_t( \underline{w}- {\hypotilde 0 w})} \ , 
\quad  
\frac{|\bz_t,\bz_{st}|}{|\bz_s,\bz_{t}|}=\pl_t\log\vf  - \frac{w_t {\hypotilde 0 w}_s }{w_s( \underline{w}- {\hypotilde 0 w})}\,. \label{eq:zzrels} 
   \end{equation} 
Taking account of \eqref{eq:weq}, this leads to the relation
\begin{equation}\label{eq:wweq}
w_t\left(\frac{|\bz_s,\bz_{st}|}{|\bz_s,\bz_{t}|}+\pl_s\log\vf\right)   
-w_s\left(\frac{|\bz_t,\bz_{st}|}{|\bz_s,\bz_{t}|}-\pl_t\log\vf\right)=w_{st}\ ,  
\end{equation}
from which \eqref{eq:PQsystemc} is derived by introducing 
the quantities $P, Q$ (which play the roles of auxiliary variables as in \eqref{eq:PQsystem}), defined as
\begin{equation}\label{eq:PQ}
   P\equiv \frac{w_s}{\vf^2}\,,\quad  Q\equiv \frac{w_t}{\vf^2}\, . 
\end{equation}
\item We prove the following relation
  \begin{equation}\label{eq:21}
\frac{\underline{w}_s w_s}{ {\hypotilde 0 w}_t w_t } =    \frac{s^2}{t^2}\frac{|\bz_s,\bz_{ss}|}{|\bz_t,\bz_{tt}|} \frac{w_t}{w_s}\,.
\end{equation}
Using \eqref{eq:11} and \eqref{eq:12}, one gets induced expressions of $\vf_s, \vf_{ss} $ (respectively $\vf_t, \vf_{tt} $) in terms of $\vf, \wb{\vf}, \wb{\wb{\vf}}$  (respectively  $\vf, \wt{\vf}, \wt{\wt{\vf}}$). Inversely, this allows us to express $\wb{\vf}, \wb{\wb{\vf}}$  (respectively $\wt{\vf}, \wt{\wt{\vf}} $) in terms of $\vf, \vf_s, \vf_{ss}$  (respectively  $\vf, \vf_t, \vf_{tt}$). Combining with \eqref{eq:13} and \eqref{eq:14} yields two linear equations in $\vf$ involving $\vf_{ss}, \vf_{tt}, \vf_s, \vf_t, \vf$. Inserting  $z_1  = \vf_1/\vf$, $z_2 = \vf_2/\vf$ into these equations leads to \eqref{eq:21}. 
\item We have from the double back-shifted discrete spectral problems \eqref{eq:11} and \eqref{eq:12}:  
\bse
\begin{align}
\wb{\vf}+\underline{\underline{\fh}}\vf   
 - \left(\frac{1}{w_s}\underline{\underline{\fg}}+s\frac{ w_{ss}}{\underline{w}_sw_s^2}\right)\vf_s &=- 
\frac{s}{\underline{w}_sw_s } \vf_{ss}\  ,   \\
\wt{\vf}+ {\hypotilde 0{\hypotilde 3 \fp}}\vf   
 - \left(\frac{1}{w_t}{\hypotilde 0{\hypotilde 3 \fq}}+t\frac{ w_{tt}}{  {\hypotilde 0 w}_tw_t^2}\right)\vf_t &=-
\frac{t}{{\hypotilde 0 w}_tw_t } \vf_{tt} \,.\end{align}
\ese
Applying $z_1  = \vf_1/\vf$, $z_2 = \vf_2/\vf$ again yields 
\bse\begin{align}\label{eq:z-vecta}
(\wb{\bz}-\bz)\frac{\wb{\vf}}{\vf}+
      \frac{1}{w_s\underline{ w}_s}
      \left( 1-n -s\, \pl_s\log\left(\frac{w_s}{\vf^2}\right)\right)       \bz_s& = -\frac{s}{w_s\underline{ w}_s}\bz_{ss}\ ,  \\
      \label{eq:z-vectb}
(\wt{\bz}-\bz)\frac{\wt{\vf}}{\vf}
+\frac{1}{w_t{\hypotilde 0 w}_t}\left( 1-m -t\,\pl_t\log\left(\frac{w_t}{\vf^2}\right)
\right) \bz_t &= -\frac{t}{w_t{\hypohat 0 w}_t}\bz_{tt}\ .
\end{align}
\ese 
Subtracting these from each other and using \eqref{eq:zz2eq}, we get the vectorial relation 
\begin{equation} 
\frac{w_t{\hypotilde 0 w}_t}{w_s\underline{ w}_s}\,\left[ \left(\pl_s\log\,P
+\frac{n-1}{s}\right) \bz_s -\bz_{ss}\right]=  \frac{t}{s}\,\left[\left( \pl_t\log\,Q
+\frac{m-1}{t}\right) \bz_t - \bz_{tt}\right] , \label{eq:zvecteq} 
\end{equation}
between $\bz_s,\bz_t,\bz_{ss}$ and $\bz_{tt}$, with   
$P, Q$ given in \eqref{eq:PQ}.  
\item Final expressions: on the one hand, combining \eqref{eq:21}, \eqref{eq:zvecteq} and \eqref{eq:PQ},  we find the vectorial equation
\begin{equation} 
\frac{s}{\left|\bz_t,\bz_{tt}\right|}\left[ \left(Q_t+\frac{m-1}{t}Q \right)\bz_t-Q\bz_{tt}\right]
= \frac{t}{\left|\bz_s,\bz_{ss}\right|}\left[ 
\left(P_s+\frac{n-1}{s}P \right)\bz_s-P\bz_{ss}\right]\,.\label{eq:altPQsystemvect} 
\end{equation} 
On the other hand, making contractions   with $\bz_s$ and  $\bz_t$ respectively using $2\times 2$ determinants leads to \eqref{eq:PQsystema} and \eqref{eq:PQsystemb}.  Conversely, inserting the expressions for $P_s$ and $Q_t$ from \eqref{eq:PQsystema} and \eqref{eq:PQsystemb} into \eqref{eq:altPQsystemvect} we get 
a combination between two vectorial identities between $\bz_s,\bz_t,\bz_{ss},\bz_{tt}$ and $\bz_{st}$, showing 
that the system \eqref{eq:PQsystem} is fully equivalent to the vectorial equation 
combined with \eqref{eq:PQsystemc}. 
 
 \end{enumerate}

Let us comment on the $\mathrm{PGL}(3)$-invariance of the equations \eqref{eq:PQsystem} and \eqref{eq:altPQsystemvect}.
  Recall $\bz = (z_1, z_2)^\intercal$.  The $\mathrm{PGL}(3)$ transformation, \cf \eqref{eq:projSL3},
\begin{equation}
   \label{eq:bzpgl3} \bz^\intercal\,\mapsto\,\bz'^\intercal\,, \quad \bz'^\intercal  =\left(
\frac{m_{11}z_1+m_{21}z_2+m_{31}}{m_{13}z_1+m_{23}z_2+m_{33}}, 
\frac{m_{12}z_1+m_{22}z_2+m_{32}}{m_{13}z_1+m_{23}z_2+m_{33}}\right) \, ,
\end{equation}  
can be alternatively written as
\begin{equation}\label{eq:3dtrans}
\begin{aligned}
    (\bz^\intercal,1) \mapsto (\bz'^\intercal,1) &= (\bz^\intercal,1)\frac{1}{\mu}\bM \\
    & =\left(
\frac{m_{11}z_1+m_{21}z_2+m_{31}}{m_{13}z_1+m_{23}z_2+m_{33}}, 
\frac{m_{12}z_1+m_{22}z_2+m_{32}}{m_{13}z_1+m_{23}z_2+m_{33}},1\right)\, ,
    \end{aligned}
\end{equation}
with $ \mu=m_{13}z_1+m_{23}z_2+m_{33}$. Here and until the end of this section, primes temporarily denote the coordinates after the transformation.
Using \eqref{eq:PQ} and the fact that $w$ is $\mathrm{PGL}(3)$-invariant, one has 
\begin{equation}
P\,\mapsto\, P'=\frac{1}{\mu^2}P\ , \qquad   
Q\,\mapsto\,Q'=\frac{1}{\mu^2}Q\,,
\end{equation}
which can be easily verified by recognising that $\vf=\vf_3$ as we did in Section \ref{sec:pgl3}. In fact, the transformation implies the invariance of the ratios 
\begin{equation}
    \frac{|\bz_s,\bz_{ss}|}{|\bz_t,\bz_s|} \ , \quad \frac{|\bz_t,\bz_{tt}|}{|\bz_t,\bz_s|}\,,
\end{equation}  
while 
\begin{equation}
    |\bz'_t,\bz'_{s}|= \frac{\det(\bM)}{\mu^3}|\bz_t,\bz_s|\, ,
\end{equation}
and we have the following transformation rules for the following ratios: 
\[ \frac{|\bz'_t,\bz'_{st}|}{|\bz'_t,\bz'_s|}=\frac{|\bz_t,\bz_{st}|}{|\bz_t,\bz_s|}-\pl_t\log\mu\ , \quad {\rm and}\quad \frac{|\bz'_t,\bz'_{ss}|}{|\bz'_t,\bz'_s|} 
=\frac{|\bz_t,\bz_{ss}|}{|\bz_t,\bz_s|}-2\pl_s\log\mu\ , \]
and similarly for the ratios of determinants with $s$ and $t$ interchanged. 
With these relations, the invariance of \eqref{eq:PQsystem} is easily established. 

One can show the $\mathrm{PGL}(3)$-invariance of the vectorial-relation \eqref{eq:altPQsystemvect} as well. This is 
verified on the basis of the following transformation rules for  
the vectors $\bz_s,\bz_t,\bz_{ss}$ and $\bz_{tt}$ following \eqref{eq:bzpgl3} and \eqref{eq:3dtrans}: 
\begin{equation*}
    \begin{aligned}
 (\bz^\intercal,1)\, &\mapsto \,(\bz'^\intercal,1) = (\bz^\intercal,1)\frac{1}{\mu}\bM\, ,\\
(\bz^\intercal_s,0)\,&\mapsto\,(\bz'^\intercal_s,0)=\left[(\bz^\intercal_s,0)-\frac{\mu_s}{\mu}(\bz^\intercal,1)\right]\frac{1}{\mu}\bM \ , \\ 
(\bz^\intercal_{st},0)\,&\mapsto\,(\bz'^\intercal_{st},0)= \left[(\bz^\intercal_{st},0)-\frac{\mu_t}{\mu}(\bz^\intercal_{s},0)-\frac{\mu_s}{\mu} 
(\bz^\intercal_{t},0)+\mu\left(\pl_s\pl_t \frac{1}{\mu}\right)(\bz^\intercal,1)\right] \frac{1}{\mu}\bM\ ,  \\ 
(\bz^\intercal_{ss},0)\,&\mapsto\,(\bz'^\intercal_{ss},0)= \left[(\bz^\intercal_{ss},0)-2\frac{\mu_s}{\mu}(\bz^\intercal_{s},0)
+\mu\left(\pl_s^2\frac{1}{\mu}\right)(\bz^\intercal,1)\right] \frac{1}{\mu}\bM\ , 
\end{aligned}
\end{equation*}
and similar rules with $s$ and $t$ interchanged. In verifying the invariance of the 
vector form one uses also that the quantity $\mu$, due to the fact that it is a linear 
combination of the components $z_1$ and $z_2$, also satisfies the relation 
\eqref{eq:altPQsystemvect} albeit in component form.

Finally, we finish by presenting the vector form of the generating PDE.  
As explained at the end of Appendix \ref{app:genPDEproofs}, the determinantal form of the 
generating PDE \eqref{eq:fullgenPDEs} can be cast into a vectorial form. The actual 
computation is quite long, so we only present the result, namely 
\begin{equation}\label{eq:genPDEs}
  \begin{aligned}
  \partial_{s}  \left(\frac{\left| \bz_t,\bz_{st} \right|}{\left| \bz_s,\bz_{t} \right|}\left[
    \frac{1}{\left|\bz_s,\bz_t\right|}\left(\frac{n}{s^2}\bz_t +\frac{m}{t^2}\bz_s\right) -\frac{(s-t)\left(\left|\bz_t, \bz_{st}\right|  \bz_s-\left|\bz_s, \bz_{st}\right|\bz_t\right)}{s\,t\left|\bz_s, \bz_{t}\right|^2}    \right]
\right) &  \\ 
-\partial_{t}  \left(\frac{\left| \bz_s,\bz_{st} \right|}{\left| \bz_s,\bz_{t} \right|}
  \left[
    \frac{1}{\left|\bz_s,\bz_t\right|}\left(\frac{n}{s^2}\bz_t +\frac{m}{t^2}\bz_s\right) -\frac{(s-t)\left(\left|\bz_t, \bz_{st}\right|  \bz_s-\left|\bz_s, \bz_{st}\right|\bz_t\right)}{s\,t\left|\bz_s, \bz_{t}\right|^2}    \right]
\right) & \\ = \partial_{s}\partial_t \left[
    \frac{1}{\left|\bz_s,\bz_t\right|}\left(\frac{n}{s^2}\bz_t +\frac{m}{t^2}\bz_s\right) -\frac{(s-t)\left(\left|\bz_t, \bz_{st}\right|  \bz_s-\left|\bz_s, \bz_{st}\right|\bz_t\right)}{s\,t\left|\bz_s, \bz_{t}\right|^2}
  \right]\,.\, &      \end{aligned}
\end{equation}
We prefer to refer to this system, rather than \eqref{eq:fullgenPDEs}, as the {\em $\mathrm{PGL}(3)$-invariant generating PDE}, since the vectorial form is more compact and exhibits the $
(s,n)\leftrightarrow (t,m)$ symmetry manifestly.  Based on the analogy with the rank-$2$ case \cite{NHJ}, the generating PDEs \eqref{eq:genPDEs} are expected to encode the complete hierarchy of  $\mathrm{PGL}(3)$-invariant BSQ equations through a systematic expansion. A detailed verification of this hierarchy-generating mechanism will be pursued in future investigations. 
A detailed verification of this hierarchy-generating mechanism will be pursued in future investigations. 

This system actually possesses a Lagrangian 
formulation, which is given by the following Lagrangian
\begin{equation}\label{eq:Lag}
  \cL(\bz,s,t;n,m) =\frac{n}{s^2}\frac{\left| \bz_t,\bz_{st} \right|}{\left|\bz_s,\bz_{t} \right|}+\frac{m}{t^2}\frac{\left| \bz_s,\bz_{st} \right|}{\left|\bz_s,\bz_{t} \right|}-\frac{s-t}{s\,t}\frac{\left|\bz_s,\bz_{st} \right|\left|\bz_t,\bz_{st} \right|}{\left|\bz_s,\bz_{t} \right|^2}\,, 
\end{equation}
from which \eqref{eq:genPDEs} arises as the Euler-Lagrange equation. 
Their $\mathrm{PGL}(3)$-invariance of the latter is clear from the following Proposition. 

\begin{proposition}
The Lagrangian \eqref{eq:Lag} is invariant under $\mathrm{PGL}(3)$ 
transformations up to null Lagrangian terms, \ie up to divergence terms; this can be, for instance, verified by showing that, for all the eight evolutionary infinitesimal generators $X$ given by \eqref{3DIG},
\begin{equation}
    \mathbf{pr}^{(2)}X(\cL) = \operatorname{Div}{\operatorname{F}}
\end{equation}
holds for some $2$-tuple $\operatorname{F}=(\operatorname{F}^1(s,t,\bz,\bz_s,\bz_t,\ldots),\operatorname{F}^2(s,t,\bz,\bz_s,\bz_t,\ldots))$, where $\operatorname{Div}$ denotes the continuous divergence with the independent variables $s,t$, and $\mathbf{pr}$ denotes the prolongation of vector fields (see, \eg \eqref{eq:pro}).
Namely, the $\mathrm{PGL}(3)$ action is a divergence 
symmetry of the associated variational problem and therefore ensures the $\mathrm{PGL}(3)$-invariance of the Euler--Lagrange equations \eqref{eq:genPDEs} (see, \eg \cite{Olverb}). 
\end{proposition}
 
 One can prove this statement through straightforward computations. Here, we remark two curious aspects of the Lagrangian \eqref{eq:Lag}.  
\begin{rmk}\label{rem:inversion}
A $\mathrm{PGL}(3)$-invariant Lagrangian (modulo null Lagrangian terms) yielding a generating system for the BSQ hierarchy was first given in \cite{TN1, TN2}. Remarkably, the Lagrangian \eqref{eq:Lag} is related to the one given in \cite{TN1, TN2} by an inversion of the independent variables
\begin{equation}
    (s,t ) \mapsto (p,q) = \left(\frac{1}{s}, \frac{1}{t}\right)\,. 
  \end{equation}
The same transformation was known in the rank-$2$ KdV case, connecting the generating PDE for the KdV hierarchy and its Schwarzian version \cite{NHJ}. This remarkable feature suggests a unified hierarchy-generating structure for both integrable PDEs of arbitrary rank and their projective formulations. The meaning of this inversion and its role in connecting integrable hierarchies remains curious, and warrants further investigation.
 \end{rmk}

 \begin{rmk}\label{rmk:multi}
    We conjecture that the Lagrangian \eqref{eq:Lag}, like its $\mathrm{PGL}(2)$ analogue, \cf \cite{NHJ}, possesses the aspect of a Lagrangian multiform in the sense of \cite{LN}, \ie a differential $2$-form in a space of many variables of the type $s,t$ whose components are given by this Lagrangian, and which is closed on solutions of the Euler--Lagrange equations.
 \end{rmk}

 \section{Concluding remarks}
Based on third-order linear spectral problems and their factorised forms, this paper established a unified framework for projective formulations of $\mathrm{PGL}(3)$ differential and difference invariants, as well as $\mathrm{PGL}(3)$-invariant BSQ-type equations, in terms of a common set of dependent
variables $z_1, z_2$ that are projective variables of the spectral problems. This framework suggests a rich interplay between invariant theory, classical projective geometry, and integrable systems. Several promising directions for future investigation emerge from this work.

\begin{itemize}
    \item  Extending $3$D-consistent systems: the three-component quad-system \eqref{eq:eq:3com3} constructed in Section \ref{sec:42} represents a natural rank-$3$ extension of the cross-ratio equation. This provides a concrete example for extending known examples of multi-component $3$D-consistent systems \cite{H, GD, TN1, TN2, ZZN} beyond scalar quad-equations \cite{Nijh2, NW, ABS1, ABS3}. Further investigation may reveal new families of BSQ-type lattice equations within this framework.  It would also be interesting to clarify the relationship between the discrete $\mathrm{PGL}(3)$-invariant BSQ system and  the multi-ratio systems, considered for instance, in \cite{BHS, KS}.
    \item Several aspects of the $\mathrm{PGL}(3)$-invariant BSQ systems derived in Sections \ref{sec:41}-\ref{sec:42} need to be clarified,  such as their Lax pairs and explicit multi-soliton solutions. In particular, following \cite{Wilson1988, STS}, 
    one could consider the ``Miura-tower" associated with $\mathrm{PGL}(3)$ within the Poisson--Lie theory, where one expects a richer structure with extra invariant subfields compared to the rank-$2$ case.  This could also leads to Hamiltonian structures for the $\mathrm{PGL}(3)$-invariant BSQ systems in both continuous and discrete cases.    
    \item Similar to how the Schwarzian KdV equation reduces to Painlevé-type equations \cite{N1, NW, NHJ}, it is of interest to investigate similarity reductions of the $\mathrm{PGL}(3)$-invariant BSQ systems and their generating PDEs. Such reductions are expected to yield higher-rank Painlevé systems, such as Garnier-type equations, establishing a concrete link between the projective BSQ hierarchy and the theory of isomonodromic deformations for higher-order linear problems. 

    \item The framework can be readily extended to arbitrary rank $N$, yielding $\mathrm{PGL}(N)$-invariant integrable systems associated with $N$th-order spectral problems. 
    Further exploration may extend the framework to other geometric settings associated with systems of linear problems, where the underlying geometry is related to matrix Schwarzian derivatives and ``Grassmannian curves"  \cite{O1, Retakh, Pickering}. 
Such extensions would provide new examples of integrable models beyond the projective case.
    
    \item While we have provided the explicit forms of the $\mathrm{PGL}(3)$-invariant generating PDEs in Section \ref{sec:genPDEs}, several aspects of this system remain to be understood. 
    For example, their connection to the Einstein--Maxwell--Weyl theory \cite{TN1, TN2}, geometric interpretations (which extend beyond the projective setting), the curious parameter inversion properties discussed in Remark \ref{rem:inversion}, and possible Lagrangian multiform structures \cite{LN} (see Remark \ref{rmk:multi}).

\end{itemize}


\subsection*{Acknowledgments}
FN is grateful for the frequent hospitality of the Department of Mathematics of Shanghai University since 2017 when the project 
was initiated, and was partially supported in this period by the National
Foreign Expert Program of China (No. G2022172028L). 
LP was partially supported by JSPS KAKENHI (JP24K06852), JST CREST (JPMJCR24Q5), and Keio University (Academic Development Fund, Fukuzawa Fund). 
CZ was supported by NSFC (No. 12171306). DZ was supported by NSFC (Nos. 12671301, 12271334 and 12411540016). 

\appendices

\section{The generating $\mathrm{PGL}(2)$- and $\mathrm{PGL}(3)$-invariants}
\label{app:A}

In this appendix, we show how both the differential and difference generating invariants of 
$\mathrm{PGL}(2)$ and $\mathrm{PGL}(3)$ actions that shown in Section \ref{sec:2} can be computed using their corresponding infinitesimal generators.

\paragraph{Notation convention:}  In Appendices \ref{app:A} and \ref{app:B}, primes
  denote the coordinates after transformations, rather than derivatives.

\subsection{The differential case}
\label{appa}
Recall the $\mathrm{PGL}(2)$-action about $(x,z)$ defined in \eqref{eq:sl2transfz}:
\begin{equation}
(x,z)\mapsto (x,z')\ , \text{ where }  z'=\frac{m_{11}z+m_{21}}{m_{12}z+m_{22}} \text{ with } \det \bM\neq 0\ .
\end{equation}
It is well known that the $\mathrm{PGL}(N)$ group is linearisable and the corresponding Lie algebra is $\mathfrak{sl}(N)$.
Consequently, we only have to focus on $\mathrm{SL}(N)$- and $\mathrm{PSL}(N)$-actions for finding the corresponding differential invariants. Namely, in computing the infinitesimal generators, we assume $\det\bM=1$. 
The corresponding infinitesimal generators are $X_i=\eta_i(x,z)\partial_z$ ($i=1,2,3$) with the coefficients  $\eta_i(x,z)$ given by 
 \begin{equation}\label{eq:ap1}
 \begin{aligned}
 \eta_1(x,z)&:=\frac{\operatorname{d}\!}{\operatorname{d}\!m_{11}}\Big|_{\bM=\text{id}}z'=2z\ ,  \\
 \eta_2(x,z)&:=\frac{\operatorname{d}\!}{\operatorname{d}\!m_{12}}\Big|_{\bM=\text{id}}z'=-z^2\ ,\\
  \eta_3(x,z)&:=\frac{\operatorname{d}\!}{\operatorname{d}\!m_{21}}\Big|_{\bM=\text{id}}z'=1\ .
  \end{aligned}
 \end{equation}
Note that $m_{22}$ is replaced by a function of $m_{11}$, $m_{12}$ and $m_{21}$ around $\bM=\text{id}$ using the condition $\det\bM=1$. Their prolongations to the $k$th-order $\mathbf{pr}^{(k)}X_i$ are (see, \eg \cite{Olverb})
 \begin{equation}\label{eq:pro}
 \mathbf{pr}^{(k)}X_i=\eta_i(x,z)\partial_z+\left(D_x\eta_i(x,z)\right)\partial_{z_x}+\cdots+\left(D_x^k\eta_i(x,z)\right)\partial_{z^{(k)}}\ .
 \end{equation}
 A function $I(x,z,z_x,\ldots)$ is a  differential invariant involving up to $k$th-order derivatives of $z$ for some positive integer $k$  if and only if it satisfies the  infinitesimal invariance condition (\cf \cite{Olvera}) that 
\begin{equation}
\mathbf{pr}^{(k)}X_i(I)=0
\end{equation}
holds for every infinitesimal generator $X_i$. This yields  the following system of linear PDEs 
 \begin{equation}
     \bma 0 & 2z & 2z_x & \dots & 2z^{(k)} \vspace{0.2cm}\\
     0 & -z^2 & -2zz_x & \dots & -(z^2)^{(k)} \vspace{0.2cm} \\
    0 & 1 & 0 & \dots & 0  
     \ema \bma  I_x \\ I_z \\ I_{z_x} \\ \vdots \\ I_{z^{(k)}} \ema = \bma  0 \\ 0 \\ 0 \ema  \, .
   \end{equation}
 Let $k=3$; using Gaussian elimination and properly arranging rows of the coefficient matrix, the above system becomes
  \begin{equation}
     \bma 0 & 1 & 0 & 0 & 0 \vspace{0.2cm} \\
     0 & 0 & z_x & 0 & z_{xxx}-\frac{3z_{xx}^2}{z_x} \vspace{0.2cm} \\
    0 & 0 & 0 & z_x &  3z_{xx}
     \ema \bma  I_x \\ I_z \\ I_{z_x} \\ I_{z_{xx}} \\ I_{z_{xxx}}\ema = \bma  0 \\ 0 \\ 0 \ema  \, .
   \end{equation}
   This linear system of PDEs can be solved by the characteristic method. The first row gives $I=I(x,z_x,z_{xx},z_{xxx})$; substituting it to the last row, \ie $z_xI_{z_{xx}}+3z_{xx}I_{z_{xxx}}=0$, we obtain 
   \begin{equation}
   I=I\left(x,z_x,z_{xxx}-\frac{3z_{xx}^2}{2z_x}\right)\ .
   \end{equation}
  This is then substituted to the final equation $z_xI_{z_x}+\left(z_{xxx}-\frac{3z_{xx}^2}{z_x}\right)I_{z_{xxx}}=0$, and we get
  \begin{equation}
  I=I\left(x, \frac{z_{xxx}}{z_x}-\frac{3z_{xx}^2}{2z_x^2} \right)\ .
  \end{equation}
This amounts to the generating differential invariants $x$ and the Schwarzian derivative
 \begin{equation}
 \mathbf{S}[z]=\frac{z_{xxx}}{z_x}-\frac{3z_{xx}^2}{2z_x^2} .
 \end{equation}
 There exist no other differential invariants with order lower than the Schwarzian derivative; all higher-order differential invariants are functions of the generating invariants $x$, the Schwarzian derivative $\mathbf{S}[z]$ and invariant derivatives of  $\mathbf{S}[z]$ with respect to $x$.

Generating differential invariants with respect to the $\mathrm{PGL}(3)$-action defined in \eqref{eq:projSL3}, namely
\begin{equation}
\left\{ \begin{array}{l} 
z_1'=\frac{\textstyle m_{11}z_1+m_{21}z_2+m_{31}}{\textstyle m_{13}z_1+m_{23}z_2+m_{33}}\   ,  \vspace{0.3cm}  \\ 
z_2'=\frac{\textstyle m_{12}z_1+m_{22}z_2+m_{32}}{\textstyle m_{13}z_1+m_{23}z_2+m_{33}}\   ,  \vspace{0.2cm}
\end{array}\right. 
\end{equation}
  can be calculated using the same method. They are the $\mathbf{S}_1[z_1,z_2]$ and $\mathbf{S}_2[z_1,z_2]$ given by \eqref{eq:BSQdiffin1} and $x$, which form the set of generating differential invariants.  We omit the details here. Note that the eight infinitesimal generators are
  \begin{equation}\label{3DIG}
\begin{aligned}
&2z_1\partial_{z_1}+z_2\partial_{z_2}\ , \quad z_1\partial_{z_2}\ ,\quad -z_1^2\partial_{z_1}-z_1z_2\partial_{z_2}\ , \quad z_2\partial_{z_1}\ ,\\
&z_1\partial_{z_1}+2z_2\partial_{z_2}\ , \quad -z_1z_2\partial_{z_1}-z_2^2\partial_{z_2}\ , \quad \partial_{z_1}\ , \quad \partial_{z_2}\ ,
\end{aligned}
\end{equation}
corresponding to the parameters $m_{11},m_{12},m_{13},m_{21},m_{22},m_{23},m_{31}$ and $m_{32}$, respectively. 
\subsection{The difference case}
\label{appb}

Similarly as in Appendix \ref{appa}, a function is a difference invariant if and only if it satisfies the discrete version of infinitesimal invariance condition, which is proposed below. 
In the discrete case, prolongations of infinitesimal generators  $X_i=\eta_i(n,z)\partial_z$ are given by (see, \eg \cite{Hydon})
\begin{equation}
\mathbf{pr}X_i=\cdots+\eta_i(n,z)\partial_z+\eta_i(n+1,\wb{z})\partial_{\wb{z}}+\eta_i\left(n+2,\wb{\wb{z}}\right)\partial_{\wb{\wb{z}}}+\eta_i\left(n+3,\wb{\wb{\wb{z}}}\right)\partial_{\wb{\wb{\wb{z}}}}+\cdots\ ,
\end{equation}
and recall $\eta_1(n,z)=2z$, $\eta_2(n,z)=-z^2$, and $\eta_3(n,z)=1$ for the $\mathrm{PGL}(2)$-action (see \eqref{eq:ap1}). A  difference invariant  $I=I\left(n,z,\wb{z},\wb{\wb{z}},\wb{\wb{\wb{z}}}\right)$ satisfies the infinitesimal invariance condition $\mathbf{pr}X_i(I)=0$ for every $X_i$, which leads to a system of linear PDEs 
  \begin{equation}\label{eq:invariantpde}
     \bma  2z &  2\wb{z} & 2\wb{\wb{z}} & 2\wb{\wb{\wb{z}}} \vspace{0.2cm} \\
       -z^2 & -\wb{z}^2 & -\left(\wb{\wb{z}}\right)^2 & -\left(\wb{\wb{\wb{z}}}\right)^2  \vspace{0.2cm}\\
    1 & 1 & 1 &  1
     \ema \bma   I_{z} \\ I_{\wb{z}} \\ I_{\wb{\wb{z}}}\\
     I_{\wb{\wb{\wb{z}}}}
     \ema = \bma  0 \\ 0 \\ 0 \ema  \, .
   \end{equation}
   This can be solved similarly using Gaussian elimination and characteristic method for linear PDEs. We first obtain an equivalent system
      \begin{equation}
     \bma 1 & 0 & 0 & \frac{\left(\wb{\wb{\wb{z}}}-\wb{\wb{z}}\right)\left(\wb{\wb{\wb{z}}}-\wb{z}\right)}{\left(\wb{\wb{z}}-{z}\right)\left(\wb{z}-{z}\right)} \vspace{0.2cm}\\
      0 & 1 & 0 &   -\frac{\left(\wb{\wb{\wb{z}}}-\wb{\wb{z}}\right)\left(\wb{\wb{\wb{z}}}-{z}\right)}{\left(\wb{\wb{z}}-\wb{z}\right)\left(\wb{z}-{z}\right)}  \vspace{0.2cm} \\
    0 & 0 & 1 &  \frac{\left(\wb{\wb{\wb{z}}}-{z}\right)\left(\wb{\wb{\wb{z}}}-\wb{z}\right)}{\left(\wb{\wb{z}}-{z}\right)\left(\wb{\wb{z}}-\wb{z}\right)} 
     \ema 
     \bma   I_{z} \\ I_{\wb{z}} \\ I_{\wb{\wb{z}}}\\
     I_{\wb{\wb{\wb{z}}}}
     \ema = \bma  0 \\ 0 \\ 0 \ema  \, .
   \end{equation}
  Solving the last row by the method of characteristic, we obtain 
  \begin{equation}
  I=I\left(n,z,\wb{z}, \frac{\left(\wb{\wb{\wb{z}}}-\wb{z}\right)\left(\wb{\wb{z}}-z\right)}{\left(\wb{\wb{\wb{z}}}-z\right)\left(\wb{\wb{z}}-\wb{z}\right)}\right)\ .
  \end{equation}
  Substituting this back to the other two equations gives 
  \begin{equation}
  I=I\left(n, \frac{\left(\wb{\wb{\wb{z}}}-\wb{z}\right)\left(\wb{\wb{z}}-z\right)}{\left(\wb{\wb{\wb{z}}}-z\right)\left(\wb{\wb{z}}-\wb{z}\right)}\right)\  ,
  \end{equation}
  and hence we get the generating difference invariants $n$ and 
  \begin{equation}
  \label{eq:2ndinv2}
 I_1=\frac{\left(\wb{\wb{\wb{z}}}-\wb{z}\right)\left(\wb{\wb{z}}-z\right)}{\left(\wb{\wb{\wb{z}}}-z\right)\left(\wb{\wb{z}}-\wb{z}\right)}\ .
  \end{equation}
  From a first glance, it looks different from the cross-ratio invariant \eqref{eq:2ndinv}, temporarily denoted by $I_0$; in fact, they are related via the relation
    \begin{equation}
  I_0+I_1=I_0I_1\ . 
  \end{equation}
  
 One can similarly prove that the $ {\mathbf I}_1[z_1,z_2]$ and ${\mathbf  I}_2[z_1,z_2]$ defined by \eqref{eq:disBSQdiffin} (or \eqref{eq:3dinv}) are indeed invariant corresponding to the $\mathrm{PGL}(3)$-action by showing that they satisfy the following system of linear PDEs
\begin{equation}
\mathbf{pr}X_i I\left(n,z_1,z_2,\wb{z}_1,\wb{z}_2,\wb{\wb{z}}_1,\wb{\wb{z}}_2,\wb{\wb{\wb{z}}}_1,\wb{\wb{\wb{z}}}_2,\wb{\wb{\wb{\wb{z}}}}_1,\wb{\wb{\wb{\wb{z}}}}_2\right)=0\ .
\end{equation}
Here, the eight infinitesimal generators are in the same form as the continuous $\mathrm{PGL}(3)$-action analysed in Appendix \ref{appa} but with discrete prolongations.
Substitution of the functions \eqref{eq:disBSQdiffin} into the above system immediately shows that both of them are invariant, and coefficient matrix of the above system is of maximal rank $8$. Hence, these two functions and $n$ serve as generating invariants of the $\mathrm{PGL}(3)$-action about $(n,z_1,z_2)$. All other difference invariants are functions of generating invariants and their shifts. 

\section{$\mathrm{PGL}(N)$-invariants from linear spectral problems}
\label{app:B}

In this appendix, we show how the method introduced in Section \ref{sec:2} for obtaining invariants from linear spectral problems can be generalized to $\mathrm{PGL}(N)$ actions, in both the differential and difference settings.

\subsection{The differential case}
\label{appgln1}

Consider an $N$th-order linear spectral problem 
\begin{equation}\label{eq:NthSP}
\partial_x^N\varphi+u_{N-2}\partial_x^{N-2}\varphi+\cdots+u_1\partial_x\varphi+u_0\varphi=\lambda\varphi\ ,
\end{equation}
and its independent solutions $\varphi_1$, $\varphi_2$, \ldots, $\varphi_N$. The natural $\mathrm{GL}(N)$-action on solutions
\begin{equation}
\left(\varphi_1,\varphi_2,\ldots,\varphi_N\right)\mapsto (\vf_1',\vf_2',\ldots,\vf_N')=(\vf_1,\vf_2,\ldots,\vf_N)\bM
\end{equation}
induces the $\mathrm{PGL}(N)$-action on the inhomogeneous coordinates
\begin{equation}
z_i=\frac{\varphi_i}{\varphi_N}\ , \quad i=1,2,\ldots, N-1
\end{equation}
as follows
\begin{equation}\label{eq:PGLN}
z_i'=\frac{\sum\limits_{j=1}^{N-1}m_{ji}z_j+m_{Ni}}{\sum\limits_{j=1}^{N-1}m_{jN}z_j+m_{NN}}\ , \quad i=1,2,\ldots, N-1\ , \text{ where } \det \bM\neq 0\ .
\end{equation}
We introduce the notations $\varphi=\varphi^{(0)}$, $\partial_x\varphi=\varphi^{(1)}$, $\partial_x^2\varphi=\varphi^{(2)}$ and so forth. Note that they stand for shifts in the discrete case. Writing the equations for $\vf=\vf_N$ and $\vf_i=z_i\vf$ ($i=1,2,\ldots,N-1$), we obtain the following system:
\bse\label{eq:PGLNsym}
\begin{eqnarray}
\varphi^{(N)}+u_{N-2}\varphi^{(N-2)}+\cdots+u_1\varphi^{(1)}+\left(u_0-\lambda\right)\varphi
&=&0\ , \label{eq:PGLNsyma}  \\ 
C_N^{N-1}z_i^{(1)}\varphi^{(N-1)}+\sum_{k=0}^{N-2}\left(C_N^kz_i^{(N-k)}+\sum_{l=k+1}^{N-2}C_l^k u_lz_i^{(l-k)}\right)\varphi^{(k)}&=&0\ .\label{eq:PGLNsymb}
\end{eqnarray}
\ese 
Here, we use the notation
\begin{equation}
C_k^l=\frac{k!}{l!(k-l)!}\ .
\end{equation}
Dividing by $\varphi$ on both sides of  \eqref{eq:PGLNsymb}, the resulting equations can be written in matrix form 
\begin{equation}\label{eq:wron}
\left(
\begin{array}{cccc}
z_1^{(N-1)} & z_1^{(N-2)} & \ldots & z_1^{(1)} \vspace{0.2cm} \\
\vdots    & \vdots & \ddots & \vdots  \vspace{0.2cm} \\
z_i^{(N-1)} & z_i^{(N-2)} & \ldots & z_i^{(1)} \vspace{0.2cm} \\
\vdots    & \vdots & \ddots & \vdots  \vspace{0.2cm} \\
z_{N-1}^{(N-1)} & z_{N-1}^{(N-2)} & \ldots & z_{N-1}^{(1)} 
\end{array}
\right)\Phi=-\left(\begin{array}{c}
z_1^{(N)} \vspace{0.2cm}\\
\vdots \vspace{0.2cm} \\
z_i^{(N)} \vspace{0.2cm} \\
\vdots \vspace{0.2cm} \\
z_{N-1}^{(N)}
\end{array}\right)\ ,
\end{equation}
where the vector $\Phi$ is given by ($l=N-2,N-1,\ldots,1$)
\begin{equation}
\Phi=\left(C_N^1\frac{\varphi^{(1)}}{\varphi },
\ldots,
 C_N^{N-l}\frac{\varphi^{(N-l)}}{\varphi}+\sum_{k=1}^{N-2-l}C_{k+l}^ku_{k+l}\frac{\varphi^{(k)}}{\varphi}+u_l,
\ldots\right)^\intercal\ .
\end{equation}
Defining $\boldsymbol{z}=\left(z_1,z_2,\ldots,z_{N-1}\right)^\intercal$,  the system \eqref{eq:wron} can be solved by Cramer's rule. In particular, we obtain a higher-order Hopf--Cole-type transformation
\begin{equation}\label{eq:NHCtra}
\frac{\varphi^{(1)}}{\varphi}=-\frac{1}{N}\frac{\left|\boldsymbol{z}^{(N)},\boldsymbol{z}^{(N-2)},\ldots,\boldsymbol{z}^{(1)}  \right|}{\left|\boldsymbol{z}^{(N-1)},\boldsymbol{z}^{(N-2)},\ldots,\boldsymbol{z}^{(1)}  \right|}\ .
\end{equation} 
 In fact, $\varphi$ can be solved from \eqref{eq:NHCtra}:
\begin{equation}
\varphi=c\left|\boldsymbol{z}^{(N-1)},\boldsymbol{z}^{(N-2)},\ldots,\boldsymbol{z}^{(1)}  \right|^{-\frac{1}{N}}
\end{equation}
and without loss of generality, the integration constant can be chosen as $c=1$. By doing so, we are released from deriving $\varphi^{(N)}/\varphi$ from the recursive relation for large $N$:
\begin{equation}\label{recrelcon}
\frac{\varphi^{(N)}}{\varphi}=\partial_x\left(\frac{\varphi^{(N-1)}}{\varphi}\right)+\frac{\varphi^{(1)}}{\varphi}\frac{\varphi^{(N-1)}}{\varphi}\ .
\end{equation}

The function $\varphi$ and its derivatives are  then substituted back to the linear system \eqref{eq:PGLNsym}, whose solution about the potentials will give us  $(N-1)$-number of differential invariants. In fact, the system \eqref{eq:PGLNsym} is now equivalent to combing solution of \eqref{eq:wron} and \eqref{eq:PGLNsyma},  which can be written in matrix form for the potentials as
\begin{equation}
\boldsymbol{A}\left(\begin{array}{c}
u_{N-2} \vspace{0.4cm} \\
u_{N-3} \vspace{0.4cm} \\
\vdots \vspace{0.4cm} \\
u_1  \vspace{0.4cm} \\
u_0-\lambda
\end{array}\right)=-\left(\begin{array}{c}
\frac{\left|\boldsymbol{z}^{(N-1)},\boldsymbol{z}^{(N)},\boldsymbol{z}^{(N-3)},\ldots,\boldsymbol{z}^{(1)}  \right|}{\left|\boldsymbol{z}^{(N-1)},\boldsymbol{z}^{(N-2)},\ldots,\boldsymbol{z}^{(1)}  \right|}+C_N^2\frac{\varphi^{(2)}}{\varphi} \vspace{0.2cm} \\
\frac{\left|\boldsymbol{z}^{(N-1)},\boldsymbol{z}^{(N-2)},\boldsymbol{z}^{(N)},\ldots,\boldsymbol{z}^{(1)}  \right|}{\left|\boldsymbol{z}^{(N-1)},\boldsymbol{z}^{(N-2)},\ldots,\boldsymbol{z}^{(1)}  \right|}+C_N^3\frac{\varphi^{(3)}}{\varphi} \vspace{0.2cm} \\
\vdots \vspace{0.2cm} \\
\frac{\left|\boldsymbol{z}^{(N-1)},\boldsymbol{z}^{(N-2)},\ldots,\boldsymbol{z}^{(2)},\boldsymbol{z}^{(N)}  \right|}{\left|\boldsymbol{z}^{(N-1)},\boldsymbol{z}^{(N-2)},\ldots,\boldsymbol{z}^{(1)}  \right|}+C_N^{N-1}\frac{\varphi^{(N-1)}}{\varphi} \vspace{0.2cm} \\
\frac{\varphi^{(N)}}{\varphi}
\end{array}\right) \ ,
\end{equation}
where the coefficient matrix is a triangular matrix 
\begin{equation}
\boldsymbol{A}=\left(\begin{array}{ccccccc}
1 & 0 & 0 & \ldots & 0 & 0 & 0 \vspace{0.2cm} \\
C_{N-2}^1\frac{\varphi^{(1)}}{\varphi} & 1 & 0 & \ldots & 0 & 0 & 0 \vspace{0.2cm} \\
C_{N-2}^2\frac{\varphi^{(2)}}{\varphi} & C_{N-3}^{1}\frac{\varphi^{(1)}}{\varphi} & 1 & \ldots & 0 & 0  & 0 \vspace{0.2cm} \\
\vdots & \vdots & \vdots & \ddots & \vdots  & \vdots& \vdots  \vspace{0.2cm} \\
\vdots & \vdots & \vdots & \ddots & \vdots  & \vdots& \vdots  \vspace{0.2cm} \\
C_{N-2}^{N-3}\frac{\varphi^{(N-3)}}{\varphi} & C_{N-3}^{N-4}\frac{\varphi^{(N-4)}}{\varphi} & C_{N-4}^{N-5}\frac{\varphi^{(N-5)}}{\varphi} & \ldots & C_2^1\frac{\varphi^{(1)}}{\varphi} & 1 & 0  \vspace{0.2cm} \\
\frac{\varphi^{(N-2)}}{\varphi} & \frac{\varphi^{(N-3)}}{\varphi} & \frac{\varphi^{(N-4)}}{\varphi} & \ldots & \frac{\varphi^{(2)}}{\varphi} & \frac{\varphi^{(1)}}{\varphi} & 1
\end{array}
\right)\ .
\end{equation}
The above system can be solved via Cramer's rule or via the following recursive relation:
\begin{equation}\label{GLN-inv}
\begin{aligned}
u_{N-2}&=-\frac{\left|\boldsymbol{z}^{(N-1)},\boldsymbol{z}^{(N)},\boldsymbol{z}^{(N-3)},\ldots,\boldsymbol{z}^{(1)}  \right|}{\left|\boldsymbol{z}^{(N-1)},\boldsymbol{z}^{(N-2)},\ldots,\boldsymbol{z}^{(1)}  \right|}-C_N^2\frac{\varphi^{(2)}}{\varphi}\ , \\
u_k&=-\frac{\left|\boldsymbol{z}^{(N-1)},\ldots,\boldsymbol{z}^{(k+1)},\boldsymbol{z}^{(N)},\boldsymbol{z}^{(k-1)},\ldots,\boldsymbol{z}^{(1)}  \right|}{\left|\boldsymbol{z}^{(N-1)},\boldsymbol{z}^{(N-2)},\ldots,\boldsymbol{z}^{(1)}  \right|}\\
&~~~~-C_N^{N-k}\frac{\varphi^{(N-k)}}{\varphi}-\sum_{l=1}^{N-2-k}C_{k+l}^lu_{k+l}\frac{\varphi^{(l)}}{\varphi}\ , \quad k=N-3,N-4,\ldots,1\ , \\
u_0-\lambda &= -\frac{\varphi^{(N)}}{\varphi}-\sum_{l=1}^{N-2}u_{l}\frac{\varphi^{(l)}}{\varphi} \ .
\end{aligned}
\end{equation}
Due to the independence of the potentials $u_{N-2},u_{N-1},\ldots,u_0$, these invariants are functionally independent from each other. 


\begin{rmk}\label{rmk:GLN}
It is clear that the highest orders of derivatives involved in the invariants \eqref{GLN-inv} are not the same, but rather it increases one each time. For instance, $u_{N-2}$ includes derivatives of $\bz$ up to the order $N+1$, while $u_{N-3}$ includes derivatives of $\bz$ up to the order $N+2$. However, these invariants can be re-arranged to a set of fundamental invariants, which all depend on derivatives of $\bz$ up to the order $N+1$; this can be done by taking the recursive relation \eqref{recrelcon} into consideration. For instance, a new invariant 
\begin{equation}
C_N^2u_{N-3}-C_{N}^3\partial_xu_{N-2}\ ,
\end{equation}
that involves derivatives up to order $N+1$, can be used to replace $u_{N-3}$, that involves derivatives up to order $N+2$.
\end{rmk}

The case $N=2$ is straightforward. Let us consider $N=3$ as an example to illustrate how the invariants \eqref{eq:BSQdiffin1} can be obtained accordingly.
For the third-order spectral problem \eqref{eq:3rdord-diff}, the system \eqref{eq:wron} becomes 
\begin{equation}
\left(\begin{array}{cc}
z_1^{(2)} & z_1^{(1)} \vspace{0.2cm} \\
z_2^{(2)} & z_2^{(1)} \end{array}\right)\left(
\begin{array}{c}
3\frac{\varphi^{(1)}}{\varphi} \vspace{0.2cm} \\
3\frac{\varphi^{(2)}}{\varphi} +u
\end{array}
\right)=-\left(\begin{array}{c}
z_1^{(3)} \vspace{0.2cm} \\
z_2^{(3)}\end{array}\right)\ .
\end{equation}
Solving it, we obtain the Hopf--Cole transformation \eqref{eq:dlogvf}  and 
\begin{equation}
u=-\frac{\left|\boldsymbol{z}^{(2)},\boldsymbol{z}^{(3)} \right|}{\left|\boldsymbol{z}^{(2)},\boldsymbol{z}^{(1)}\right|}-3\frac{\varphi^{(2)}}{\varphi}\ .
\end{equation}
Solving for $\varphi$ from the Hopf--Cole transformation leads to
\begin{equation}
\varphi=\left|\boldsymbol{z}^{(2)},\boldsymbol{z}^{(1)}\right|^{-\frac13}\ ,
\end{equation}
which is substituted back to $u$. 
This gives the first invariant $\mathbf{S}_1[z_1,z_2]$ in \eqref{eq:BSQdiffin1}. The other one, \ie $-3(v-\lambda)+u_x=\mathbf{S}_2[z_1,z_2]$, is obtained by substituting the results back to the spectral problem \eqref{eq:3rdord-diff}.

\subsection{The difference case}
\label{appgln2}

Consider an $N$th-order linear difference equation of the form 
\begin{equation}\label{eq:NthDE} 
\Lambda\vf=\lambda \vf\ ,\quad \Lambda=T^N+\mathfrak{h}_{N-1}T^{N-1}+\dots+ \mathfrak{h}_1T +\mathfrak{h}_0 \   , 
\end{equation}
with  $\lambda$, $\mathfrak{h}_0,\mathfrak{h}_1,\dots,\mathfrak{h}_{N-1}$ a collection of potential functions which can be realised as 
diagonal matrices. We have a natural $\mathrm{GL}(N)$-action on $(\vf_1,\vf_2,\ldots,\vf_N)$, \ie
the vector of independent solutions. Introducing 
\begin{equation}
z_i=\frac{\vf_i}{\vf_N}\ ,\quad i=1,2,\dots,N-1\   , 
\end{equation}
and denoting $\varphi=\varphi_N$, we get the set of equations:
\bse\label{eq:Nthsyst}
\bea 
 \vf^{(N)}+\mathfrak{h}_{N-1}\vf^{(N-1)}+\cdots+\mathfrak{h}_1\vf^{(1)}-(\ld-\mathfrak{h}_0)
\vf&=&0 \   ,  \\  
z_i^{(N)}\vf^{(N)}+\mathfrak{h}_{N-1}z_i^{(N-1)}\vf^{(N-1)}+\cdots+\mathfrak{h}_1 z_i^{(1)}\vf^{(1)}-(\ld-\mathfrak{h}_0)z_i 
\vf&=&0 \   , 
\eea\ese 
where the superscript denotes the order of the shift: ~$z^{(j)}_i=T^jz_i$. The above system  can be rewritten in matrix form 
\begin{equation}
\left(\begin{array}{cccc}
1 & 1 & \ldots & 1 \vspace{0.2cm}\\
z_1^{(N)} & z_1^{(N-1)} & \ldots & z_1^{(1)} \vspace{0.2cm} \\
\vdots & \vdots & \ddots  & \vdots \vspace{0.2cm} \\
z_{N-1}^{(N)} & z_{N-1}^{(N-1)} & \ldots & z_{N-1}^{(1)}  \\
\end{array}\right)
\left(
\begin{array}{c}
\frac{\varphi^{(N)}}{\varphi} \vspace{0.2cm} \\
\mathfrak{h}_{N-1}\frac{\varphi^{(N-1)}}{\varphi} \vspace{0.2cm} \\
\vdots \vspace{0.2cm} \\
\mathfrak{h}_1 \frac{\varphi^{(1)}}{\varphi}
\end{array}
\right)=(\lambda-\mathfrak{h}_0) \left(\begin{array}{c}
1 \vspace{0.2cm} \\
z_1 \vspace{0.2cm}\\
\vdots \vspace{0.2cm} \\
z_{N-1}
\end{array}\right)\ .
\end{equation}
Let us introduce the notation $\boldsymbol{z}=\left(1,z_1,z_2,\ldots,z_{N-1}\right)^\intercal$ temporarily. The system can be solved using Cramer's rule, namely
\begin{equation}
\begin{aligned}
\frac{\varphi^{(N)}}{\varphi}&=(\lambda-\mathfrak{h}_0) \frac{\left|\boldsymbol{z},\boldsymbol{z}^{(N-1)},\ldots,\boldsymbol{z}^{(1)}\right|}{\left|\boldsymbol{z}^{(N)},\boldsymbol{z}^{(N-1)},\ldots,\boldsymbol{z}^{(1)}\right|}\ , \\
 \frac{\varphi^{(i)}}{\varphi}&=\frac{\lambda-\mathfrak{h}_0}{\mathfrak{h}_i}\frac{\left|\boldsymbol{z}^{(N)},,\ldots,\boldsymbol{z}^{(i+1)},\boldsymbol{z},\boldsymbol{z}^{(i-1)},\ldots,\boldsymbol{z}^{(1)}\right|}{\left|\boldsymbol{z}^{(N)},\boldsymbol{z}^{(N-1)},\ldots,\boldsymbol{z}^{(1)}\right|}\ , \quad i=1,2,\ldots,N-1\ .
 \end{aligned}
\end{equation}
By eliminating $\varphi$ and its shifts using 
\begin{equation}
 \frac{\varphi^{(i)}}{\varphi}\cdot  \frac{\varphi}{\varphi^{(i-1)}}\cdot \left( \frac{\varphi}{\varphi^{(1)}}\right)^{(i-1)}=1\ , \quad i=2,3,\ldots,N\ ,
\end{equation} 
we obtain the following $(N-1)$-number of difference invariants:
\begin{equation}\label{eq:PGLNdifferenceinv}
\frac{\mathfrak{h}_{i-1} \mathfrak{h}_1^{(i-1)}}{(\lambda-\mathfrak{h}_0) \mathfrak{h}_i}=\frac{\left|\boldsymbol{z}^{(N)},\ldots,\boldsymbol{z}^{(i)},\boldsymbol{z},\boldsymbol{z}^{(i-2)},\ldots,\boldsymbol{z}^{(1)}\right|
}{\left|\boldsymbol{z}^{(N)},\ldots,\boldsymbol{z}^{(i+1)},\boldsymbol{z},\boldsymbol{z}^{(i-1)},\ldots,\boldsymbol{z}^{(1)}\right|
} \left(\frac{\left|\boldsymbol{z}^{(N)},\boldsymbol{z}^{(N-1)},\ldots,\boldsymbol{z}^{(2)},\boldsymbol{z}\right|
}{\left|\boldsymbol{z}^{(N)},\boldsymbol{z}^{(N-1)},\ldots,\bz^{(2)},\boldsymbol{z}^{(1)}\right|
}\right)^{(i-1)}
\end{equation}
for $i=2,3,\ldots,N$, where we temporarily define $\mathfrak{h}_N=1$.
Similarly to the continuous case, each invariant of \eqref{eq:PGLNdifferenceinv} is written in terms of different numbers of variables, and the highest order of changes increases by $1$ as the index $i$ increases by $1$.
A set of difference invariants, that depend on the same number of variables, \ie $\bz,\bz^{(1)},\ldots,\bz^{(N)},\bz^{(N+1)}$, can be derived as follows:
\begin{equation*}
\begin{aligned}
\frac{\mathfrak{h}_{1} \mathfrak{h}_1^{(1)}}{(\lambda-\mathfrak{h}_0) \mathfrak{h}_2}&=\frac{\left|\boldsymbol{z}^{(N)},\ldots,\boldsymbol{z}^{(3)},\boldsymbol{z}^{(2)},\boldsymbol{z}\right|
}{\left|\boldsymbol{z}^{(N)},\ldots,\boldsymbol{z}^{(3)},\boldsymbol{z},\boldsymbol{z}^{(1)}\right|
} \left(\frac{\left|\boldsymbol{z}^{(N)},\boldsymbol{z}^{(N-1)},\ldots,\boldsymbol{z}^{(2)},\boldsymbol{z}\right|
}{\left|\boldsymbol{z}^{(N)},\boldsymbol{z}^{(N-1)},\ldots,\bz^{(2)},\boldsymbol{z}^{(1)}\right|
}\right)^{(1)}\ , \\
\frac{\mathfrak{h}_{i-1}\mathfrak{h}_{i-1}^{(1)}}{\mathfrak{h}_i\mathfrak{h}_{i-2}^{(1)}}&=\frac{\mathfrak{h}_{i-1} \mathfrak{h}_1^{(i-1)}}{(\lambda-\mathfrak{h}_0) \mathfrak{h}_i} \Big/ \left(\frac{\mathfrak{h}_{i-2} \mathfrak{h}_1^{(i-2)}}{(\lambda-\mathfrak{h}_0) \mathfrak{h}_{i-1}}\right)^{(1)}\\
&=\frac{\left|\boldsymbol{z}^{(N)},\ldots,\boldsymbol{z}^{(i)},\boldsymbol{z},\boldsymbol{z}^{(i-2)},\ldots,\boldsymbol{z}^{(1)}\right|
}{\left|\boldsymbol{z}^{(N)},\ldots,\boldsymbol{z}^{(i+1)},\boldsymbol{z},\boldsymbol{z}^{(i-1)},\ldots,\boldsymbol{z}^{(1)}\right|}
\left(\frac{\left|\boldsymbol{z}^{(N)},\ldots,\boldsymbol{z}^{(i)},\boldsymbol{z},\boldsymbol{z}^{(i-2)},\ldots,\boldsymbol{z}^{(1)}\right|
}{\left|\boldsymbol{z}^{(N)},\ldots,\boldsymbol{z}^{(i-1)},\boldsymbol{z},\boldsymbol{z}^{(i-3)},\ldots,\boldsymbol{z}^{(1)}\right|
}\right)^{(1)}
\end{aligned}
\end{equation*}
for $i=3,4,\ldots,N$.

\section{Darboux--Crum formulae}
\label{app:DCF}

The Darboux transformations discussed in Sections \ref{sec:3} generate exact discretisations and reveal the self-dual structure underlying the BSQ equations. Here we present the associated Darboux--Crum formulae, along the line of \cite{ZhPeZh} where the analogous 
formulae for the KdV case of systems were developed, which serve as explicit formulae for multi-soliton solutions of the BSQ equations.

 \begin{theorem}[Continuous Darboux--Crum formulae]
Consider the continuous spectral problem \eqref{eq:3rdord-diff}. 
   Assume there exist $N$ linearly independent  solutions $\vf_j$, $j=1,2,\dots,N$, of the spectral problem  ${\cal L} \varphi = \lambda\varphi$ at $\lambda = \alpha_j$ respectively, then an $N$-step Darboux transformation amounts to  the map $\{\varphi,u,v\}\mapsto \{\varphi^{[N]},u^{[N]},v^{[N]}\}$:
\begin{equation}
  \label{eq:mBsqsol}
     \varphi^{[N]} =  \frac{W(\vf_1,\dots, \vf_N,\varphi)}{W(\vf_1,\dots, \vf_N)}\,, 
   \end{equation}
where $W(\ast, \dots, \ast)$ denotes the Wronskian, and  
\begin{equation}\label{eq:unvn}
u^{[N]} = u  -3 s'_{1}\,, ~~
      v^{[N]}= v+    N u' + 3(s_1s_1'-s_1''-s_2')\,,
\end{equation}
 where $s_{1}, s_2$ are defined as 
   \begin{equation}
     s_1= -\frac{|\Phi,\dots, \Phi^{(N-2)},\Phi^{(N)}|}{W(\vf_1,\dots, \vf_N)}\,,\quad
     s_2= \frac{|\Phi,\dots,\Phi^{(N-3)} ,\Phi^{(N-1)}, \Phi^{(N)}|}{W(\vf_1,\dots, \vf_N)}\,,
   \end{equation}
with $\Phi = (\vf_1,\dots,\vf_N)^\intercal$ , and the superscript ${}^{(n)}$ meaning the $n$th-order derivative. To be consistent with Section \ref{sec:3}, note that the primes in \eqref{eq:unvn} denote derivatives.
 \end{theorem}
 
 \prf The action of an $N$-step Darboux transformation   on $\vf$ using $\vf_j$, $j = 1,2,\dots, N$, yields 
 \begin{equation}
   \label{eq:exvp}
   \varphi^{[N]} = \varphi^{(N)}+s_1\varphi^{(N-1)}+s_2\varphi^{(N-2)}+\cdots+ s_N\varphi \, ,
 \end{equation}
 where the coefficient functions $s_j$ are determined by the linear system
   \begin{equation}
     \bma \vf_1 & \vf^{(1)}_1 & \dots & \vf_1^{(N-1)}  \\
     \vdots & \vdots & \dots & \vdots  \\
     \vf_N & \phi_N^{(1)} & \dots & \vf_N^{(N-1)}  
     \ema \bma  s_N \\ \vdots \\ s_1 \ema = -\bma  \vf_1^{(N)} \\ \vdots \\ \vf_N^{(N)} \ema  \,,
   \end{equation}
which is the consequence of $   \varphi^{[N]} \vert_{\varphi = \vf_j} =0$.  The expressions of $u^{[N]}$ and $v^{[N]}$ are obtained by equating the coefficients of ${\cal L}^{[N]} \varphi^{[N]}= \lambda \varphi^{[N]} $ with ${\cal L}^{[N]} := \partial_x^3 +u^{[N]} \partial_x+v^{[N]}$. \finprf
\medskip

\begin{theorem}[Discrete Darboux--Crum formulae]
   Consider the discrete spectral problem  \eqref{eq:3dDE}.   Assume there exist $M$ linearly independent  solutions $\vf_j$, $j=1,2,\dots,M$, of the spectral problem $\Lambda \vf =\lambda\vf$ at $\lambda = \beta_j$ respectively, then a $M$-step Darboux transformation amounts to the  map  $\{\varphi,\fh,\fg\}\mapsto \{\varphi^{\langle M \rangle},\fh^{\langle M \rangle},\fg^{\langle M \rangle}\}$:
   \begin{equation}
     \varphi^{\langle M \rangle} =  \frac{C(\vf_1,\dots, \vf_M,\varphi)}{C(\vf_1,\dots, \vf_M)}\,, 
   \end{equation}
where $C(\ast, \dots, \ast)$ denotes the Casorati determinant, and  
\begin{equation}
\fh^{\langle M \rangle} = \fh^{[M]}  -\sigma_1\,,\quad \fg^{\langle M \rangle}=\fg^{[M]}-\sigma_2+\overline{\overline{\fs}}_1\sigma_1+\fs_1 \fh^{[M-1]}-\overline{\overline{\fs}}_1 \fh^{[M]}\,,
\end{equation}
where  $\sigma_\ell =\overline{\overline{\overline{\fs}}}_\ell-\fs_\ell$, $\ell =1,2$, and $\fs_\ell$ are defined as 
   \begin{equation}
    \fs_1= -\frac{|\Phi,\dots, \Phi^{[M-2]},\Phi^{[M]}|}{C(\vf_1,\dots, \vf_M)}\,,\quad
     \fs_2= \frac{|\Phi,\dots,\Phi^{[M-3]} ,\Phi^{[M-1]}, \Phi^{[M]}|}{C(\vf_1,\dots, \vf_M)}\,, 
   \end{equation}
with  $\Phi = (\vf_1,\dots,\vf_M)^\intercal$ , and the superscript ${}^{[n]}$ meaning the $n$th-order shift. 
\end{theorem}
\prf
The proof is in complete analogy with the continuous case. 
\finprf

\section{Elimination of $P,Q$ in \eqref{eq:PQsystem} and connected vectorial form}
\label{app:genPDEproofs}
We rewrite  the system \eqref{eq:PQsystem}, which is linear in $P,Q$, as follows 
\bse \label{eq:QPsyst} \begin{align}
  P_s &= N\,P +K\,Q\,,  \label{eq:QPsystema} \\
  Q_t &= M\,Q +L\,P  \,,  \label{eq:QPsystemb} \\ 
  P_t+Q_s &= A\,P+B\,Q\ , \label{eq:QPsystemc} 
  \end{align}
  \ese
  with the abbreviations: 
\bse\label{eq:NMKLAB}
\begin{align}
N:=  \frac{1-n}{s} +\frac{|\bz_t,\bz_{ss}|}{|\bz_t,\bz_{s}|}\ ,& \quad 
M:= \frac{1-m}{t} +\frac{|\bz_s,\bz_{tt}|}{|\bz_s,\bz_{t}|} \ , \\ 
K:=  \frac{s}{t} \frac{|\bz_s,\bz_{ss}|}{|\bz_s,\bz_{t}|}\ ,& \quad 
L:= \frac{t}{s} \frac{|\bz_t,\bz_{tt}|}{|\bz_t,\bz_{s}|} \ , \\ 
A:=  2\frac{|\bz_t,\bz_{st}|}{|\bz_t,\bz_{s}|}\ , &\quad 
B:= 2\frac{|\bz_s,\bz_{st}|}{|\bz_s,\bz_{t}|}\  .  
\end{align}
\ese 
We now eliminate the quantities $P$ and $Q$ as follows. First, from \eqref{eq:QPsystemb} 
and \eqref{eq:QPsystemc}, using $(Q_t)_s=(Q_s)_t$ we find (after back-substituting 
the expressions for $Q_t,Q_s$ and $P_s$ from the system):
\begin{align}  
& \left( MQ+LP\right)_s=\left(AP+BQ-P_t\right)_t \nonumber \\ 
 \Rightarrow \hspace{0.2cm} & P_{tt}=(A+M)P_t+[A_t-L_s+(B-N)L-AM]P+(B_t-M_s-KL)Q\  . 
\end{align}
Second, from \eqref{eq:QPsystema} by differentiating with respect to $t$ and back-substituting $Q_t$ 
from \eqref{eq:QPsystemb}, we get 
\begin{equation}
    P_{st}=NP_t+(N_t+KL)P+(K_t+MK) Q\  .
\end{equation}    
Using now the equality $(P_{tt})_s=(P_{st})_t$ through the above expressions for 
$P_{tt}$ and $P_{st}$, and performing the various back-substitutions for 
$Q_s,Q_t$ and $P_s$ from \eqref{eq:QPsyst} and the already obtained expressions for 
$P_{tt}$ and $P_{st}$, we arrive at: 
\begin{equation}
\begin{aligned} 
&\left( 2N_t+B_t-2M_s-A_s\right)P_t  \\ 
&+ \Big[ N_{tt}+(KL)_t+LK_t-(A+M)N_t   \\ 
& \qquad +L_{ss}-A_{st}+(AM)_s-((B-N)L)_s+A(M_s-B_t) \Big]P \\ 
&+ \Big[ K_{tt}+(MK)_t -AK_t-K(A_t-L_s) \\
& \qquad +M_{ss}+(KL)_s-B_{st}+(B-N)(M_s-B_t)\Big] Q=0\ .  
\end{aligned}
\end{equation}
Here the coefficient of $P_t$ vanishes identically, using the expressions \eqref{eq:NMKLAB}: 
\begin{equation}\label{eq:NBMA} 
2N_t+B_t= 2\pl_t( \pl_s \log \left|\bz_t,\bz_s\right|)=2M_s+A_s\ . 
\end{equation}
Setting the coefficients of $P$ and $Q$ in the remaining terms equal to zero, we get the 
two equations:
\bse\label{eq:NLMK}
\begin{align} 
& N_{tt}+L_{ss}-A_{st}+(KL)_t+(AM)_s+AM_s+ LK_t-(A+M)N_t \nonumber\\
& \qquad   -((B-N)L)_s-AB_t=0 \  , \label{eq:NL}\\ 
& M_{ss}+K_{tt}-B_{st}+(KL)_s+(BN)_t+BN_t+KL_s-(B+N)M_s \nonumber\\
& \qquad -((A-M)K)_t-A_sB=0\  , \label{eq:MK}
\end{align}
\ese 
where \eqref{eq:NLMK} is used to bring the coefficients in the symmetric form. 

First, by substituting the 
expressions \eqref{eq:NMKLAB} into \eqref{eq:NL} and \eqref{eq:MK}, we obtain the
equations \eqref{eq:fullgenPDEs}
expressed in terms of the $2\times2$ determinants of the derivatives of the vector $\bz$.

Second, to obtain the vectorial equations \eqref{eq:genPDEs} from the system \eqref{eq:NLMK}, we note that the 
terms with the highest derivative $\bz_{sstt}$ in \eqref{eq:NL} and \eqref{eq:MK} 
are
\begin{equation}
\left(\frac{t}{s}-1\right)\frac{|\bz_t,\bz_{ttss}|}{|\bz_t,\bz_{s}|}\quad {\rm and}\quad 
\left(\frac{s}{t}-1\right)\frac{|\bz_s,\bz_{sstt}|}{|\bz_s,\bz_{t}|} \ ,
\end{equation}
respectively. Thus, we can extract the vector equation from this coupled system by 
considering the combination of these equations of the form: 
\begin{equation}
s\eqref{eq:NL}\bz_s-t\eqref{eq:MK}\bz_t=(t-s)\bz_{ttss}+\cdots\ ,  
\end{equation}
where $\cdots$ stands for the remainder of \eqref{eq:genPDEs} if we normalise the equations 
such that the right-hand side is the leading term.



\medskip

\medskip

\noindent 
\noindent Frank Nijhoff

\noindent {\rm School of Mathematics, University of Leeds, Leeds LS2 9JT, United Kingdom}

\noindent  \url{f.w.nijhoff@leeds.ac.uk}

~~~~~

\noindent Linyu Peng

\noindent {\rm Department of Mechanical Engineering, Keio University, Yokohama 223-8522, Japan}

\noindent \url{l.peng@mech.keio.ac.jp}


~~~~~

\noindent Cheng Zhang

\noindent {\rm Department of Mathematics, Shanghai University, Shanghai 200444, China}

\noindent {\rm Newtouch Center for Mathematics of Shanghai University, Shanghai 200444, China}

\noindent \url{ch.zhang.maths@gmail.com}

~~~~~

\noindent Da-jun Zhang

\noindent {\rm Department of Mathematics, Shanghai University, Shanghai 200444, China}

\noindent {\rm Newtouch Center for Mathematics of Shanghai University, Shanghai 200444, China}

\noindent \url{djzhang@staff.shu.edu.cn}

\end{document}